\documentclass{aa}

\newcommand{\adam}{\texttt{ADAM}}
\newcommand{\mpcd}{\texttt{MPCD}}
\newcommand{\sage}{\texttt{SAGE}}
\newcommand{\oasis}{\texttt{OASIS}}
\newcommand{\mistral}{\texttt{Mistral}}

\usepackage{upgreek}
\usepackage{graphicx}
\usepackage{subfig}
\usepackage{threeparttable}
\usepackage{txfonts}
\usepackage[allcolors=blue]{hyperref}
\begin{document} 

   \title{The equilibrium shape of (65)~Cybele: primordial or relic of a large impact? \thanks{Based on observations made with ESO Telescopes at the Paranal Observatory under programme ID 107.22QN.001 (PI: M. Marsset)}\fnmsep\thanks{Reduced and deconvolved images listed in Table \ref{tab:ao} are available in electronic form at the CDS via anonymous ftp to cdsarc.cds.unistra.fr (130.79.128.5) or via \url{https://cdsarc.cds.unistra.fr/cgi-bin/qcat?J/A+A/}}}

   \author{M. Marsset\inst{1,2}
          \and
          M. Bro\v z\inst{3}
          \and
          J. Vermersch\inst{4}          
          \and
          N. Rambaux\inst{4}
          \and
          M. Ferrais\inst{5}
          \and
          M. Viikinkoski \inst{6}
          \and
          J. Hanu\v s\inst{3}
          \and
          E. Jehin\inst{7}
          \and
          E. Podlewska-Gaca\inst{8}
          \and
          P. Bartczak\inst{8}
        \and
        G. Dudziński\inst{8}
          \and 
          B. Carry\inst{9}
          \and
          P. Vernazza\inst{5}
        \and
        R. Szak{\'a}ts\inst{10} 
        \and
        R. Duffard\inst{11}
        \and
        A. Jones\inst{12}
        \and
        D. Molina\inst{13}
        \and
        T. Santana-Ros\inst{14,15}
        \and
        Z. Benkhaldoun\inst{16}
        \and
        M. Birlan\inst{4,17}
        \and
        C. Dumas\inst{18}
        \and
        R. F\'etick\inst{5,19}
        \and
        T. Fusco\inst{5,19}
        \and
        L. Jorda\inst{5}
        \and
        F. Marchis\inst{5,20}
        \and
        F. Vachier\inst{4}
        \and
        B. Yang\inst{1,21}
        }

   \institute{European Southern Observatory (ESO), Alonso de Cordova 3107, 1900 Casilla Vitacura, Santiago, Chile\\
              \email{mmarsset@eso.org}
         \and%2
             Department of Earth, Atmospheric and Planetary Sciences, MIT, 77 Massachusetts Avenue, Cambridge, MA 02139, USA
        \and%3
             Charles University, Faculty of Mathematics and Physics, Institute of Astronomy, V Hole\v sovi\v ck\'ach 2, CZ-18000, Prague 8, Czech Republic
        \and%4
            IMCCE, CNRS, Observatoire de Paris, PSL Université, Sorbonne Université, 77 Ave. Denfert-Rochereau, 75014 Paris, France
         \and%5
             Aix Marseille Univ, CNRS, CNES, Laboratoire d'Astrophysique de Marseille, Marseille, France
        \and%6
            Mathematics and Statistics, Tampere University, 33720 Tampere, Finland
        \and%7
            Space sciences, Technologies and Astrophysics Research Institute, Universit\'e de Li\`ege, All\'ee du 6 Ao\'ut 17, 4000 Li\`ege, Belgium
        \and%8
            Astronomical Observatory Institute, Faculty of Physics, Adam Mickiewicz University, S{\l}oneczna 36, 60-286 Pozna{\'n}, Poland
        \and%9
            Universit\'e C\^ote d'Azur, Observatoire de la C\^ote d'Azur, CNRS, Laboratoire Lagrange, France
         \and%10
             Konkoly Observatory, Research Centre for Astronomy and Earth Sciences, E\"otv\"os Lor\'and Research Network (ELKH), H-1121 Budapest, Konkoly Thege Mikl\'os \'ut 15-17, Hungary
        \and%11
            Instituto de Astrofísica de Andalucía (CSIC), Glorieta de la Astronomía s/n, 18008
             Granada, Spain.
        \and%12
            I64, SL6 1XE Maidenhead, UK
        \and%13
            Anunaki Observatory, Calle de los Llanos, 28410 Manzanares el Real, Spain
        \and%14
            Departamento de F\'isica, Ingenier\'ia de Sistemas y Teor\'ia de la Se\~nal, Universidad de Alicante,
            E-03080 Alicante, Spain
        \and%15
            Institut de Ci\`encies del Cosmos (ICCUB), Universitat de Barcelona (IEEC-UB), Carrer de Mart\'i i Franqu\`es, 1, 08028 Barcelona, Spain
        \and%16
            Oukaimeden Observatory, High Energy Physics and Astrophysics Laboratory, Cadi Ayyad University, Marrakech, Morocco
        \and%17
            Astronomical Institute of the Romanian Academy, 5-Cu\c titul de Argint, 040557 Bucharest, Romania
        \and%18
            TMT Observatory, 100 W. Walnut Street, Suite 300, Pasadena, CA 91124, USA
        \and%19
            DOTA, ONERA, Universit\'e Paris Saclay, F-91123 Palaiseau, France
        \and%20
            SETI Institute, Carl Sagan Center, 189 Bernado Avenue, Mountain View CA 94043, USA
        \and%21
        N\'ucleo de Astronom\'ia, Facultad de Ingenier\'ia y Ciencias, Universidad Diego Portales
      }

   \date{Received XXX; accepted XXX}

 \abstract{Cybele asteroids constitute an appealing reservoir of primitive material genetically linked to the outer Solar System, and the physical properties (size and shape) of the largest members can be readily accessed by large (8m class) telescopes.}
 {We took advantage of the bright apparition of the most iconic member of the Cybele population, (65)~Cybele, in July and August 2021 to acquire high-angular-resolution images and optical light curves of the asteroid with which we aim  to analyse its shape and bulk properties.}
 {Eight series of images were acquired with VLT/SPHERE+ZIMPOL, seven of which were combined with optical light curves to reconstruct the shape of the asteroid using the \adam, \mpcd,{} and \sage{} algorithms. The origin of the shape was investigated by means of N-body simulations.}
 {Cybele has a volume-equivalent diameter of {\rm 263$\pm$\,3\,km} and a bulk density of ${\rm 1.55\,\pm\,0.19\,g.cm^{-3}}$. Notably, its shape and rotation state are closely compatible with those of a Maclaurin equilibrium figure. The lack of a collisional family associated with Cybele and the higher bulk density of that body with respect to other large P-type asteroids suggest that it never experienced any large disruptive impact followed by rapid re-accumulation. This would imply that its present-day shape represents the original one. However, numerical integration of the long-term dynamical evolution of a hypothetical family of Cybele shows that it is dispersed by gravitational perturbations and chaotic diffusion over gigayears of evolution.} 
 {The very close match between Cybele and an equilibrium figure opens up the possibility that $D\,\geq\,260$\,km ($M\,\geq\,1.5\,\times10^{19}$\,kg) small bodies from the outer Solar System all formed at equilibrium. However, we cannot currently rule out an old impact as the origin of the equilibrium shape of Cybele. 
 Cybele itself is found to be dynamically unstable, implying that it was `recently' ($<$1\,Gyr ago) placed on its current orbit either through slow diffusion from a relatively stable orbit in the Cybele region or, less likely, from an unstable, Jupiter-family-comet orbit in the planet-crossing region.} 
 

   \titlerunning{Did (65)~Cybele preserve its primordial shape?}
    \authorrunning{Marsset et al.}
   \maketitle

%
%-------------------------------------------------------------------

\section{Introduction}

The Cybele region, with semi-major axis between the 2:1 (at 3.27\,au) and 5:3 (3.70\,au) mean-motion resonances (MMRs) with Jupiter,
is populated by compositionally primitive (non-igneous; C, P, and D-type) asteroids \citep{Demeo:2013}. 
Like Jupiter Trojans and Hilda asteroids (e.g. \citealt{Morbidelli:2005, Nesvorny:2013}), these objects are
thought to have formed in the outer Solar System ($>$10\,au) among the progenitors of the Kuiper Belt before being implanted in the inner Solar System during the early phase of giant planet migrations \citep{Levison:2009,Vokrouhlicky:2016}. This implies
that Cybeles could be genetically related to comets and small Kuiper Belt objects (KBOs). 
This dynamical scenario
is currently supported by the similarity in the size distributions of Trojans and small KBOs \citep{Fraser:2014} and the
similarity in the spectral properties and bulk densities between P/D-type asteroids, Trojans, and comets \citep{Emery:2006,Emery:2011, Vernazza:2015,Vernazza:2021}.

The relatively close distance of Cybeles to the Earth makes them an appealing reservoir of primitive material genetically
linked to the outer Solar System whose physical properties (size and shape) can be directly measured by large (8m class) ground-based telescopes.
The ability of the adaptive-optics (AO) Spectro-Polarimetric High-Contrast Exoplanet Research (SPHERE) instrument \citep{Beuzit:2019} on the Very Large Telescope (VLT) to decipher
the origin and thermal history of Cybeles was recently demonstrated by observations of two of the largest of these bodies: (87) Sylvia (D$\simeq$274\,km; \citealt{Carry:2021}) and (107) Camilla (D$\simeq$254\,km; \citealt{Pajuelo:2018}). 
These observations led to measurements of low bulk densities ($\simeq$1.3\,g.cm$^{-3}$)
consistent with a very pristine composition. 
However, they did not allow an assessment of the original shape of the planetesimals formed in the outer Solar System. 
Indeed, both Camilla and Sylvia experienced a violent collisional past, which was revealed by the existence of two small satellites orbiting each of these objects \citep{Marchis:2005,Pajuelo:2018}, as well as of a collisional family in the case of Sylvia \citep{Vokrouhlicky:2010}. 
As such, their present-day shapes are most likely collisionally evolved and do not represent the original ones. 

With a measured diameter from 240 to 300 km (e.g. \citealt{Muller:2004, Nugent:2016, Viikinkoski:2017}), (65)~Cybele is another large member of the Cybele population (as its name subtly suggests). 
Unlike Camilla and Sylvia, Cybele is not known to host any satellite\footnote{In 1979, a hint of a possible 11km wide companion at 917\,km from Cybele was reported during an occultation
(IAUC 3439), but this was never corroborated.}, nor is it associated with any collisional family. 
As such, it is possible that Cybele has maintained its original shape to this day. 
Cybele is also one of the first main-belt asteroids, along with (24)~Themis \citep{Campins:2010, Rivkin:2010}, where water ice and organics have been claimed to be detected on the surface \citep{Licandro:2011}. 
However, this claim currently remains a matter of debate (e.g. \citealt{Beck:2011, Orourke:2020}). 

July 2021 offered the best opportunity within the subsequent 6 years to scrutinise Cybele in unprecedented detail. 
We took advantage of ESO's P107 Special Call for the submission of time-critical projects ---which followed the suspension of the regular Call for Proposals due to the COVID-19 pandemic--- 
to acquire high-angular-resolution images of Cybele in order to reconstruct the shape and bulk properties of the asteroid. 

In Section\,\ref{sec:obs}, we present our observations of new images and optical light curves acquired for Cybele. In Section\,\ref{sec:shape}, we describe the shape reconstruction achieved using our observations and archival data available for the asteroid. 
In Section\,\ref{sec:results}, we demonstrate that the shape and rotation state of Cybele match an equilibrium figure and we investigate possible origins for this equilibrium state by means of N-body simulations. 
Finally, we present our conclusions in Section\,\ref{sec:conclusion}.

\section{Observations}
\label{sec:obs}

\subsection{Disk-resolved images}
\label{sec:ao}

With an apparent V-band magnitude of 11.1 and an angular diameter of 0.16", the conditions of Cybele's opposition on July 2021 were ideal for conducting AO imaging of the large asteroid with the VLT. 
Eight series of five 240 s images were acquired with the SPHERE+Zurich Imaging Polarimeter (ZIMPOL) instrument \citep{Thalmann:2008, Beuzit:2019} on July 3, 21, and August 8, 2021. 
All observations were obtained in the mode of classical imaging with the N$\_$R narrow band filter (central wavelength = 645.9 nm, width = 56.7 nm), using Cybele as a natural guide star for real-time AO corrections. 
All images were collected under seeing conditions of better than 0.8\arcsec and an airmass of below 1.5. 
The apparent geometry of Cybele during our observations was almost equator-on (aspect angle $\simeq$90--98\degr), enabling accurate measurements of its three-dimensional axis and overall shape. 
See Appendix~\ref{sec:appA} for a complete description of the observing circumstance of the AO images. 

\begin{figure*}
   \centering
   \includegraphics[width=\hsize]{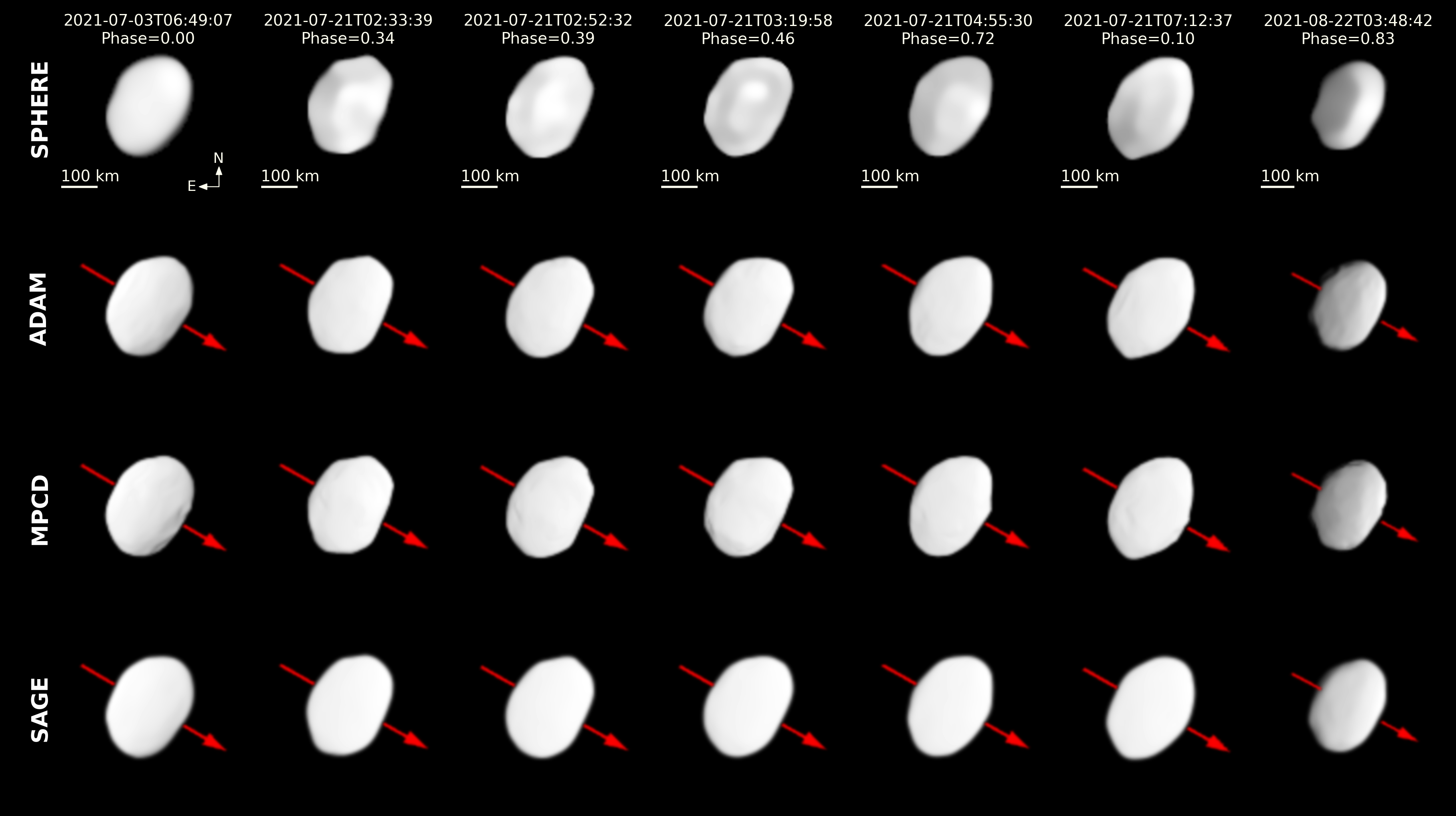}
      \caption{Comparison between the deconvolved images of (65)~Cybele (top row) and the corresponding synthetic images generated by the \oasis{} tool \citep{Jorda:2010} of the \adam{}, \mpcd,{} and \sage{} shape models. 
      The red arrows indicate the direction of the spin axis. Observing conditions for the images are listed in Appendix\,\ref{sec:appA}. Residuals between the observed and synthetic images are shown in Appendix~\ref{sec:appB}.}
         \label{fig:mosaic}
   \end{figure*}

We reduced our data with ESO's pipeline Esorex following the steps described in \citet{Vernazza:2018}. Image deconvolution was then performed using the \mistral~deconvolution algorithm \citep{Fusco:2003, Mugnier:2004}, and a parametric point-spread function with a Moffat profile \citep{Moffat:1969}. We refer the reader to \citet{Fetick:2019} for information about the reliability of this method. One series of lower-quality images acquired on July 3 was discarded from our analysis as it could have compromised the reconstruction of the shape of the  asteroid. The deconvolved images acquired in the other seven epochs are shown in Fig.~\ref{fig:mosaic}.

%%%%%%%%%% Start Figure %%%%%%%%%%%%%%%%%%%%%%%%
%%%%%%%%%% End Figure %%%%%%%%%%%%%%%%%%%%%%%%

\subsection{Optical disk-integrated photometry}\label{sec:photometry}

\begin{figure*}
   \centering
   \includegraphics[width=0.8\hsize]{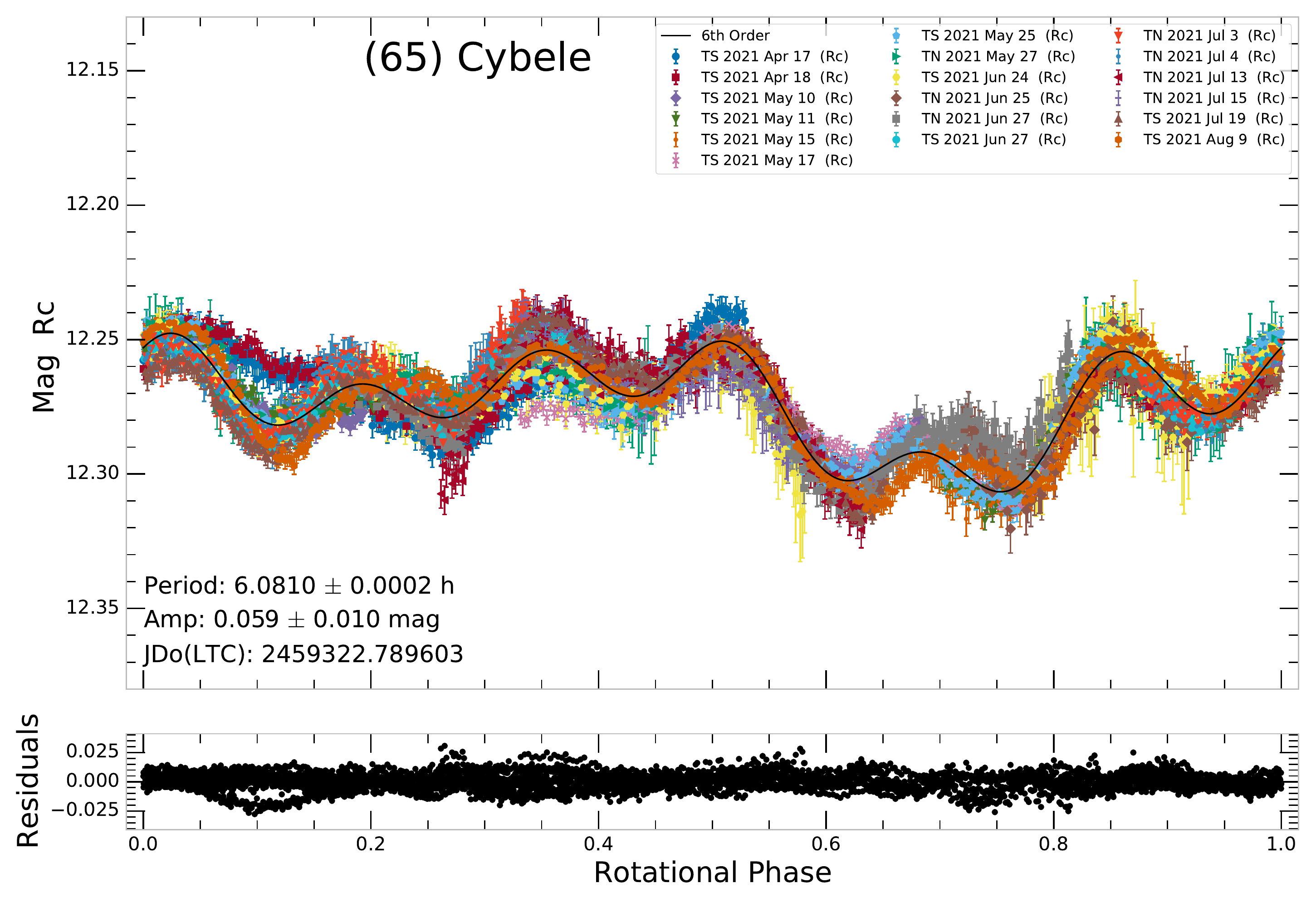}
      \caption{Composite light curve of (65)~Cybele obtained between April and August 2021 with the TRAPPIST-North and South telescopes. The solid line shows a sixth-order polynomial fitted to the data. The residuals of the fit are shown in the bottom panel. The large residuals are due to the long time span of the photometric measurements, during which the orientation of the object significantly varied with respect to the Earth. 
      Observing conditions for these data and additional light curves obtained with other telescopes are described in Appendix\,\ref{sec:appA}.}
         \label{fig:LCs}
   \end{figure*}

Optical light curves are particularly important for the spin period determination and a proper phasing of the AO images.   Figure \ref{fig:LCs}  shows the   18 light curves  that we acquired during Cybele's apparition between April and August 2021 with the 60-cm TRAPPIST-North and South telescopes \citep{Jehin:2011}. 
We also used 16 additional light curves obtained by amateur observers with small telescopes via the GaiaGOSA 
service \citep{Santana:2016}. The data gathered during two oppositions were supported by
 observations taken in La Sagra (IAA CSIC, Spain) and Piszk\'estet\H{o} (Hungary) observatories. 
We also made use of archival data obtained by \citet{Schober1980}, \citet{Weidenschilling1987, Weidenschilling1990}, \citet{Hutton1990}, \citet{Lagerkvist1995},  \citet{Shevchenko1996}, and  \citet{Pilcher2010e, Pilcher2011f, Pilcher2012c} and data available on the CdR database\footnote{\url{https://obswww.unige.ch/~behrend/page\_cou.html}}. These archival data were previously used for shape modelling of Cybele by \citet{Franco2015} and \citet{Viikinkoski:2017}.
Information about the photometric data are provided in Appendix~\ref{sec:appA}.

%%%%%%%%%% Start Figure %%%%%%%%%%%%%%%%%%%%%%%%
%%%%%%%%%% End Figure %%%%%%%%%%%%%%%%%%%%%%%%

\section{Three-dimensional shape reconstruction}
\label{sec:shape}

We first used the All-Data Asteroid Modelling (\adam{}) inversion technique \citep{Viikinkoski:2015, Viikinkoski:2016} for the reconstruction of the 3D shape model and the spin of Cybele using our disk-resolved (images) and disk-integrated (optical light curves) data as inputs. The \adam{} technique is a well-described inversion algorithm that has previously been applied to tens of asteroids \citep[e.g.][]{Viikinkoski:2015b, Viikinkoski:2017, Viikinkoski:2018, Hanus:2017, Hanus:2019, Marsset:2017, Marsset:2020_Pallas, Vernazza:2018, Vernazza:2020, Vernazza:2021, Carry:2019, Carry:2021, Ferrais:2020, Yang:2020}. Exhaustive information about this modelling technique can be found in these latter studies. 

The first iteration of the shape-optimisation procedure produced an equatorial indentation in the shape model of Cybele which was not clearly seen in the data. This typically indicates that the shape optimisation converged to a suboptimal solution. 
The usual regularisation methods \citep{Viikinkoski:2015b} used for alleviating surface artefacts did not lead to a satisfying outcome in this case, and so instead we increased the degrees of freedom by allowing the albedos of surface facets to vary by $\pm$20\% with respect to their nominal values. This approach allowed the optimisation process to escape from the local minimum and converge. We note that neither light curves nor AO images exhibit significant albedo variegation, and so in this case facet brightness variegation has no physical meaning beyond being a way of optimising the modelling solution.

The ADAM model was then refined using the Multi-resolution PhotoClinometry by Deformation (\mpcd) method \citep{Capanna:2013, Jorda:2016}, following the procedure described in \citet{Ferrais:2020}. \mpcd~uses the measured brightness of the AO images and gradually deforms the vertices of the 3D mesh in order to fit the synthetic images of the model to the observed ones. For the reflectance properties of Cybele, we used the geometric albedo of 0.059 from \citet{Mainzer2016} and typical Hapke parameters for C-type asteroids \citep{Helfenstein1989}. The initial \adam~model was only slightly modified by this procedure (Section\,\ref{sec:results}). 

The shape of Cybele was also independently reconstructed using the Shaping Asteroid models using Genetic Evolution (\sage{}) method \citep{Bartczak:2018, Dudzinski:2020} in order to assess the robustness of the \adam{} and \mpcd{} solutions. We used the same set of AO images and light curves as for the \adam~model. In \sage, only information about the silhouette of the body is used from the images (as opposed to \adam{} and \mpcd,{} which use image brightness). 
Uncertainties on the local shape of the model were assessed by creating clones of the nominal shape model and accepting the ones fitting the set of images and light curves within an acceptable confidence level \citep{Bartczak:2019}. 
Complementary information about the procedure is provided in Appendix\,\ref{sec:appC}. 

Table~\ref{tab:param} provides the final values for the spin-axis orientation, sidereal rotation period, volume-equivalent diameter, and dimensions along the major axes of Cybele. 
Projections of the \adam{}, \mpcd,{} and \sage{} shape models generated by the \oasis~software \citep{Jorda:2010} and with similar geometry to the SPHERE images are shown in Fig.\,\ref{fig:mosaic}. 
Images of the residuals between the observed and synthetic images are provided in Appendix~\ref{sec:appB}.

%Added by TeX Support
\begin{table}
 \caption{\label{tab:param}
 Physical properties of (65)~Cybele based on the \adam{}, \mpcd{}~and \sage{}~shape modeling methods.} %\textcolor{red}{Needs to be completed in a consistent way, using the same measurement/uncertainty method for all three models.}}
 \centering
 \begin{threeparttable}
% \centering
 \begin{tabular}{llll}
  \hline
  Parameter  & \adam{}   & \mpcd{} & \sage{} \\
  \hline
  \hline
   %$ (h)   & \multicolumn{3}{c}{4.148200(1)} \\\ ,$\pm$\,0.000005
   \\
   $P$ (h)              & 6.081433          & --``--            & 6.081433   \\              \\
   $\lambda$ ($\degr$)  & 204\,$\pm$\,3     & --``--            & 203\,$\pm$\,2            \\
   $\beta$   ($\degr$)  &--19\,$\pm$\,2     & --``--            & --20\,$\pm$\,1           \\\\
   $D$ (km)             & 263\,$\pm$\,3     & 263\,$\pm$\,3      &  {$263.9^{+4.8}_{-3.3}$}     \\\\
   $a$ (km)             & 297\,$\pm$\,3     & 296\,$\pm$\,3       & 296\,$\pm$\,7            \\
   $b$ (km)             & 291\,$\pm$\,3     & 290\,$\pm$\,3       & 292\,$\pm$\,7            \\
   $c$ (km)             & 213\,$\pm$\,3     & 213\,$\pm$\,3       & 213\,$\pm$\,7            \\
   $a/b$                & 1.02\,$\pm$\,0.01 & 1.02\,$\pm$\,0.02   & 1.01\,$\pm$\,0.04        \\
   $b/c$                & 1.37\,$\pm$\,0.02 & 1.36\,$\pm$\,0.04   & 1.46\,$\pm$\,0.06        \\
   $a/c$                & 1.39\,$\pm$\,0.02 & 1.39\,$\pm$\,0.04   & 1.46\,$\pm$\,0.06        \\
%   $\rho$ (g.cm$^{-3}$) & 1.46\,$\pm$\,0.36 & \,$\pm$\, & 1.46\,$\pm$\,0.37 \\
   $\rho$ (g.cm$^{-3}$) & 1.55\,$\pm$\,0.19 & 1.55\,$\pm$\,0.19 & 1.54\,$\pm$\,0.25 \\
  \hline
 \end{tabular}
 \end{threeparttable}
 \tablefoot{The listed parameters are : sidereal rotation period $P$, spin-axis ecliptic J2000 coordinates $\lambda$ and $\beta$, volume-equivalent diameter $D$, dimensions along the major axis of the best-fit triaxial ellipsoids ($a$ > $b$ > $c$), their ratios $a/b$, $b/c$ and $a/c$, and bulk density $\rho$. Uncertainties correspond to 1-$\sigma$ values.
 }
\end{table}

\section{Results}
\label{sec:results}

\subsection{Overall shape and putative impact features}
\label{sec:craters}

The three shape models of Cybele are very close to an oblate spheroid with two nearly equal-size equatorial radii (Table\,\ref{tab:param}). 
The volume-equivalent diameter (${\rm D_{eq}}$) of ${\rm 263\pm\,3\,km}$ of the \adam{} and \mpcd{} models is very close to that obtained with \sage{}, with ${\rm D_{eq}\,\simeq\,263.9^{+4.8}_{-3.3}}$\,km. 

No obvious large excavations could be identified in the images and in the shape model of Cybele. However, localised flattened areas that may correspond to craters can be seen in some images, as well as in the elevation map of the asteroid (Appendix\,\ref{sec:elev}). 
Future pole-on observations of Cybele are needed in order to investigate the presence of craters near its equator.

The elevation map of Cybele further reveals a hemispherical asymmetry, with the northern hemisphere being flatter than the southern one. 
This feature could be the remnant of a non-disruptive polar impact. 
Future higher-resolution images taken with 30 to 40m class telescopes may help in further  investigations of the morphology and possible collisional origin of the flattened north pole of Cybele.

\subsection{Mass, density, and interior}
\label{sec:density}

To further explore the internal structure and equilibrium shape of Cybele, we needed to assess its bulk density. The estimated mass of Cybele was retrieved by compiling available measurements from the literature (18 in total; Table\,\ref{tab:masses}), obtained via studies of planetary ephemeris and orbital deflections during close encounters (e.g. \citealt{Carry:2012}). 

\begin{figure}
   \centering
   \includegraphics[width=\hsize, trim={0cm 0 0 0cm}, clip]{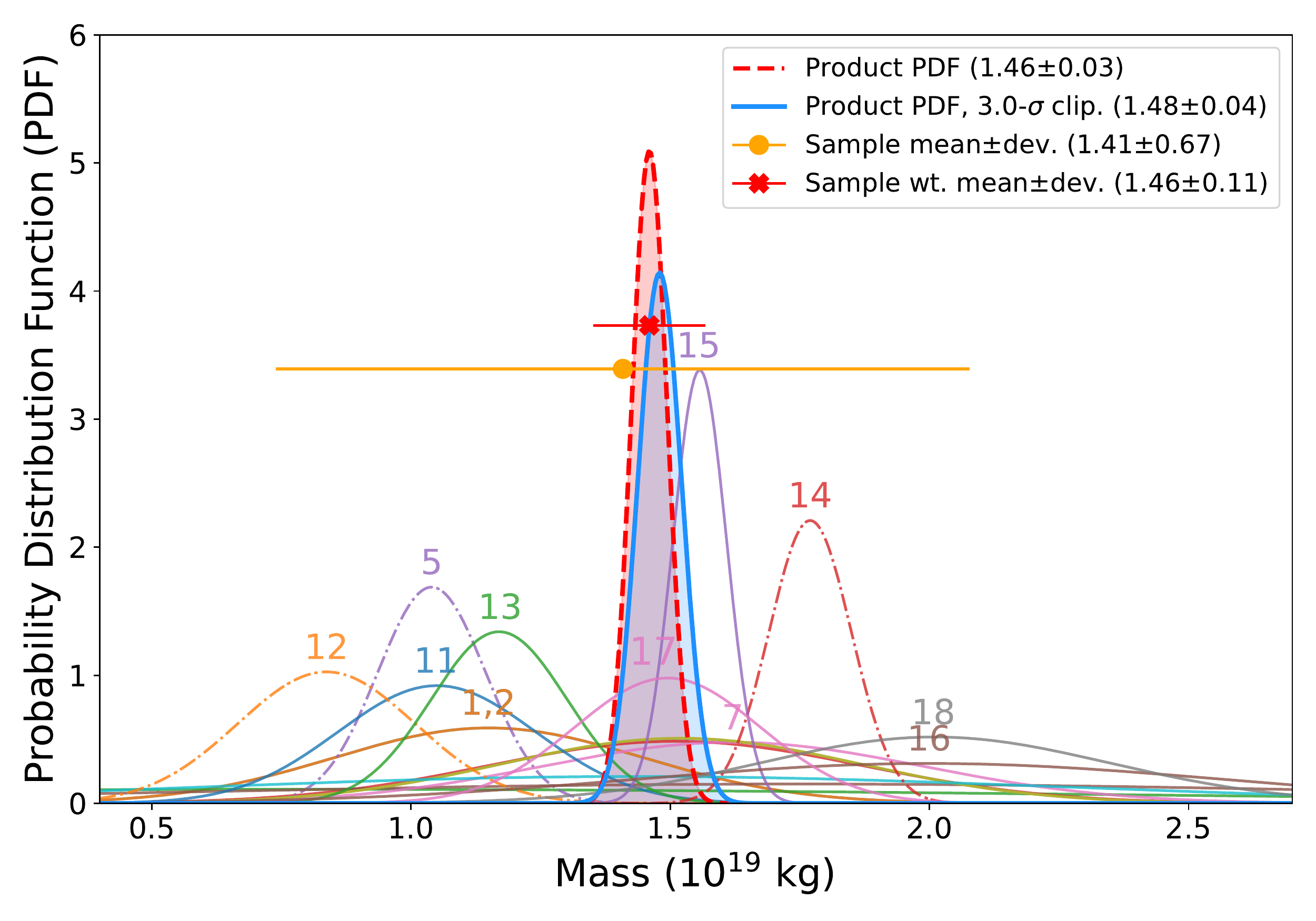}
      \caption{Mass measurements of (65)~Cybele from Table\,\ref{tab:masses} converted to Gaussian probability distribution functions (PDFs). Numbers correspond to references listed in the table (only a subset of reference numbers is displayed for improved readability). 
      The adopted mass estimate was obtained by computing the product PDF as described in Section\,\ref{sec:density}. Continuous lines indicate measurements used to compute the mass estimate, and dashed lines correspond to discarded measurements. The filled curves are the product PDFs before (dashed red) and after (blue) rejection of the discarded measurements. The orange and red circles show the unweighted and weighted sample averages and standard deviations, respectively.} 
         \label{fig:masses}
   \end{figure}

First, three mass estimates found to be inconsistent with the weighted sample average at the 3-$\sigma$ level were discarded (where $\sigma$ corresponds to the reported uncertainty on the estimate). The remaining 15 measurements were converted into Gaussian probability distribution functions (PDFs) with mean and standard deviation equal to the value and uncertainty on the measurement, respectively (Fig.\,\ref{fig:masses}). Next, the combined mass estimate was obtained by multiplying the distributions by one another, $\prod_{i=1}^{15} {PDF}_i$, and measuring the mean and deviation of the resulting product PDF. Following this approach, we obtained a mass of (1.48$\pm$0.04)$\,\times\,10^{19}$\,kg for Cybele. 

However, considering the scatter of reported mass values from the literature, we suspected that our uncertainty might be underestimated, possibly due to incorrectly derived statistical uncertainties and/or unaccounted-for systematic uncertainties affecting the literature measurements. Therefore, in order to assess the reliability of our estimate, we compiled mass measurements for a number of large asteroids harbouring one or several satellite(s), and applied the same method to derive their mass as that applied to Cybele (Appendix\,\ref{sec:appBC}). By doing so, we find that the derived masses for these objects agree on average to within 12\% with the value measured from the orbital study of the satellite(s). Adopting this uncertainty leads to a final mass estimate of (1.48$\pm$0.18)$\,\times\,10^{19}$\,kg for Cybele. 

Combined with our measurement of its diameter, the estimated mass of Cybele yields a bulk density of 1.55$\pm$0.19\,g.cm$^{-3}$. This low value is comparable to those measured for other large C- and P-type asteroids \citep{Carry:2012, Vernazza:2021}, and is compatible with a water-rich interior. This agrees with Cybele's apparently low surface topography, which is suggestive of a relaxed, water-rich subsurface. Cybele's bulk density is nevertheless slightly larger than that of the other two large (D$\geq$250 km) P-type Cybele asteroids with accurately measured densities: (87)~Sylvia and (107)~Camilla, with $\rho\,\simeq$\,1.3--1.4\,g.cm$^{-3}$ \citep{Pajuelo:2018, Carry:2021}.

\begin{table}
\caption{Mass estimates of (65)~Cybele from the literature.}         
\label{tab:masses}      
\small  
\centering          
\begin{tabular}{c c l c}     % 7 columns 
\hline       
 Mass  &    Method  & Reference & \#   \\
 ($\times{10}^{19}$\,kg)  &   &   &   \\
 \hline\hline
 1.15  $\pm$ 0.30    & DEFLECT & \citealt{Chernetenko:2002} & 1 \\
 1.15  $\pm$ 0.30    & DEFLECT & \citealt{Kochetova:2004}   & 2 \\
 0.80  $\pm$ 1.59    & DEFLECT & \citealt{Ivantsov:2007}    & 3 \\
 1.51  $\pm$ 0.36    & DEFLECT & \citealt{Baer:2008}        & 4 \\
 1.04  $\pm$ 0.10    & EPHEM   & \citealt{Folkner:2009}     & 5 \\
 1.43  $\pm$ 0.85    & EPHEM   & \citealt{Fienga:2011}      & 6 \\
 1.75  $\pm$ 1.16    & DEFLECT & \citealt{Zielenbach:2011}  & 7 \\
 1.62  $\pm$ 0.37    & DEFLECT & \citealt{Zielenbach:2011}  & 8 \\
 1.52  $\pm$ 0.34    & DEFLECT & \citealt{Zielenbach:2011}  & 9 \\
 1.52  $\pm$ 0.35    & DEFLECT & \citealt{Zielenbach:2011}  & 10 \\
 1.05  $\pm$ 0.19    & DEFLECT & \citealt{Baer:2011}        & 11 \\
 0.84  $\pm$ 0.17    & EPHEM   & \citealt{Fienga:2013}      & 12 \\
 1.17  $\pm$ 0.13    & DEFLECT & \citealt{Kochetova:2014}   & 13 \\
 1.77  $\pm$ 0.08    & DEFLECT & \citealt{Goffin:2014}      & 14 \\
 1.56  $\pm$ 0.05    & EPHEM   & \citealt{Baer:2017}        & 15 \\
 1.50  $\pm$ 0.18    & EPHEM   & \citealt{Baer:2017}        & 16 \\
 2.00  $\pm$ 0.56    & EPHEM   & \citealt{Viswanathan:2017} & 17 \\
 2.01  $\pm$ 0.34    & EPHEM   & \citealt{Fienga:2019}      & 18 \\
 \hline\hline
1.48  $\pm$ 0.18    &         & Adopted estimate (see Section\,\ref{sec:density}) \\
\hline                  
\end{tabular}
\end{table}

%%%%%%%%%% Start Figure %%%%%%%%%%%%%%%%%%%%%%%%
%%%%%%%%%% End Figure %%%%%%%%%%%%%%%%%%%%%%%%

\subsection{Maclaurin equilibrium shape}
\label{sec:equilibrium}

Next, we tested the compatibility of the 3D shape of Cybele with a hydrostatic equilibrium figure by following the approach developed in \citet{Rambaux:2015, Rambaux:2017} and presented in some of our previous works (e.g. \citealt{Hanus:2020, Marsset:2020_Pallas, Yang:2020, Vernazza:2020, Vernazza:2021}).

\begin{figure}
\centering
\includegraphics[width=\hsize, trim={3cm 0cm 3cm 0cm}, clip]{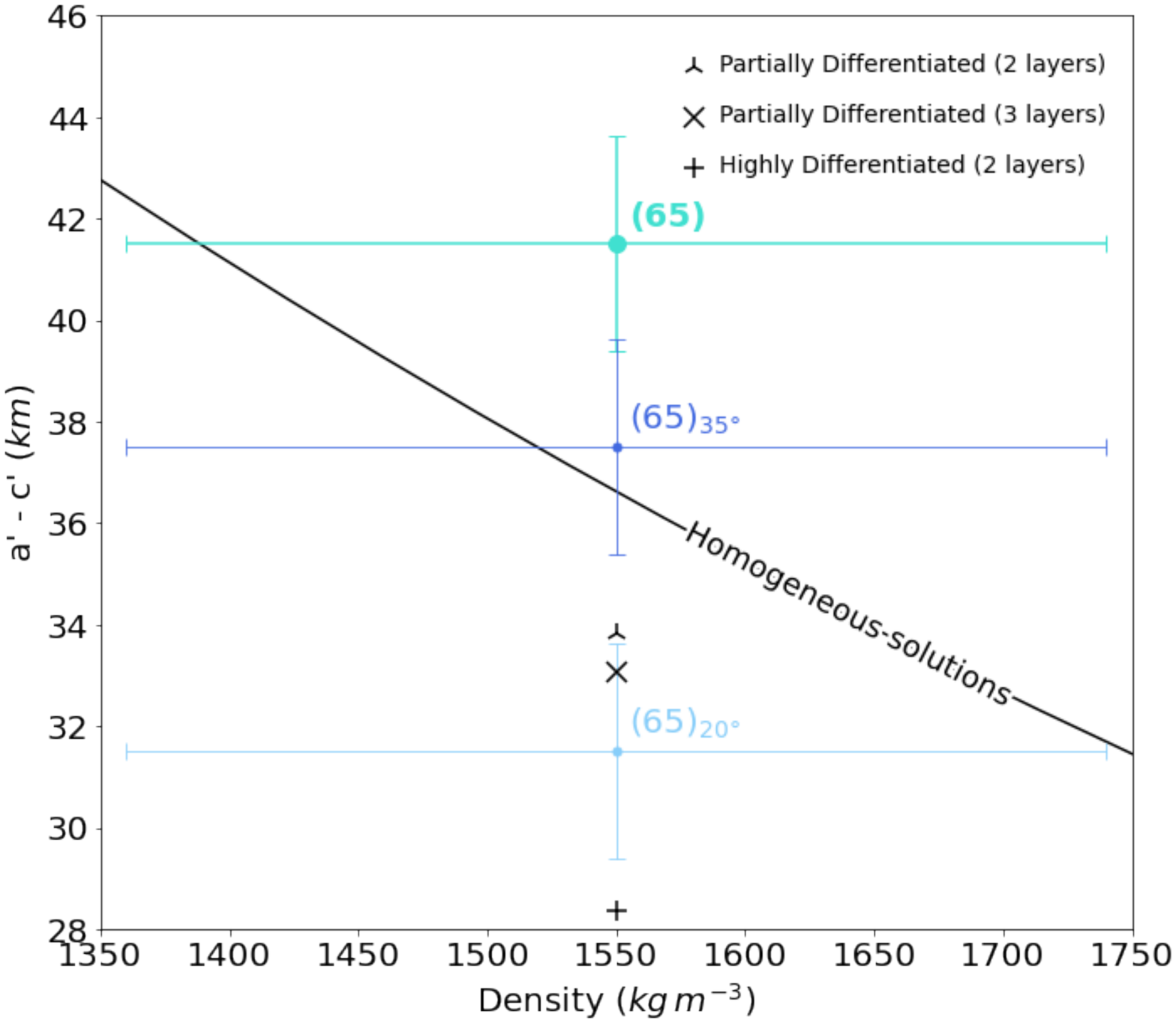}
\caption{(a'-c') dimension of (65)~Cybele as a function of density (top cyan circle). The polar oblateness of the asteroid is close to the expected values for a homogeneous body with a similar rotation period (black line). The expected oblateness of bodies with increasingly differentiated interiors (`X' symbols) gradually depart from the measured value. A highly differentiated interior (plus symbol) is unlikely, unless Cybele's rotation state significantly changed after it acquired its shape. Considering the flattening of the north pole of Cybele, we also computed best-fit ellipsoids obtained by fitting the southern pole of the 3D model only. The resulting (a'-c') dimensions for fits up to planetocentric latitudes of 20$^{\circ}$ south and 35$^{\circ}$ south are shown by the middle blue and bottom light-blue circles.}
\label{fig:amc}
\end{figure}

First, the equilibrium figure was calculated using the Clairaut equation \citep{Rambaux:2015}, which provides the Maclaurin solution for homogeneous bodies. 
By doing so, we find that Cybele's present-day shape is compatible with hydrostatic equilibrium, although only with the lower end of the error bar on our density measurement (Fig.\,\ref{fig:amc}). A homogeneous interior provides the closest match to the measured properties. A partially differentiated interior, analogous to that proposed for (87)~Sylvia \citep{Carry:2021}  for
example,
cannot be ruled out, but is more unlikely. This is illustrated in Fig.\,\ref{fig:amc}, where we report the expected (a'-c') dimension of a differentiated body with similar size, bulk density, and angular velocity as Cybele (here, the primes indicate the body's \textit{radius} along its main axis). 
Specifically, we explored three possibilities: a partially differentiated two-layer body with layer densities of 1.60 and 1.22\,g.cm$^{-3}$, a partially differentiated three-layer body with layer densities of 1.80, 1.60, and 1.22\,g.cm$^{-3}$ similar to Sylvia \citep{Carry:2021}, and a highly differentiated two-layer body with layer densities of 3.00 and 1.22\,g.cm$^{-3}$, corresponding to a rocky core and an icy shell. 
By doing so, we find that increasingly differentiated solutions gradually depart from the measurements, making them more unlikely than a homogeneous interior, unless the rotation state of Cybele was significantly altered in the past. From a thermophysical modelling point of view, both an undifferentiated and a differentiated interior are possible for an object of the size of Cybele, the outcome of the modelling being strongly dependent on the formation time and initial composition of the body (Castillo-Rogez, priv. comm.).

We then considered the fact that Cybele exhibits a hemispherical asymmetry (Section\,\ref{sec:craters}), with the northern hemisphere being flatter than the southern one (Appendix\,\ref{sec:elev}). This may indicate that the southern hemisphere represents the fossil shape of Cybele, while the northern hemisphere is collisionally evolved. To explore this possibility, we calculated the best-fit ellipsoid of Cybele by fitting the southern hemisphere only, up to planetocentric latitudes of 20$^{\circ}$ south and 35$^{\circ}$ south. The results are shown in Fig.\,\ref{fig:amc}. The 35$^{\circ}$ south best-fit ellipsoid provides a perfect match to the expected oblateness for an homogeneous interior. The 20$^{\circ}$ south best-fit ellipsoid, on the other hand, offers a closer match to the solutions computed for a partially differentiated interior. Therefore, it is currently hard to conclude on the internal structure of Cybele. Ideally, we would need to study its gravitational field by means of an orbiter.

Next, we computed the radial differences between
the \mpcd{} shape model of Cybele and a best-fitting ellipsoid,
and derived the (model - ellipsoid) average residual (in \%). By doing so, we find that the residual for Cybele (2.5\,\%) is very close to that of other objects with equilibrium shapes: (10)~Hygiea (1.3\,\%; \citealt{Vernazza:2020}) and (704)~Interamnia (2.3\,\%; \citealt{Hanus:2020}).

Finally, we compared the specific angular momentum ({\it \^L}) and
the normalised angular velocity ($\hat{\omega}$) of Cybele 
with the expected values for Maclaurin and Jacobi ellipsoids (Appendix\,\ref{sec:w_L}). Cybele falls exactly along the Maclaurin sequence, like most water-rich asteroids observed with SPHERE \citep{Vernazza:2021}. In short, the similar equatorial dimensions (a$\simeq$b) of Cybele, its polar oblateness, the good adjustment of its 3D shape to an oblate ellipsoid, and its rotation state are all highly compatible with a Maclaurin equilibrium shape. 

%%%%%%%%%% Start Figure %%%%%%%%%%%%%%%%%%%%%%%%
%%%%%%%%%% End Figure %%%%%%%%%%%%%%%%%%%%%%%%

\subsection{Nature or nurture?}

Two scenarios may explain the equilibrium shape of Cybele. 
In the first scenario, this shape was acquired shortly after the formation of the asteroid, about 5 Myr after the condensation of calcium-aluminium-rich inclusions (CAIs; \citealt{Neveu:2019,Carry:2021}). Radioactive decay of short- and long-lived radionuclides allowed partial melting of the interior of Cybele and mass redistribution through water percolation and/or relaxation of hot silicates. 
No large, subcatastrophic impact subsequently altered the shape of Cybele in a significant way. 
In that scenario, the present-day shape of Cybele  would be primordial. 

In the second scenario,  the equilibrium shape of Cybele was acquired following a giant impact, similar to the cases of (10)~Hygiea \citep{Vernazza:2020} and (31)~Euphrosyne \citep{Yang:2020}. 
Following the impact, macroscopic oscillations drove the material to behave like a fluid \citep{Tanga:2009}, naturally resulting in the formation of a nearly oblate object in rotational equilibrium. 
Fragmentation and re-accumulation further erased any previous impact feature from the surface of the asteroid.

A possible hint in favour of the primordial shape hypothesis may come from Cybele's higher bulk density with respect to Sylvia and Camilla (Section\,\ref{sec:density}). This difference could be explained by the fact that both Sylvia and Camilla experienced a large impact as evidenced by the existence of their satellites and, in the case of Sylvia, its associated collisional family. Such impact would have increased the internal porosity of these bodies following partial or complete fragmentation and subsequent re-accumulation. The higher density of Cybele, on the other hand, may indicate that this body never experienced such a large impact and therefore that it may have preserved its primordial shape.

In an attempt to further distinguish between the two scenarios mentioned above, we investigated the existence of a collisional family associated to Cybele. If such a family ever existed, this would be evidence that the shape of Cybele was reset by a large collision. Otherwise, its shape would most likely be primordial.

\subsubsection{The lack of observed Cybele family}

\begin{figure}
\centering
\includegraphics[width=9cm]{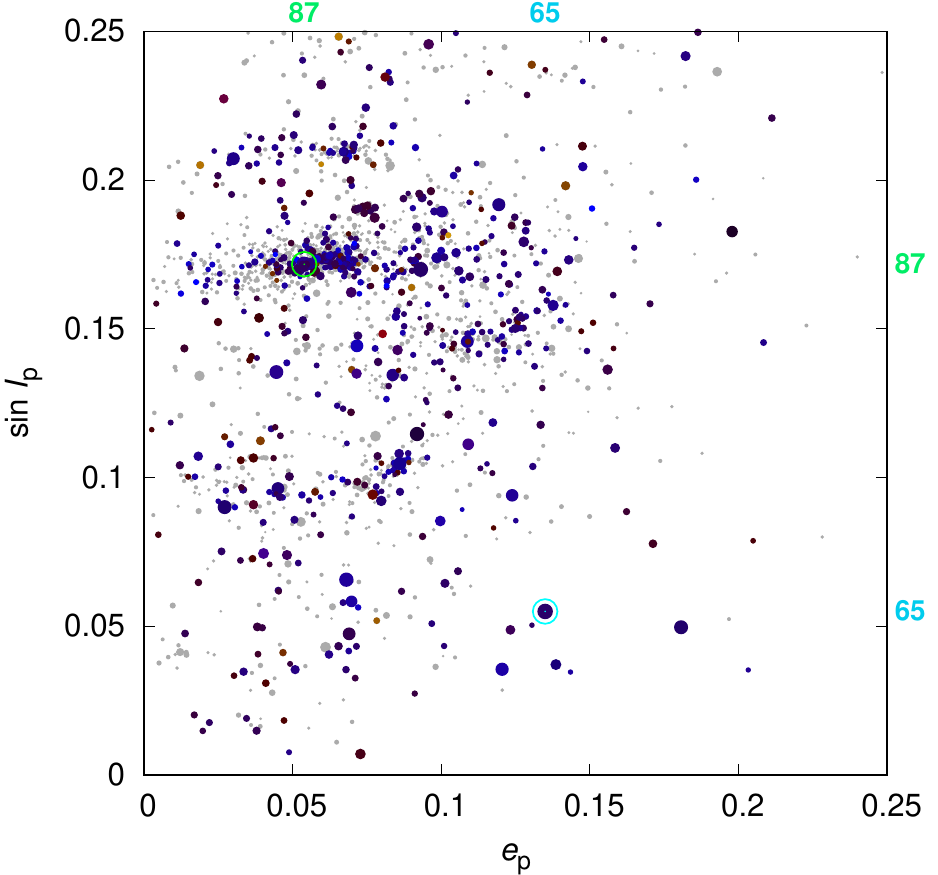}
\hbox to 9cm{\hfil\includegraphics[width=7.45cm]{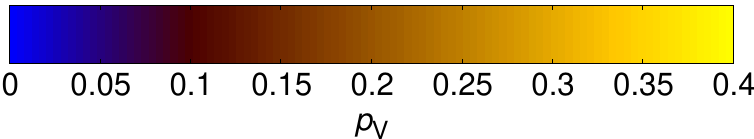}\kern.25cm}
\caption{
Observed proper eccentricities and proper inclinations of (65) Cybele
and other asteroids in its vicinity.
We do not find evidence for the existence of a family related to Cybele located at $e_{\rm p} = 0.13$ and $\sin I_{\rm p} = 0.05$ (number `65' along the axis).
For comparison, (87) Sylvia and its well-known family are clearly identifiable at $e_{\rm p} = 0.06$ and $\sin I_{\rm p} = 0.17$ (number `87' along the axis). 
Colours correspond to albedo values from the WISE and AKARI catalogues. Asteroids without albedo measurements are plotted in grey.
Symbol sizes are proportional to the logarithm of the diameter. 
}
\label{fig:ei_wise}
\end{figure}

To study the area of space surrounding Cybele, we used
a recent version of the catalogue of proper elements
\citep{Knezevic_Milani_2003A&A...403.1165K,Novakovic_2019EPSC...13.1671N}
as well as available size and albedo measurements from WISE \citep{Nugent_2015ApJ...814..117N} and Akari \citep{Usui_2011PASJ...63.1117U}.
The orbital region of Cybele is plotted in Fig.~\ref{fig:ei_wise}.
The well-known Sylvia family \citep{Vokrouhlicky_2010AJ....139.2148V}
is clearly seen around proper eccentricity and proper inclination values of $e_p = 0.06$ and $sin\,I_p = 0.17$.
No comparable clustering is found near Cybele, which is 
located at $e = 0.13$ and $sin\,I_p = 0.05$. 

Considering the completeness limit of small bodies in the outer belt (H$\simeq$16; \citealt{Hendler:2020}), we cannot rule out the existence of a family of small (subkilometre sized) asteroids linked to Cybele. However, even subcatastrophic impacts are sufficient to produce tens of multi-kilometre sized fragments (e.g. \citealt{Broz:2022}). A small family hidden below the size completeness limit would therefore certainly correspond to a small cratering impact, and not to a disruptive and re-accumulating event.

More generally, we note the scarcity of objects in the direct neighbourhood of Cybele, which is most likely due to the presence of a series of MMRs with Jupiter (21:11, 19:10, 17:9, 15:8, 13:7, 11:6, 20:11, 9:5) and three-body resonances with Jupiter and Saturn
($5{+}2{-}3$, $5{-}3{-}2$, $6{-}1{-}3$) located between $a = 3.38$ and $3.52\,{\rm au}$.
Closer to the orbit of Jupiter, the resonances get closer to each other and, at $e \gtrsim 0.2$, they begin to overlap.

The outer limit of the main belt is usually considered to be the 2:1 MMR at $a = 3.27\,{\rm au}$, but the actual border is located farther out, where resonance overlapping and chaotic diffusion dominate\footnote{An aphelion condition for the Hill sphere of Jupiter, $Q = a(1+e) = 4.61\,{\rm au}$, is at even larger eccentricities ($e \gtrsim 0.4$).}.
Cybele is located very close to this limit, meaning that if a family ever existed in its vicinity, it may have been scattered beyond recognition over time spans significantly shorter than the age of the Solar System. 
To investigate this possibility, we tested the orbital stability of a putative family near the current location of Cybele in the outer belt.

\subsubsection{Stability of a putative Cybele family}

In order to address orbital stability, we numerically integrated a synthetic family of orbits located in the vicinity of Cybele.
Initial conditions for the planets were taken from the JPL DE405 ephemerides
\citep{Giorgini_1996DPS....28.2504G}.
We included only the four giant planets in our simulation,
and applied a barycentric correction
for the terrestrial planets, Ceres, and Vesta.
We rotated our reference frame so that it coincides with the Laplace plane.

We used the SWIFT integrator \citep{Levison_Duncan_1994Icar..108...18L},
namely the symplectic algorithm with non-symplectic close encounters (`RMVS3').
In addition to gravitational perturbations, our dynamical model, which is described in \cite{Broz_2011MNRAS.414.2716B}, incorporates  
the Yarkovsky effect,
the YORP effect,
collisional reorientations, and
critical rotation
with size-dependent tensile strength
\citep{Holsapple_2007Icar..187..500H}.

Family members in our simulation were ejected from Cybele
assuming only one of the possible geometries:
true anomaly $f = 100^\circ$ and 
argument of pericentre $\omega = 330^\circ$.
The velocity field was isotropic and size dependent such that 
$v(D) \propto D^{-1}$, 
with a velocity distribution peak close to the escape speed from Cybele,
$v_{\rm esc} = 110\,{\rm m}\,{\rm s}^{-1}$,
and including outliers up to
$v_{\rm max} \simeq 400\,{\rm m}\,{\rm s}^{-1}$.

The initial size distribution of the family is shallow and 
composed mostly of $D > 1\,{\rm km}$ bodies,
similar to the Kalliope family \citep{Broz:2022}.
The total number of bodies was set to 3000 ---which is ten times larger than expected--- in order to improve statistics.
Indeed, it was important that a fraction of the bodies remained at the end of the simulation in order to investigate the possibility of a very old dispersed family.

We assumed the following values of thermal parameters for the family members:
a bulk density of $\rho_{\rm b} = 1300\,{\rm kg}\,{\rm m}^{-3}$, a surface density of $\rho_{\rm s} = \rho_{\rm b}$, a thermal conductivity of $K = 0.001\,{\rm W}\,{\rm m}^{-1}\,{\rm K}^{-1}$, a heat capacity of $C = 680\,{\rm J}\,{\rm kg}^{-1}\,{\rm m}\,{\rm K}^{-1}$, a Bond albedo of $A = 0.1$,
an infrared emissivity of $\epsilon = 0.9$, and
a YORP scaling parameter of $c_{\rm YORP} = 0.33$
\citep{Hanus_2011A&A...530A.134H}.

We used a time step of 9.13125\,days%
\footnote{The usual time step of 36.525\,d used in most simulations is not enough to sample orbits
within the Hill sphere of Jupiter (scaled by a factor of 3.5 in RMVS3).}
and a time span of 1\,Gyr.
Proper elements were computed with a sequence of digital filters
\citep{Quinn_1991AJ....101.2287Q,Sidlichovsky_1996CeMDA..65..137S},
set up suitably for the Cybele region.
Our adopted input sampling was 1\,yr, convolution filters were A, A, B, and B,
and decimation factors were 10, 10, 3, and 3
in order to prevent aliasing
and preserve dominant oscillations with a period of approximately 6500\,yr.
Proper frequencies and amplitudes were obtained
by frequency-modified Fourier transform,
with the exclusion of known planetary frequencies.
The output sampling was 1\,Myr.

\begin{figure}
\centering
\includegraphics[width=9cm]{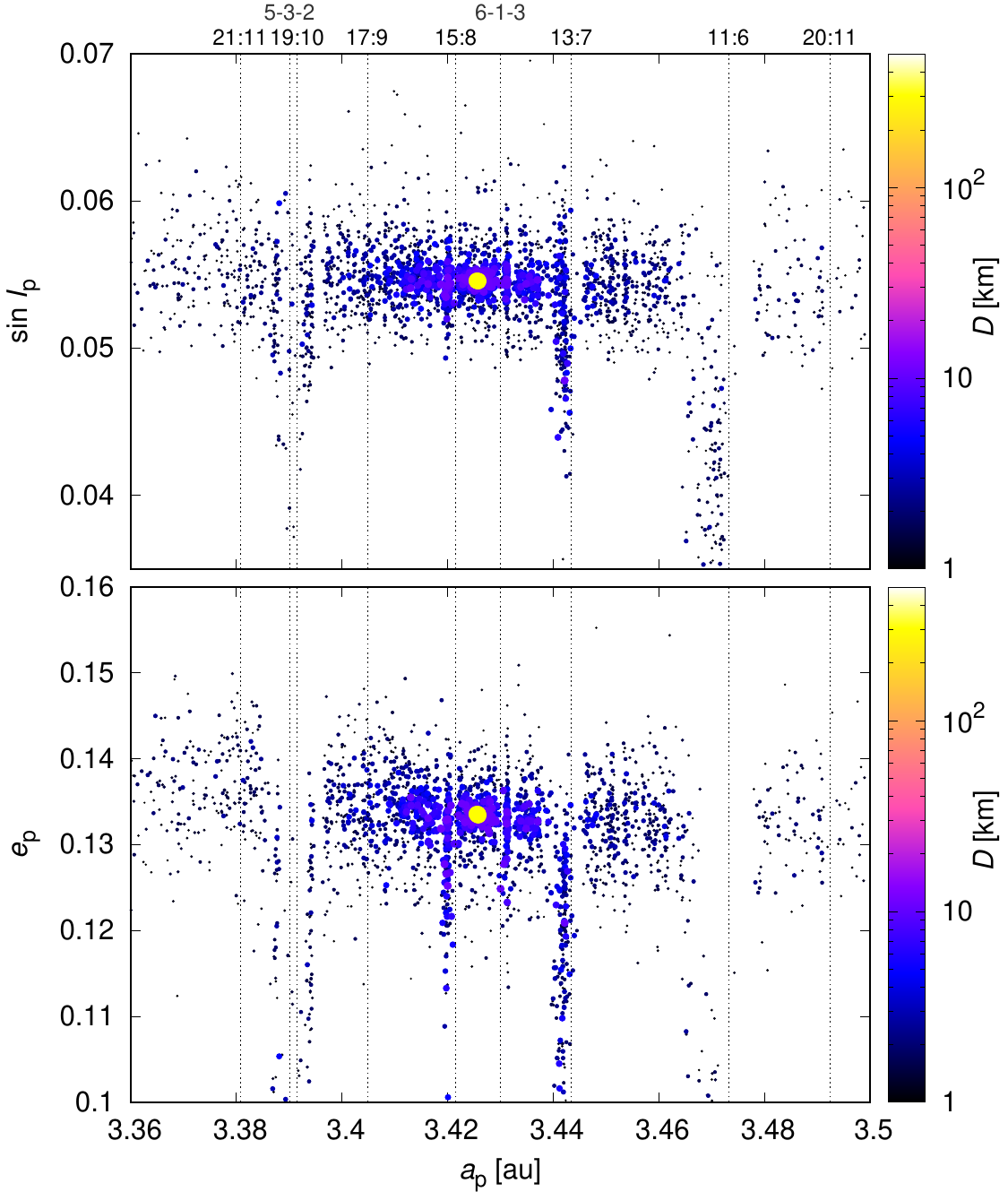}
\caption{
Initial conditions of a putative family related to (65)~Cybele plotted in proper semi-major axis $a_{\rm p}$ vs. proper inclination $\sin I_{\rm p}$ (top panel) and $a_{\rm p}$ vs. proper eccentricity $e_{\rm p}$ (bottom panel).
The vertical dotted lines indicate mean-motion and three-body resonances affecting the orbital evolution of the simulated asteroids.
The number of bodies is 3000,
corresponding to a relatively large family mostly composed of multi-kilometre-sized objects.}
\label{fig:aei_0000}
\end{figure}

Initially, the family is compact and prominent in our simulation (Fig.~\ref{fig:aei_0000}).
Its long-term orbital evolution is driven by gravitational perturbations,
inducing oscillations of the osculating eccentricity from 0.05 to 0.15,
and chaotic diffusion due to overlapping resonances.
The Yarkovsky drift is substantial for bodies of  1 to 10km in size
($5\times 10^{-4}$ to $5\times 10^{-5}\,{\rm au}\,{\rm Myr}^{-1}$), 
and delivers particles from stable orbits to unstable resonances.
The population decay is well described by an exponential function,
$N(t) = N_0\exp(-t/\tau)$,
with a timescale of $\tau \simeq 0.37\,{\rm Gyr}$ (Fig.~\ref{fig:decay}). 

Our simulation shows that the number of objects in any family older than approximately $5\tau$ ($\simeq$2.2\,Gyr)
is depleted by two orders of magnitude.
Therefore, one cannot expect that a 4Gyr-old family born from Cybele would have been preserved.
This is in contrasts to the Sylvia family, which is located in a dynamically quiescent environment \citep{Vokrouhlicky_2010AJ....139.2148V, Carruba:2015}.
On the other hand, any family younger than ${\simeq}\,3\tau$ ($\simeq$1.3\,Gyr) 
should still be recognisable (Fig.~\ref{fig:ei_0000}).

\subsubsection{Stability and origin of (65) Cybele itself}

\begin{figure}
\centering
\begin{tabular}{cc@{}c}
\multicolumn{2}{c}{$\sim\,$0\,Gyr} \\ [0.15cm]
\includegraphics[width=0.45\hsize, trim={1cm 0 0 0}]{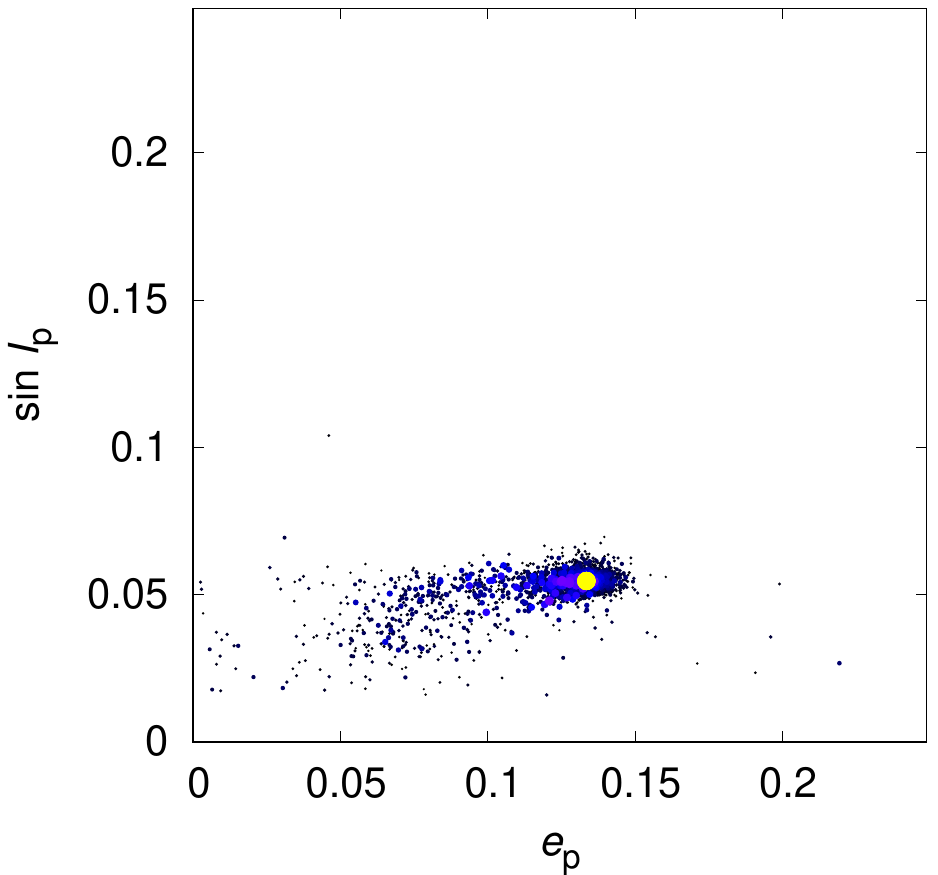} &
\includegraphics[width=0.45\hsize, trim={1cm 0 0 0}]{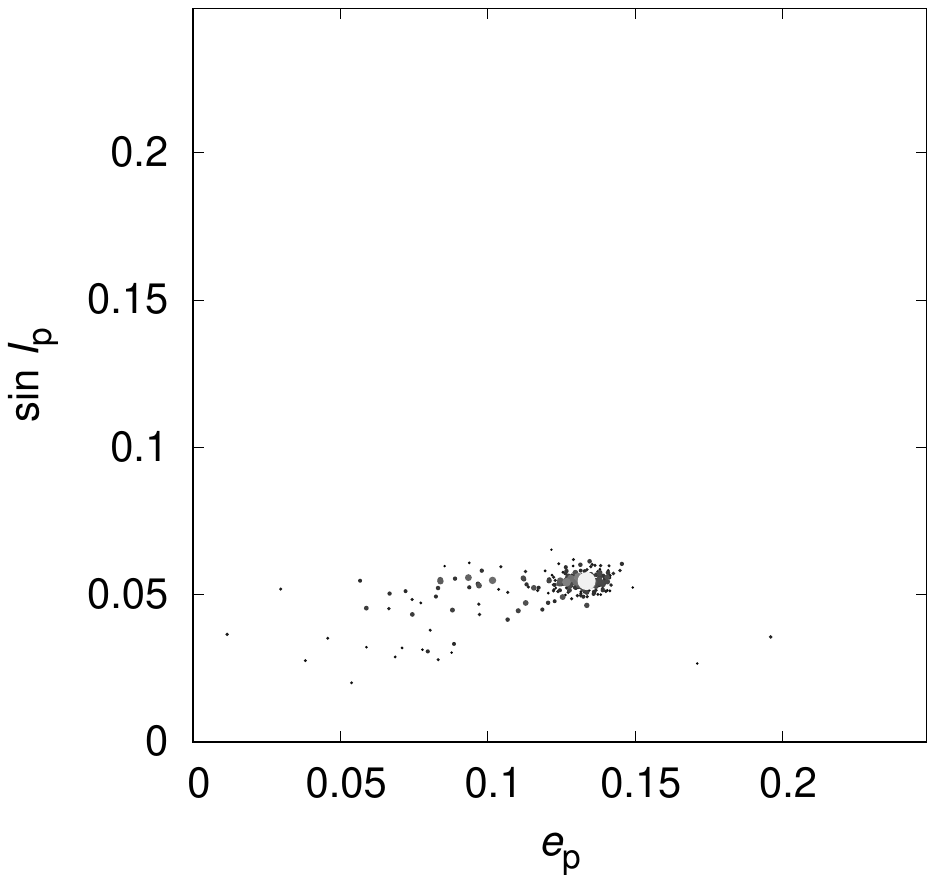} \\[-0cm]
\multicolumn{2}{c}{0.5\,Gyr} \\ [0.15cm]
\includegraphics[width=0.45\hsize, trim={1cm 0 0 0}]{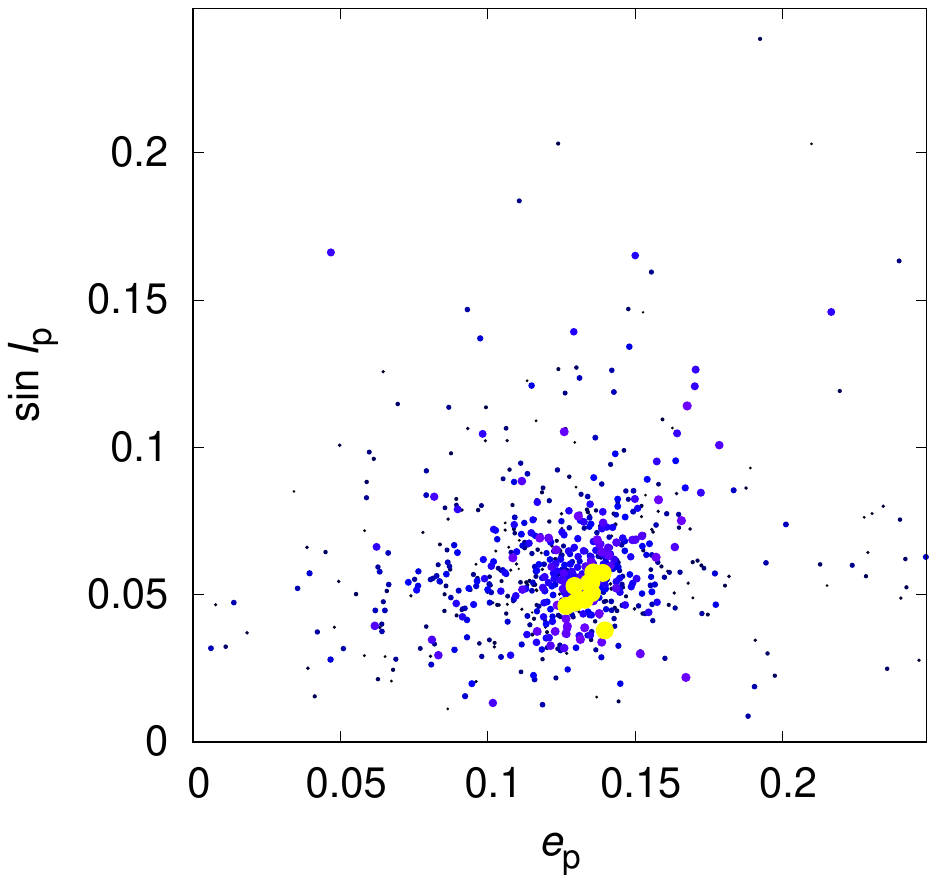} &
\includegraphics[width=0.45\hsize, trim={1cm 0 0 0}]{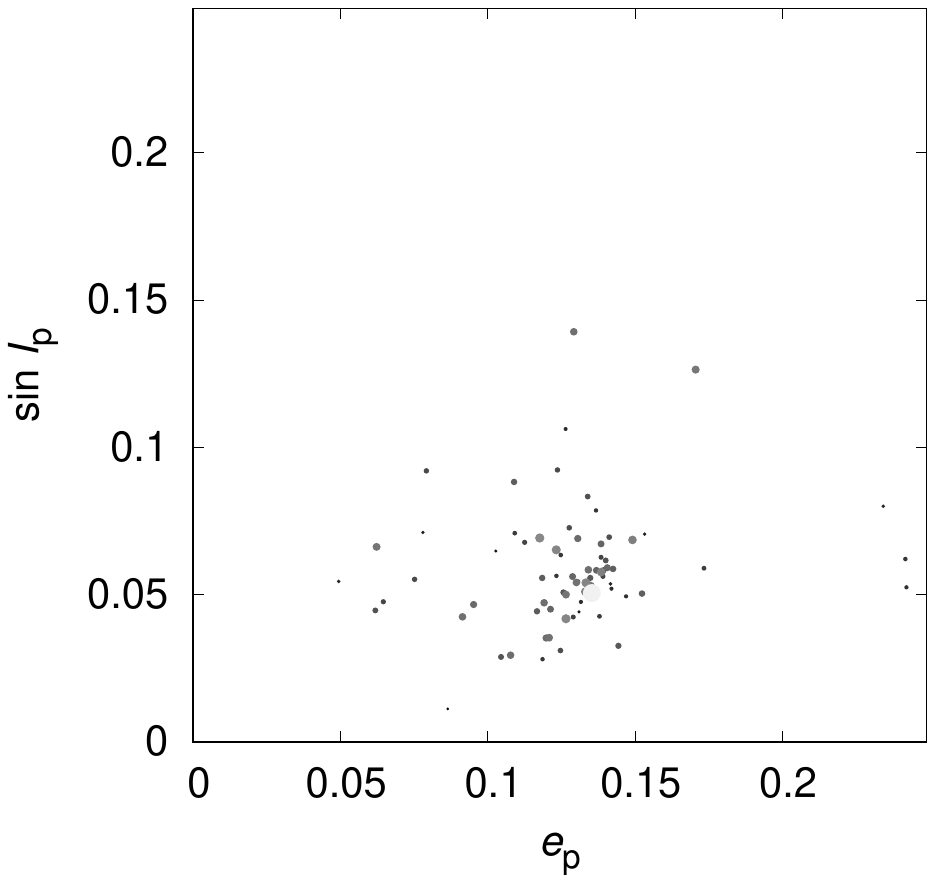} \\[-0cm]
\multicolumn{2}{c}{1\,Gyr} \\ [0.15cm]
\includegraphics[width=0.45\hsize, trim={1cm 0 0 0}]{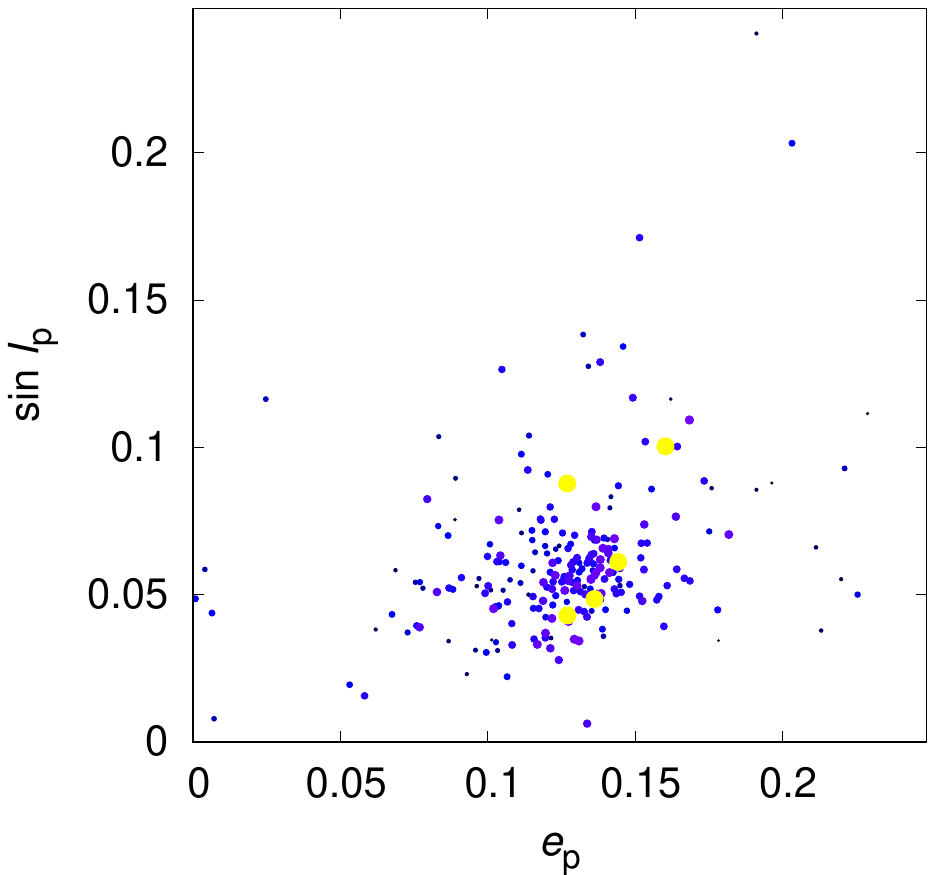} &
\includegraphics[width=0.45\hsize, trim={1cm 0 0 0}]{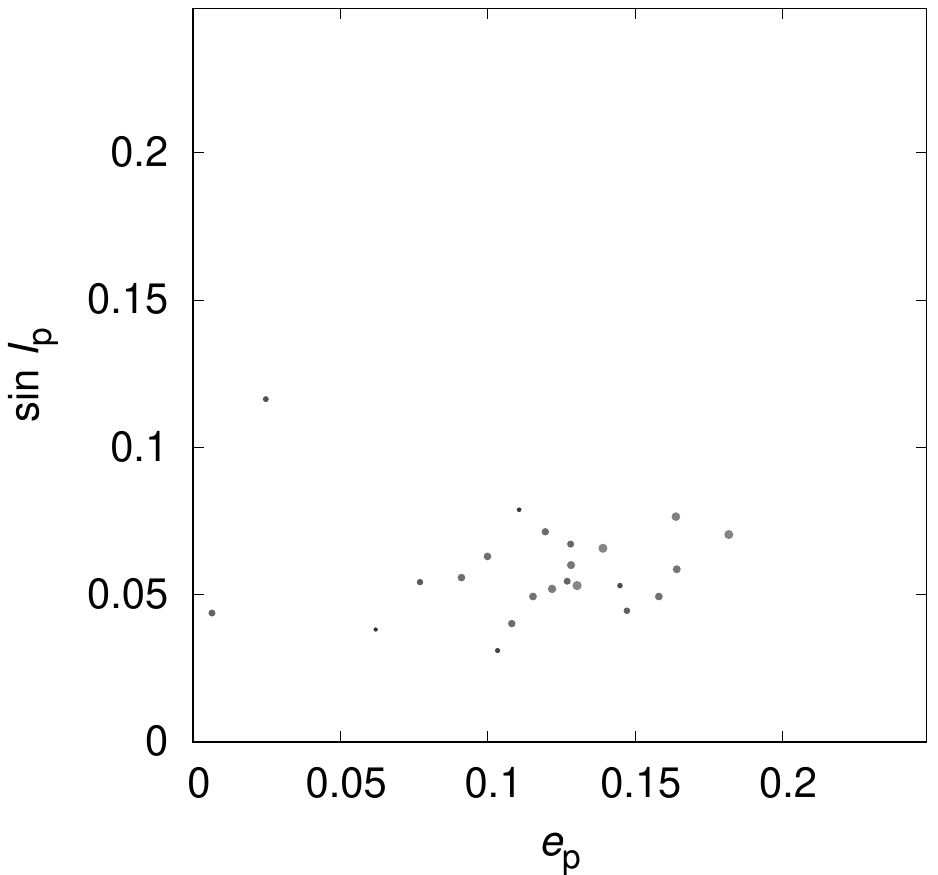} \\[-0cm]
\multicolumn{2}{c}{2\,Gyr} \\ [0.15cm]
\includegraphics[width=0.45\hsize, trim={1cm 0 0 0}]{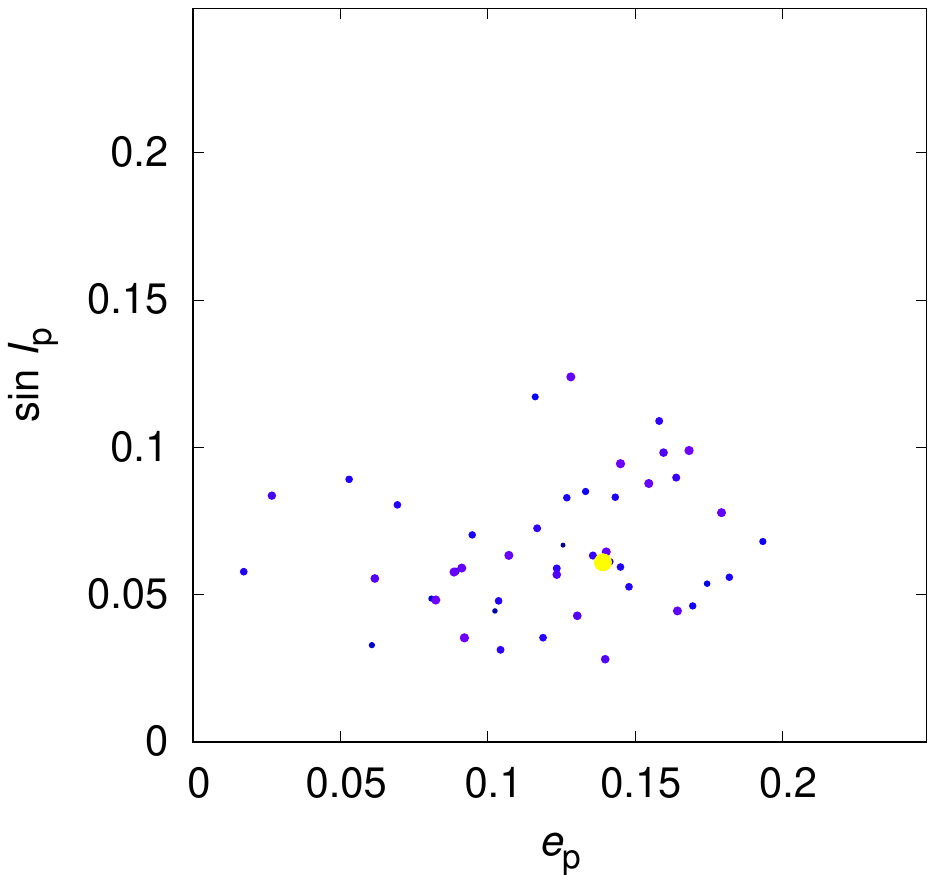} &
\includegraphics[width=0.45\hsize, trim={1cm 0 0 0}]{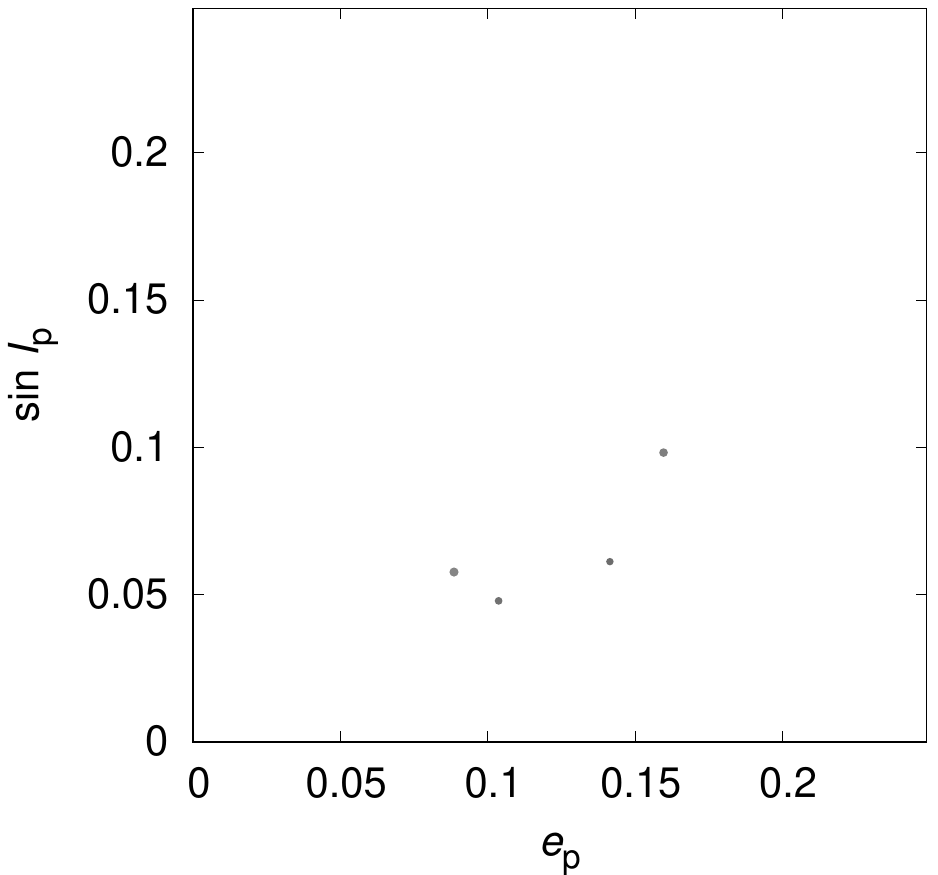} \\[-0cm]
\end{tabular}
\caption{
Proper eccentricity $e_{\rm p}$ vs. proper inclination $\sin I_{\rm p}$ distribution of a simulated family linked to (65) Cybele.
The dynamical evolution of the family was computed from 0 to 4\,Gyr.
{\it Left panels:}
Large family made of 3000 multi-kilometre-sized bodies ejected from (65) Cybele.
Ten clones of Cybele itself were also produced (yellow).
The colours and symbol sizes correspond to the logarithm of the diameter of the objects.
{\it Right panels:}
Smaller subset of 300 bodies drawn from the large family.
The axis ranges correspond 1:1 to Fig.~\ref{fig:ei_wise}.
After 1 or 2\,Gyr, depending on the initial population,
the family is no longer observable.
}
\label{fig:ei_0000}
\end{figure}

Interestingly, our simulation shows that
the clones of Cybele itself (yellow circles in Fig.~\ref{fig:aei_0000} and Fig.~\ref{fig:ei_0000}) are also affected by chaotic diffusion.
Out of ten clones, none were preserved after 0.8\,Gyr of dynamical evolution.
This is not surprising considering Cybele's present location close to the outer boundary of the Cybele region, which is sculpted by chaotic diffusion.

Evolution due to gravitational perturbations is a reversible process. 
Therefore, chaotic diffusion may have acted on Cybele in the past in the opposite direction compared to our simulation. 
If true, this implies that Cybele may have temporarily been on a chaotic orbit, crossing the orbits of giant planets. 

This opens the interesting possibility that Cybele is a recently ($<$1\,Gyr ago) implanted Jupiter-family comet (JFC). 
These objects originating from the Kuiper belt (e.g. \citealt{Lowry:2008}) are present virtually everywhere beyond $Q\,=\,4.61$\,au,
corresponding to the edge of the Hill sphere of Jupiter.
Several comets may have diffused from highly eccentric orbits towards lower eccentricities. 
A well-known example is 133P/Elst-Pizarro, which may be a JFC that evolved into an asteroid-like orbit \citep{Hsieh:2004}.
Such diffusion can be driven by low-order resonances like the 2:1 MMR at 3.27 au, or to high-order series (e.g. the 9:4, 11:5, 13:6, 15:7, 17:8, 19:9 series, or the 19:10, 17:9, 15:8, 13:7, 11:6 series, which is within the Cybele region), which often overlap with three-body resonances.
If JFCs were indeed implanted in the Cybele region, they would most likely be extinct today considering the very long diffusion timescale of $\simeq$0.5\,Gyr in these resonances. 

\begin{figure}
\centering
\includegraphics[width=9cm]{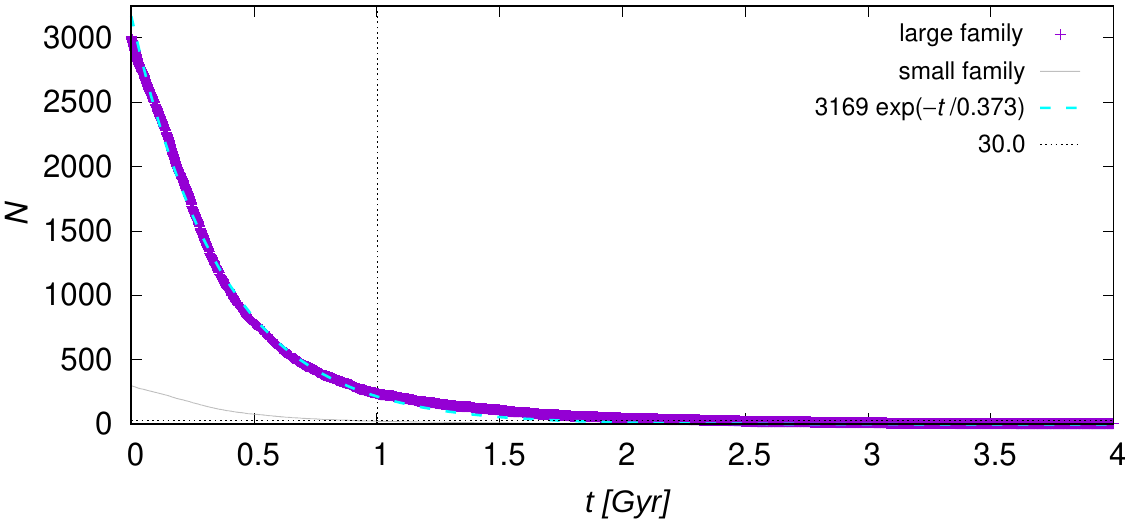}
\caption{
Dynamical decay of the putative Cybele family.
The number of objects ($N$) decreases with time~$t$.
An approximating exponential function
with timescale $\tau = 0.37\,{\rm Gyr}$ is plotted.
We consider the family
to be definitely no longer observable when $N$ decreases below about 30.
This is especially true considering that the orbits of the family members are dispersed in the same way as those of the background population.
}
\label{fig:decay}
\end{figure}

The hypothesis of Cybele being an implanted JFC however faces several issues. 
The first counter-argument is that many more smaller implanted JFCs would be expected to be found on similar orbits owing to their steep size--frequency distribution \citep{Granvik:2018}. Although most of these small objects would be eliminated on a short timescale (Fig.\,\ref{fig:decay}) due to the fact that  the Yarkovsky effect delivers them to nearby resonances, it seems strange that very few such bodies are found in the vicinity of Cybele. 
It is possible that, in the near future, large surveys such as the Legacy Survey of Space and Time (LSST) of the Vera C. Rubin Observatory \citep{Ivezic:2019} will be able to detect a population of implanted JFCs in the main belt.

Secondly, previous simulations of comet-like starting orbital elements transitioning onto main-belt orbits have shown to be largely prevented from reaching low-eccentricity and low-inclination orbits \citep{Haghighipour:2009, Hsieh:2016}. According to these works, the real-world population of main-belt comets with low eccentricities and inclinations is likely made up of fragments of larger icy asteroids. 
Therefore, it would be very difficult for a JFC to evolve into Cybele's low-inclination orbit and it appears more plausible that the asteroid came to its current orbit through a slow diffusion from a nearby, more stable orbit billions of years after its implantation in the outer main belt during the early phase of planetary migrations \citep{Levison:2009,Vokrouhlicky:2016}.

A possible genetically related compositional analogue for Cybele may be the irregular Saturnian moon Phoebe, which is believed to be a captured Centaur formed beyond the orbit of Saturn \citep{Clark:2019}, possibly within the original reservoir of C-type asteroids \citep{Castillo:2019}. Phoebe has a similar size (D$\simeq$212\,km) and bulk density (1.64$\pm$0.02\,g.cm$^{-3}$) to Cybele and was found to be at hydrostatic equilibrium \citep{Rambaux:2020}. However, unlike Cybele, Phoebe exhibits a strongly differentiated interior \citep{Rambaux:2020} and a surface dominated by water ice \citep{Fraser:2018}, which might point to a distinct initial composition (ice-to-rock ratio) and/or early thermal evolution.

\section{Conclusion}
\label{sec:conclusion}

In contrast to other large (D$>$200\,km) asteroids located in the Cybele region ((87)~Sylvia and (107)~Camilla), (65)~Cybele exhibits a perfect hydrostatic equilibrium shape. 
A previous study of (704)~Interamnia led by \citet{Hanus:2020} placed the transition mass limit between irregularly shaped small asteroids and larger bodies at equilibrium near $\simeq$3.5\,$\times10^{19}$\,kg. Here, our study suggests an even lower limit of $\simeq$1.5\,$\times10^{19}$\,kg, and opens the possibility that $D\,\geq\,260$\,km small bodies from the outer Solar System all formed at equilibrium. 

However, the origin of Cybele's shape currently remains a mystery as it could either be a relic of its original shape, or the result of a large impact as in the cases of (10)~Hygiea \citep{Vernazza:2020} and (31)~Euphrosyne \citep{Yang:2020}. 
In particular, {despite the detection of a present-day family associated to Cybele,} we show that the hypothesis of an old fragmenting impact cannot be ruled out: in the unstable orbital region of Cybele, a collisional family would be totally dispersed over $\simeq$2\,Gyr of dynamical evolution, thereby erasing any evidence of such a large impact. 
Cybele's higher bulk density (${\rm 1.55\,\pm\,0.19\,g.cm^{-3}}$) with respect to Sylvia and Camilla (${\rm 1.3-1.4\,g.cm^{-3}}$; \citealt{Pajuelo:2018, Carry:2021}) may hint at a structurally intact interior for Cybele, which would favour the primordial shape hypothesis.

The orbit of Cybele itself is long-term unstable, implying that it was recently ($<$1\,Gyr ago) placed on its current orbit, likely through slow diffusion from a relatively stable nearby orbit in the outer main belt or, less likely, from the population of JFCs in the planet-crossing region.

\begin{acknowledgements}
Based on observations collected at the European Organisation for Astronomical Research in the Southern Hemisphere under ESO program 107.22QN.001 (PI: Marsset). This research has made use of IMCCE's SsODNet VO tool (\url{https://ssp.imcce.fr/webservices/ssodnet/}). This work has been supported by the Czech Science Foundation through grants 20-08218S (J. Hanu\v s) and 21-11058S (M. Bro\v z), as well as by the National Science Foundation under Grant No. 1743015 (F. Marchis). T. Santana-Ros acknowledges funding from the NEO-MAPP project (H2020-EU-2-1-6/870377). In addition, this work was partially funded by the Spanish MICIN/AEI/10.13039/501100011033 and by “ERDF A way of making Europe” by the “European Union” through grant RTI2018-095076-B-C21, and the Institute of Cosmos Sciences University of Barcelona (ICCUB, Unidad de Excelencia ‘María de Maeztu’) through grant CEX2019-000918-M. This research has made use of the Asteroid Families Portal maintained at the Department of Astronomy, University of Belgrade. TRAPPIST is a project funded by the Belgian Fonds (National) de la Recherche Scientifique (F.R.S.-FNRS) under grant PDR T.0120.21. TRAPPIST-North is a project funded by the University of Liège, in collaboration with the Cadi Ayyad University of Marrakech (Morocco). E. Jehin is F.R.S.-FNRS Senior Research Associate. 
\end{acknowledgements}

\bibliographystyle{aa}
\bibliography{references, references_mira, mybib, ssodnet}

\begin{thebibliography}{123}
\expandafter\ifx\csname natexlab\endcsname\relax\def\natexlab#1{#1}\fi

\bibitem[{{Aslan} {et~al.}(2007){Aslan}, {Gumerov}, {Hudkova}, {Ivantsov},
  {Khamitov}, \& {Pinigin}}]{Aslan:2007}
{Aslan}, Z., {Gumerov}, R., {Hudkova}, L., {et~al.} 2007, in Astronomical
  Society of the Pacific Conference Series, Vol. 370, Solar and Stellar Physics
  Through Eclipses, ed. O.~{Demircan}, S.~O. {Selam}, \& B.~{Albayrak}, 52

\bibitem[{{Baer} \& {Chesley}(2008)}]{Baer:2008}
{Baer}, J. \& {Chesley}, S.~R. 2008, Celestial Mechanics and Dynamical
  Astronomy, 100, 27

\bibitem[{{Baer} \& {Chesley}(2017)}]{Baer:2017}
{Baer}, J. \& {Chesley}, S.~R. 2017, \aj, 154, 76

\bibitem[{{Baer} {et~al.}(2011){Baer}, {Chesley}, \& {Matson}}]{Baer:2011}
{Baer}, J., {Chesley}, S.~R., \& {Matson}, R.~D. 2011, \aj, 141, 143

\bibitem[{{Bartczak} \& {Dudzi{\'n}ski}(2018)}]{Bartczak:2018}
{Bartczak}, P. \& {Dudzi{\'n}ski}, G. 2018, \mnras, 473, 5050

\bibitem[{{Bartczak} \& {Dudzi{\'n}ski}(2019)}]{Bartczak:2019}
{Bartczak}, P. \& {Dudzi{\'n}ski}, G. 2019, \mnras, 485, 2431

\bibitem[{{Beck} {et~al.}(2011){Beck}, {Quirico}, {Sevestre},
  {Montes-Hernandez}, {Pommerol}, \& {Schmitt}}]{Beck:2011}
{Beck}, P., {Quirico}, E., {Sevestre}, D., {et~al.} 2011, \aap, 526, A85

\bibitem[{{Beuzit} {et~al.}(2019){Beuzit}, {Vigan}, {Mouillet}, {Dohlen},
  {Gratton}, {Boccaletti}, {Sauvage}, {Schmid}, {Langlois}, {Petit},
  {Baruffolo}, {Feldt}, {Milli}, {Wahhaj}, {Abe}, {Anselmi}, {Antichi},
  {Barette}, {Baudrand}, {Baudoz}, {Bazzon}, {Bernardi}, {Blanchard}, {Brast},
  {Bruno}, {Buey}, {Carbillet}, {Carle}, {Cascone}, {Chapron}, {Charton},
  {Chauvin}, {Claudi}, {Costille}, {De Caprio}, {de Boer}, {Delboulb{\'e}},
  {Desidera}, {Dominik}, {Downing}, {Dupuis}, {Fabron}, {Fantinel}, {Farisato},
  {Feautrier}, {Fedrigo}, {Fusco}, {Gigan}, {Ginski}, {Girard}, {Giro},
  {Gisler}, {Gluck}, {Gry}, {Henning}, {Hubin}, {Hugot}, {Incorvaia}, {Jaquet},
  {Kasper}, {Lagadec}, {Lagrange}, {Le Coroller}, {Le Mignant}, {Le Ruyet},
  {Lessio}, {Lizon}, {Llored}, {Lundin}, {Madec}, {Magnard}, {Marteaud},
  {Martinez}, {Maurel}, {M{\'e}nard}, {Mesa}, {M{\"o}ller-Nilsson}, {Moulin},
  {Moutou}, {Orign{\'e}}, {Parisot}, {Pavlov}, {Perret}, {Pragt}, {Puget},
  {Rabou}, {Ramos}, {Reess}, {Rigal}, {Rochat}, {Roelfsema}, {Rousset}, {Roux},
  {Saisse}, {Salasnich}, {Santambrogio}, {Scuderi}, {Segransan}, {Sevin},
  {Siebenmorgen}, {Soenke}, {Stadler}, {Suarez}, {Tiph{\`e}ne}, {Turatto},
  {Udry}, {Vakili}, {Waters}, {Weber}, {Wildi}, {Zins}, \&
  {Zurlo}}]{Beuzit:2019}
{Beuzit}, J.~L., {Vigan}, A., {Mouillet}, D., {et~al.} 2019, \aap, 631, A155

\bibitem[{{Bro{\v z}} {et~al.}(2011){Bro{\v z}}, {Vokrouhlick{\'y}},
  {Morbidelli}, {Nesvorn{\'y}}, \& {Bottke}}]{Broz_2011MNRAS.414.2716B}
{Bro{\v z}}, M., {Vokrouhlick{\'y}}, D., {Morbidelli}, A., {Nesvorn{\'y}}, D.,
  \& {Bottke}, W.~F. 2011, \mnras, 414, 2716

\bibitem[{{Bro{\v{z}}} {et~al.}(2022){Bro{\v{z}}}, {Ferrais}, {Vernazza},
  {{\v{S}}eve{\v{c}}ek}, \& {Jutzi}}]{Broz:2022}
{Bro{\v{z}}}, M., {Ferrais}, M., {Vernazza}, P., {{\v{S}}eve{\v{c}}ek}, P., \&
  {Jutzi}, M. 2022, \aap, 664, A69

\bibitem[{{Campins} {et~al.}(2010){Campins}, {Morbidelli}, {Tsiganis}, {de
  Le{\'o}n}, {Licandro}, \& {Lauretta}}]{Campins:2010}
{Campins}, H., {Morbidelli}, A., {Tsiganis}, K., {et~al.} 2010, \apjl, 721, L53

\bibitem[{{Capanna} {et~al.}(2013){Capanna}, {Gesqui\`ere}, {Jorda}, {Lamy}, \&
  {Vibert}}]{Capanna:2013}
{Capanna}, C., {Gesqui\`ere}, G., {Jorda}, L., {Lamy}, P., \& {Vibert}, D.
  2013, The Visual Computer, 29, 825

\bibitem[{{Carruba} {et~al.}(2015){Carruba}, {Nesvorn{\'y}}, {Aljbaae}, \&
  {Huaman}}]{Carruba:2015}
{Carruba}, V., {Nesvorn{\'y}}, D., {Aljbaae}, S., \& {Huaman}, M.~E. 2015,
  \mnras, 451, 244

\bibitem[{{Carry}(2012)}]{Carry:2012}
{Carry}, B. 2012, \planss, 73, 98

\bibitem[{{Carry} {et~al.}(2019){Carry}, {Vachier}, {Berthier}, {Marsset},
  {Vernazza}, {Grice}, {Merline}, {Lagadec}, {Fienga}, {Conrad},
  {Podlewska-Gaca}, {Santana-Ros}, {Viikinkoski}, {Hanu{\v{s}}}, {Dumas},
  {Drummond}, {Tamblyn}, {Chapman}, {Behrend}, {Bernasconi}, {Bartczak},
  {Benkhaldoun}, {Birlan}, {Castillo-Rogez}, {Cipriani}, {Colas}, {Drouard},
  {{\v{D}}urech}, {Enke}, {Fauvaud}, {Ferrais}, {Fetick}, {Fusco}, {Gillon},
  {Jehin}, {Jorda}, {Kaasalainen}, {Keppler}, {Kryszczynska}, {Lamy},
  {Marchis}, {Marciniak}, {Michalowski}, {Michel}, {Pajuelo}, {Tanga}, {Vigan},
  {Warner}, {Witasse}, {Yang}, \& {Zurlo}}]{Carry:2019}
{Carry}, B., {Vachier}, F., {Berthier}, J., {et~al.} 2019, \aap, 623, A132

\bibitem[{{Carry} {et~al.}(2021){Carry}, {Vernazza}, {Vachier}, {Neveu},
  {Berthier}, {Hanu{\v{s}}}, {Ferrais}, {Jorda}, {Marsset}, {Viikinkoski},
  {Bartczak}, {Behrend}, {Benkhaldoun}, {Birlan}, {Castillo-Rogez}, {Cipriani},
  {Colas}, {Drouard}, {Dudzi{\'n}ski}, {Desmars}, {Dumas}, {{\v{D}}urech},
  {Fetick}, {Fusco}, {Grice}, {Jehin}, {Kaasalainen}, {Kryszczynska}, {Lamy},
  {Marchis}, {Marciniak}, {Michalowski}, {Michel}, {Pajuelo}, {Podlewska-Gaca},
  {Rambaux}, {Santana-Ros}, {Storrs}, {Tanga}, {Vigan}, {Warner}, {Wieczorek},
  {Witasse}, \& {Yang}}]{Carry:2021}
{Carry}, B., {Vernazza}, P., {Vachier}, F., {et~al.} 2021, \aap, 650, A129

\bibitem[{{Castillo-Rogez} {et~al.}(2019){Castillo-Rogez}, {Vernazza}, \&
  {Walsh}}]{Castillo:2019}
{Castillo-Rogez}, J., {Vernazza}, P., \& {Walsh}, K. 2019, \mnras, 486, 538

\bibitem[{{Chernetenko} \& {Kochetova}(2002)}]{Chernetenko:2002}
{Chernetenko}, Y.~A. \& {Kochetova}, O.~M. 2002, in ESA Special Publication,
  Vol. 500, Asteroids, Comets, and Meteors: ACM 2002, ed. B.~{Warmbein},
  437--440

\bibitem[{{Clark} {et~al.}(2019){Clark}, {Brown}, {Cruikshank}, \&
  {Swayze}}]{Clark:2019}
{Clark}, R.~N., {Brown}, R.~H., {Cruikshank}, D.~P., \& {Swayze}, G.~A. 2019,
  \icarus, 321, 791

\bibitem[{{DeMeo} \& {Carry}(2013)}]{Demeo:2013}
{DeMeo}, F.~E. \& {Carry}, B. 2013, \icarus, 226, 723

\bibitem[{{Dudzi{\'n}ski} {et~al.}(2020){Dudzi{\'n}ski}, {Podlewska-Gaca},
  {Bartczak}, {Benseguane}, {Ferrais}, {Jorda}, {Hanu{\v{s}}}, {Vernazza},
  {Rambaux}, {Carry}, {Marchis}, {Marsset}, {Viikinkoski}, {Bro{\v{z}}},
  {Fetick}, {Drouard}, {Fusco}, {Birlan}, {Jehin}, {Berthier},
  {Castillo-Rogez}, {Cipriani}, {Colas}, {Dumas}, {Kryszczynska}, {Lamy}, {Le
  Coroller}, {Marciniak}, {Michalowski}, {Michel}, {Santana-Ros}, {Tanga},
  {Vachier}, {Vigan}, {Witasse}, \& {Yang}}]{Dudzinski:2020}
{Dudzi{\'n}ski}, G., {Podlewska-Gaca}, E., {Bartczak}, P., {et~al.} 2020,
  \mnras, 499, 4545

\bibitem[{{Emery} {et~al.}(2011){Emery}, {Burr}, \& {Cruikshank}}]{Emery:2011}
{Emery}, J.~P., {Burr}, D.~M., \& {Cruikshank}, D.~P. 2011, \aj, 141, 25

\bibitem[{{Emery} {et~al.}(2006){Emery}, {Cruikshank}, \& {Van
  Cleve}}]{Emery:2006}
{Emery}, J.~P., {Cruikshank}, D.~P., \& {Van Cleve}, J. 2006, \icarus, 182, 496

\bibitem[{{Ferrais} {et~al.}(2022){Ferrais}, {Jorda}, {Vernazza}, {Carry},
  {Bro{\v{z}}}, {Rambaux}, {Hanu{\v{s}}}, {Dudzi{\'n}ski}, {Bartczak},
  {Vachier}, {Aristidi}, {Beck}, {Marchis}, {Marsset}, {Viikinkoski}, {Fetick},
  {Drouard}, {Fusco}, {Birlan}, {Podlewska-Gaca}, {Burbine}, {Dyar},
  {Bendjoya}, {Benkhaldoun}, {Berthier}, {Castillo-Rogez}, {Cipriani}, {Colas},
  {Dumas}, {{\v{D}}urech}, {Fauvaud}, {Grice}, {Jehin}, {Kaasalainen},
  {Kryszczynska}, {Lamy}, {Le Coroller}, {Marciniak}, {Michalowski}, {Michel},
  {Prieur}, {Reddy}, {Rivet}, {Santana-Ros}, {Scardia}, {Tanga}, {Vigan},
  {Witasse}, \& {Yang}}]{Ferrais:2022}
{Ferrais}, M., {Jorda}, L., {Vernazza}, P., {et~al.} 2022, \aap, 662, A71

\bibitem[{{Ferrais} {et~al.}(2020){Ferrais}, {Vernazza}, {Jorda}, {Rambaux},
  {Hanu{\v{s}}}, {Carry}, {Marchis}, {Marsset}, {Viikinkoski}, {Bro{\v{z}}},
  {Fetick}, {Drouard}, {Fusco}, {Birlan}, {Podlewska-Gaca}, {Jehin},
  {Bartczak}, {Berthier}, {Castillo-Rogez}, {Cipriani}, {Colas},
  {Dudzi{\'n}ski}, {Dumas}, {{\v{D}}urech}, {Kaasalainen}, {Kryszczynska},
  {Lamy}, {Le Coroller}, {Marciniak}, {Michalowski}, {Michel}, {Santana-Ros},
  {Tanga}, {Vachier}, {Vigan}, {Witasse}, \& {Yang}}]{Ferrais:2020}
{Ferrais}, M., {Vernazza}, P., {Jorda}, L., {et~al.} 2020, \aap, 638, L15

\bibitem[{{F{\'e}tick} {et~al.}(2019){F{\'e}tick}, {Jorda}, {Vernazza},
  {Marsset}, {Drouard}, {Fusco}, {Carry}, {Marchis}, {Hanu{\v s}},
  {Viikinkoski}, {Birlan}, {Bartczak}, {Berthier}, {Castillo-Rogez},
  {Cipriani}, {Colas}, {Dudzi{\'n}ski}, {Dumas}, {Ferrais}, {Jehin},
  {Kaasalainen}, {Kryszczynska}, {Lamy}, {Le Coroller}, {Marciniak},
  {Michalowski}, {Michel}, {Mugnier}, {Neichel}, {Pajuelo}, {Podlewska-Gaca},
  {Santana-Ros}, {Tanga}, {Vachier}, {Vigan}, {Witasse}, \&
  {Yang}}]{Fetick:2019}
{F{\'e}tick}, R.~J., {Jorda}, L., {Vernazza}, P., {et~al.} 2019, Astron.
  Astrophys., 623, A6

\bibitem[{{Fienga} {et~al.}(2020){Fienga}, {Avdellidou}, \&
  {Hanu{\v{s}}}}]{Fienga:2020}
{Fienga}, A., {Avdellidou}, C., \& {Hanu{\v{s}}}, J. 2020, \mnras, 492, 589

\bibitem[{{Fienga} {et~al.}(2019){Fienga}, {Deram}, {Viswanathan}, {Di Ruscio},
  {Bernus}, {Durante}, {Gastineau}, \& {Laskar}}]{Fienga:2019}
{Fienga}, A., {Deram}, P., {Viswanathan}, V., {et~al.} 2019, Notes
  Scientifiques et Techniques de l'Institut de Mecanique Celeste, 109

\bibitem[{{Fienga} {et~al.}(2011){Fienga}, {Laskar}, {Kuchynka}, {Manche},
  {Desvignes}, {Gastineau}, {Cognard}, \& {Theureau}}]{Fienga:2011}
{Fienga}, A., {Laskar}, J., {Kuchynka}, P., {et~al.} 2011, Celestial Mechanics
  and Dynamical Astronomy, 111, 363

\bibitem[{{Fienga} {et~al.}(2009){Fienga}, {Laskar}, {Morley}, {Manche},
  {Kuchynka}, {Le Poncin-Lafitte}, {Budnik}, {Gastineau}, \&
  {Somenzi}}]{Fienga:2009}
{Fienga}, A., {Laskar}, J., {Morley}, T., {et~al.} 2009, \aap, 507, 1675

\bibitem[{{Fienga} {et~al.}(2013){Fienga}, {Manche}, {Laskar}, {Gastineau}, \&
  {Verma}}]{Fienga:2013}
{Fienga}, A., {Manche}, H., {Laskar}, J., {Gastineau}, M., \& {Verma}, A. 2013,
  arXiv e-prints, arXiv:1301.1510

\bibitem[{{Fienga} {et~al.}(2014){Fienga}, {Manche}, {Laskar}, {Gastineau}, \&
  {Verma}}]{Fienga:2014}
{Fienga}, A., {Manche}, H., {Laskar}, J., {Gastineau}, M., \& {Verma}, A. 2014,
  arXiv e-prints, arXiv:1405.0484

\bibitem[{{Folkner} {et~al.}(2009){Folkner}, {Williams}, \&
  {Boggs}}]{Folkner:2009}
{Folkner}, W.~M., {Williams}, J.~G., \& {Boggs}, D.~H. 2009, Interplanetary
  Network Progress Report, 42-178, 1

\bibitem[{{Folkner} {et~al.}(2014){Folkner}, {Williams}, {Boggs}, {Park}, \&
  {Kuchynka}}]{Folkner:2014}
{Folkner}, W.~M., {Williams}, J.~G., {Boggs}, D.~H., {Park}, R.~S., \&
  {Kuchynka}, P. 2014, Interplanetary Network Progress Report, 42-196, 1

\bibitem[{{Franco} \& {Pilcher}(2015)}]{Franco2015}
{Franco}, L. \& {Pilcher}, F. 2015, Minor Planet Bulletin, 42, 204

\bibitem[{{Fraser} \& {Brown}(2018)}]{Fraser:2018}
{Fraser}, W.~C. \& {Brown}, M.~E. 2018, \aj, 156, 23

\bibitem[{{Fraser} {et~al.}(2014){Fraser}, {Brown}, {Morbidelli}, {Parker}, \&
  {Batygin}}]{Fraser:2014}
{Fraser}, W.~C., {Brown}, M.~E., {Morbidelli}, A., {Parker}, A., \& {Batygin},
  K. 2014, \apj, 782, 100

\bibitem[{Fusco {et~al.}(2003)Fusco, Mugnier, Conan, Marchis, Chauvin, Rousset,
  Lagrange, Mouillet, \& Roddier}]{Fusco:2003}
Fusco, T., Mugnier, L.~M., Conan, J.-M., {et~al.} 2003, in Adaptive Optical
  System Technologies II, Vol. 4839, SPIE, 1065--1075

\bibitem[{{Giorgini} {et~al.}(1996){Giorgini}, {Yeomans}, {Chamberlin},
  {Chodas}, {Jacobson}, {Keesey}, {Lieske}, {Ostro}, {Standish}, \&
  {Wimberly}}]{Giorgini_1996DPS....28.2504G}
{Giorgini}, J.~D., {Yeomans}, D.~K., {Chamberlin}, A.~B., {et~al.} 1996, in
  AAS/Division for Planetary Sciences Meeting Abstracts, Vol.~28, AAS/Division
  for Planetary Sciences Meeting Abstracts \#28, 25.04

\bibitem[{{Goffin}(2014)}]{Goffin:2014}
{Goffin}, E. 2014, \aap, 565, A56

\bibitem[{{Granvik} {et~al.}(2018){Granvik}, {Morbidelli}, {Jedicke}, {Bolin},
  {Bottke}, {Beshore}, {Vokrouhlick{\'y}}, {Nesvorn{\'y}}, \&
  {Michel}}]{Granvik:2018}
{Granvik}, M., {Morbidelli}, A., {Jedicke}, R., {et~al.} 2018, \icarus, 312,
  181

\bibitem[{{Haghighipour}(2009)}]{Haghighipour:2009}
{Haghighipour}, N. 2009, Meteoritics \& Planetary Science, 44, 1863

\bibitem[{{Hanu{\v s}} {et~al.}(2011){Hanu{\v s}}, {{\v D}urech}, {Bro{\v z}},
  {Warner}, {Pilcher}, {Stephens}, {Oey}, {Bernasconi}, {Casulli}, {Behrend},
  {Polishook}, {Henych}, {Lehk{\'y}}, {Yoshida}, \&
  {Ito}}]{Hanus_2011A&A...530A.134H}
{Hanu{\v s}}, J., {{\v D}urech}, J., {Bro{\v z}}, M., {et~al.} 2011, \aap, 530,
  A134

\bibitem[{{Hanu{\v{s}}} {et~al.}(2019){Hanu{\v{s}}}, {Marsset}, {Vernazza},
  {Viikinkoski}, {Drouard}, {Bro{\v{z}}}, {Carry}, {Fetick}, {Marchis},
  {Jorda}, {Fusco}, {Birlan}, {Santana-Ros}, {Podlewska-Gaca}, {Jehin},
  {Ferrais}, {Grice}, {Bartczak}, {Berthier}, {Castillo-Rogez}, {Cipriani},
  {Colas}, {Dudzi{\'n}ski}, {Dumas}, {{\v{D}}urech}, {Kaasalainen},
  {Kryszczynska}, {Lamy}, {Le Coroller}, {Marciniak}, {Michalowski}, {Michel},
  {Pajuelo}, {Tanga}, {Vachier}, {Vigan}, {Witasse}, \& {Yang}}]{Hanus:2019}
{Hanu{\v{s}}}, J., {Marsset}, M., {Vernazza}, P., {et~al.} 2019, \aap, 624,
  A121

\bibitem[{{Hanu{\v{s}}} {et~al.}(2020){Hanu{\v{s}}}, {Vernazza}, {Viikinkoski},
  {Ferrais}, {Rambaux}, {Podlewska-Gaca}, {Drouard}, {Jorda}, {Jehin}, {Carry},
  {Marsset}, {Marchis}, {Warner}, {Behrend}, {Asenjo}, {Berger}, {Bronikowska},
  {Brothers}, {Charbonnel}, {Colazo}, {Coliac}, {Duffard}, {Jones}, {Leroy},
  {Marciniak}, {Melia}, {Molina}, {Nadolny}, {Person}, {Pejcha}, {Riemis},
  {Shappee}, {Sobkowiak}, {Sold{\'a}n}, {Suys}, {Szakats}, {Vantomme},
  {Birlan}, {Berthier}, {Bartczak}, {Dumas}, {Dudzi{\'n}ski}, {{\v{D}}urech},
  {Castillo-Rogez}, {Cipriani}, {Fetick}, {Fusco}, {Grice}, {Kaasalainen},
  {Kryszczynska}, {Lamy}, {Michalowski}, {Michel}, {Santana-Ros}, {Tanga},
  {Vachier}, {Vigan}, {Witasse}, \& {Yang}}]{Hanus:2020}
{Hanu{\v{s}}}, J., {Vernazza}, P., {Viikinkoski}, M., {et~al.} 2020, \aap, 633,
  A65

\bibitem[{{Hanu{\v{s}}} {et~al.}(2017){Hanu{\v{s}}}, {Viikinkoski}, {Marchis},
  {{\v{D}}urech}, {Kaasalainen}, {Delbo'}, {Herald}, {Frappa}, {Hayamizu},
  {Kerr}, {Preston}, {Timerson}, {Dunham}, \& {Talbot}}]{Hanus:2017}
{Hanu{\v{s}}}, J., {Viikinkoski}, M., {Marchis}, F., {et~al.} 2017, \aap, 601,
  A114

\bibitem[{{Helfenstein} \& {Veverka}(1989)}]{Helfenstein1989}
{Helfenstein}, P. \& {Veverka}, J. 1989, in Asteroids II, ed. R.~P. {Binzel},
  T.~{Gehrels}, \& M.~S. {Matthews}, 557--593

\bibitem[{{Hendler} \& {Malhotra}(2020)}]{Hendler:2020}
{Hendler}, N.~P. \& {Malhotra}, R. 2020, The Planetary Science Journal, 1, 75

\bibitem[{{Holsapple}(2007)}]{Holsapple_2007Icar..187..500H}
{Holsapple}, K.~A. 2007, \icarus, 187, 500

\bibitem[{{Hsieh} \& {Haghighipour}(2016)}]{Hsieh:2016}
{Hsieh}, H.~H. \& {Haghighipour}, N. 2016, \icarus, 277, 19

\bibitem[{{Hsieh} {et~al.}(2004){Hsieh}, {Jewitt}, \&
  {Fern{\'a}ndez}}]{Hsieh:2004}
{Hsieh}, H.~H., {Jewitt}, D.~C., \& {Fern{\'a}ndez}, Y.~R. 2004, \aj, 127, 2997

\bibitem[{{Hutton}(1990)}]{Hutton1990}
{Hutton}, R.~G. 1990, Minor Planet Bulletin, 17, 34

\bibitem[{{Ivantsov}(2008)}]{Ivantsov:2008}
{Ivantsov}, A. 2008, \planss, 56, 1857

\bibitem[{{Ivantsov}(2007)}]{Ivantsov:2007}
{Ivantsov}, A.~V. 2007, Kinematics and Physics of Celestial Bodies, 23, 108

\bibitem[{{Ivezi{\'c}} {et~al.}(2019){Ivezi{\'c}}, {Kahn}, {Tyson}, {Abel},
  {Acosta}, {Allsman}, {Alonso}, {AlSayyad}, {Anderson}, {Andrew}, {Angel},
  {Angeli}, {Ansari}, {Antilogus}, {Araujo}, {Armstrong}, {Arndt}, {Astier},
  {Aubourg}, {Auza}, {Axelrod}, {Bard}, {Barr}, {Barrau}, {Bartlett}, {Bauer},
  {Bauman}, {Baumont}, {Bechtol}, {Bechtol}, {Becker}, {Becla}, {Beldica},
  {Bellavia}, {Bianco}, {Biswas}, {Blanc}, {Blazek}, {Blandford}, {Bloom},
  {Bogart}, {Bond}, {Booth}, {Borgland}, {Borne}, {Bosch}, {Boutigny},
  {Brackett}, {Bradshaw}, {Brandt}, {Brown}, {Bullock}, {Burchat}, {Burke},
  {Cagnoli}, {Calabrese}, {Callahan}, {Callen}, {Carlin}, {Carlson},
  {Chandrasekharan}, {Charles-Emerson}, {Chesley}, {Cheu}, {Chiang}, {Chiang},
  {Chirino}, {Chow}, {Ciardi}, {Claver}, {Cohen-Tanugi}, {Cockrum}, {Coles},
  {Connolly}, {Cook}, {Cooray}, {Covey}, {Cribbs}, {Cui}, {Cutri}, {Daly},
  {Daniel}, {Daruich}, {Daubard}, {Daues}, {Dawson}, {Delgado}, {Dellapenna},
  {de Peyster}, {de Val-Borro}, {Digel}, {Doherty}, {Dubois},
  {Dubois-Felsmann}, {Durech}, {Economou}, {Eifler}, {Eracleous}, {Emmons},
  {Fausti Neto}, {Ferguson}, {Figueroa}, {Fisher-Levine}, {Focke}, {Foss},
  {Frank}, {Freemon}, {Gangler}, {Gawiser}, {Geary}, {Gee}, {Geha}, {Gessner},
  {Gibson}, {Gilmore}, {Glanzman}, {Glick}, {Goldina}, {Goldstein}, {Goodenow},
  {Graham}, {Gressler}, {Gris}, {Guy}, {Guyonnet}, {Haller}, {Harris},
  {Hascall}, {Haupt}, {Hernandez}, {Herrmann}, {Hileman}, {Hoblitt}, {Hodgson},
  {Hogan}, {Howard}, {Huang}, {Huffer}, {Ingraham}, {Innes}, {Jacoby}, {Jain},
  {Jammes}, {Jee}, {Jenness}, {Jernigan}, {Jevremovi{\'c}}, {Johns}, {Johnson},
  {Johnson}, {Jones}, {Juramy-Gilles}, {Juri{\'c}}, {Kalirai}, {Kallivayalil},
  {Kalmbach}, {Kantor}, {Karst}, {Kasliwal}, {Kelly}, {Kessler}, {Kinnison},
  {Kirkby}, {Knox}, {Kotov}, {Krabbendam}, {Krughoff}, {Kub{\'a}nek},
  {Kuczewski}, {Kulkarni}, {Ku}, {Kurita}, {Lage}, {Lambert}, {Lange},
  {Langton}, {Le Guillou}, {Levine}, {Liang}, {Lim}, {Lintott}, {Long},
  {Lopez}, {Lotz}, {Lupton}, {Lust}, {MacArthur}, {Mahabal}, {Mandelbaum},
  {Markiewicz}, {Marsh}, {Marshall}, {Marshall}, {May}, {McKercher}, {McQueen},
  {Meyers}, {Migliore}, {Miller}, {Mills}, {Miraval}, {Moeyens}, {Moolekamp},
  {Monet}, {Moniez}, {Monkewitz}, {Montgomery}, {Morrison}, {Mueller},
  {Muller}, {Mu{\~n}oz Arancibia}, {Neill}, {Newbry}, {Nief}, {Nomerotski},
  {Nordby}, {O'Connor}, {Oliver}, {Olivier}, {Olsen}, {O'Mullane}, {Ortiz},
  {Osier}, {Owen}, {Pain}, {Palecek}, {Parejko}, {Parsons}, {Pease},
  {Peterson}, {Peterson}, {Petravick}, {Libby Petrick}, {Petry},
  {Pierfederici}, {Pietrowicz}, {Pike}, {Pinto}, {Plante}, {Plate}, {Plutchak},
  {Price}, {Prouza}, {Radeka}, {Rajagopal}, {Rasmussen}, {Regnault}, {Reil},
  {Reiss}, {Reuter}, {Ridgway}, {Riot}, {Ritz}, {Robinson}, {Roby}, {Roodman},
  {Rosing}, {Roucelle}, {Rumore}, {Russo}, {Saha}, {Sassolas}, {Schalk},
  {Schellart}, {Schindler}, {Schmidt}, {Schneider}, {Schneider}, {Schoening},
  {Schumacher}, {Schwamb}, {Sebag}, {Selvy}, {Sembroski}, {Seppala}, {Serio},
  {Serrano}, {Shaw}, {Shipsey}, {Sick}, {Silvestri}, {Slater}, {Smith},
  {Smith}, {Sobhani}, {Soldahl}, {Storrie-Lombardi}, {Stover}, {Strauss},
  {Street}, {Stubbs}, {Sullivan}, {Sweeney}, {Swinbank}, {Szalay}, {Takacs},
  {Tether}, {Thaler}, {Thayer}, {Thomas}, {Thornton}, {Thukral}, {Tice},
  {Trilling}, {Turri}, {Van Berg}, {Vanden Berk}, {Vetter}, {Virieux},
  {Vucina}, {Wahl}, {Walkowicz}, {Walsh}, {Walter}, {Wang}, {Wang}, {Warner},
  {Wiecha}, {Willman}, {Winters}, {Wittman}, {Wolff}, {Wood-Vasey}, {Wu},
  {Xin}, {Yoachim}, \& {Zhan}}]{Ivezic:2019}
{Ivezi{\'c}}, {\v{Z}}., {Kahn}, S.~M., {Tyson}, J.~A., {et~al.} 2019, \apj,
  873, 111

\bibitem[{{Jehin} {et~al.}(2011){Jehin}, {Gillon}, {Queloz}, {Magain},
  {Manfroid}, {Chantry}, {Lendl}, {Hutsem{\'e}kers}, \& {Udry}}]{Jehin:2011}
{Jehin}, E., {Gillon}, M., {Queloz}, D., {et~al.} 2011, The Messenger, 145, 2

\bibitem[{{Jorda} {et~al.}(2016){Jorda}, {Gaskell}, {Capanna}, {Hviid}, {Lamy},
  {{\v{D}}urech}, {Faury}, {Groussin}, {Guti{\'e}rrez}, {Jackman}, {Keihm},
  {Keller}, {Knollenberg}, {K{\"u}hrt}, {Marchi}, {Mottola}, {Palmer},
  {Schloerb}, {Sierks}, {Vincent}, {A'Hearn}, {Barbieri}, {Rodrigo}, {Koschny},
  {Rickman}, {Barucci}, {Bertaux}, {Bertini}, {Cremonese}, {Da Deppo},
  {Davidsson}, {Debei}, {De Cecco}, {Fornasier}, {Fulle}, {G{\"u}ttler}, {Ip},
  {Kramm}, {K{\"u}ppers}, {Lara}, {Lazzarin}, {Lopez Moreno}, {Marzari},
  {Naletto}, {Oklay}, {Thomas}, {Tubiana}, \& {Wenzel}}]{Jorda:2016}
{Jorda}, L., {Gaskell}, R., {Capanna}, C., {et~al.} 2016, \icarus, 277, 257

\bibitem[{{Jorda} {et~al.}(2010){Jorda}, {Spjuth}, {Keller}, {Lamy}, \&
  {Llebaria}}]{Jorda:2010}
{Jorda}, L., {Spjuth}, S., {Keller}, H.~U., {Lamy}, P., \& {Llebaria}, A. 2010,
  in Society of Photo-Optical Instrumentation Engineers (SPIE) Conference
  Series, Vol. 7533, Computational Imaging VIII, ed. C.~A. {Bouman},
  I.~{Pollak}, \& P.~J. {Wolfe}, 753311

\bibitem[{{Kne{\v z}evi{\'c}} \&
  {Milani}(2003)}]{Knezevic_Milani_2003A&A...403.1165K}
{Kne{\v z}evi{\'c}}, Z. \& {Milani}, A. 2003, \aap, 403, 1165

\bibitem[{{Kochetova}(2004)}]{Kochetova:2004}
{Kochetova}, O.~M. 2004, Solar System Research, 38, 66

\bibitem[{{Kochetova} \& {Chernetenko}(2014)}]{Kochetova:2014}
{Kochetova}, O.~M. \& {Chernetenko}, Y.~A. 2014, Solar System Research, 48, 295

\bibitem[{{Konopliv} {et~al.}(2011){Konopliv}, {Asmar}, {Folkner}, {Karatekin},
  {Nunes}, {Smrekar}, {Yoder}, \& {Zuber}}]{Konopliv:2011}
{Konopliv}, A.~S., {Asmar}, S.~W., {Folkner}, W.~M., {et~al.} 2011, \icarus,
  211, 401

\bibitem[{{Krasinsky} {et~al.}(2001){Krasinsky}, {Pitjeva}, {Vasiliev}, \&
  {Yagudina}}]{Krasinsky:2001}
{Krasinsky}, G.~A., {Pitjeva}, E.~V., {Vasiliev}, M.~V., \& {Yagudina}, E.~I.
  2001, Communications of IAA of RAS.

\bibitem[{{Kretlow}(2014)}]{Kretlow:2014}
{Kretlow}, M. 2014, Minor Planet Bulletin, 41, 194

\bibitem[{{Kuchynka} \& {Folkner}(2013)}]{Kuchynka:2013}
{Kuchynka}, P. \& {Folkner}, W.~M. 2013, \icarus, 222, 243

\bibitem[{{Lagerkvist} {et~al.}(1995){Lagerkvist}, {Erikson}, {Debehogne},
  {Festin}, {Magnusson}, {Mottola}, {Oja}, {de Angelis}, {Belskaya},
  {Dahlgren}, {Gonano-Beurer}, {Lagerros}, {Lumme}, \&
  {Pohjolainen}}]{Lagerkvist1995}
{Lagerkvist}, C.~I., {Erikson}, A., {Debehogne}, H., {et~al.} 1995, \aaps, 113,
  115

\bibitem[{{Levison} {et~al.}(2009){Levison}, {Bottke}, {Gounelle},
  {Morbidelli}, {Nesvorn{\'y}}, \& {Tsiganis}}]{Levison:2009}
{Levison}, H.~F., {Bottke}, W.~F., {Gounelle}, M., {et~al.} 2009, \nat, 460,
  364

\bibitem[{{Levison} \& {Duncan}(1994)}]{Levison_Duncan_1994Icar..108...18L}
{Levison}, H.~F. \& {Duncan}, M.~J. 1994, \icarus, 108, 18

\bibitem[{{Licandro} {et~al.}(2011){Licandro}, {Campins}, {Kelley}, {Hargrove},
  {Pinilla-Alonso}, {Cruikshank}, {Rivkin}, \& {Emery}}]{Licandro:2011}
{Licandro}, J., {Campins}, H., {Kelley}, M., {et~al.} 2011, \aap, 525, A34

\bibitem[{{Lowry} {et~al.}(2008){Lowry}, {Fitzsimmons}, {Lamy}, \&
  {Weissman}}]{Lowry:2008}
{Lowry}, S., {Fitzsimmons}, A., {Lamy}, P., \& {Weissman}, P. 2008, in The
  Solar System Beyond Neptune, ed. M.~A. {Barucci}, H.~{Boehnhardt}, D.~P.
  {Cruikshank}, A.~{Morbidelli}, \& R.~{Dotson} (The University of Arizona
  Press Tucson), 397

\bibitem[{{Mainzer} {et~al.}(2016){Mainzer}, {Bauer}, {Cutri}, {Grav},
  {Kramer}, {Masiero}, {Nugent}, {Sonnett}, {Stevenson}, \&
  {Wright}}]{Mainzer2016}
{Mainzer}, A.~K., {Bauer}, J.~M., {Cutri}, R.~M., {et~al.} 2016, NASA Planetary
  Data System, 247

\bibitem[{{Marchis} {et~al.}(2005){Marchis}, {Descamps}, {Hestroffer}, \&
  {Berthier}}]{Marchis:2005}
{Marchis}, F., {Descamps}, P., {Hestroffer}, D., \& {Berthier}, J. 2005, \nat,
  436, 822

\bibitem[{{Marchis} {et~al.}(2021){Marchis}, {Jorda}, {Vernazza}, {Bro{\v{z}}},
  {Hanu{\v{s}}}, {Ferrais}, {Vachier}, {Rambaux}, {Marsset}, {Viikinkoski},
  {Jehin}, {Benseguane}, {Podlewska-Gaca}, {Carry}, {Drouard}, {Fauvaud},
  {Birlan}, {Berthier}, {Bartczak}, {Dumas}, {Dudzi{\'n}ski}, {{\v{D}}urech},
  {Castillo-Rogez}, {Cipriani}, {Colas}, {Fetick}, {Fusco}, {Grice},
  {Kryszczynska}, {Lamy}, {Marciniak}, {Michalowski}, {Michel}, {Pajuelo},
  {Santana-Ros}, {Tanga}, {Vigan}, {Witasse}, \& {Yang}}]{Marchis:2021}
{Marchis}, F., {Jorda}, L., {Vernazza}, P., {et~al.} 2021, \aap, 653, A57

\bibitem[{{Marsset} {et~al.}(2020){Marsset}, {Bro{\v{z}}}, {Vernazza},
  {Drouard}, {Castillo-Rogez}, {Hanu{\v{s}}}, {Viikinkoski}, {Rambaux},
  {Carry}, {Jorda}, {{\v{S}}eve{\v{c}}ek}, {Birlan}, {Marchis},
  {Podlewska-Gaca}, {Asphaug}, {Bartczak}, {Berthier}, {Cipriani}, {Colas},
  {Dudzi{\'n}ski}, {Dumas}, {Durech}, {Ferrais}, {F{\'e}tick}, {Fusco},
  {Jehin}, {Kaasalainen}, {Kryszczynska}, {Lamy}, {Le Coroller}, {Marciniak},
  {Michalowski}, {Michel}, {Richardson}, {Santana-Ros}, {Tanga}, {Vachier},
  {Vigan}, {Witasse}, \& {Yang}}]{Marsset:2020_Pallas}
{Marsset}, M., {Bro{\v{z}}}, M., {Vernazza}, P., {et~al.} 2020, Nature
  Astronomy, 4, 569

\bibitem[{{Marsset} {et~al.}(2017){Marsset}, {Carry}, {Dumas}, {Hanu{\v{s}}},
  {Viikinkoski}, {Vernazza}, {M{\"u}ller}, {Delbo}, {Jehin}, {Gillon}, {Grice},
  {Yang}, {Fusco}, {Berthier}, {Sonnett}, {Kugel}, {Caron}, \&
  {Behrend}}]{Marsset:2017}
{Marsset}, M., {Carry}, B., {Dumas}, C., {et~al.} 2017, \aap, 604, A64

\bibitem[{{Moffat}(1969)}]{Moffat:1969}
{Moffat}, A.~F.~J. 1969, \aap, 3, 455

\bibitem[{{Morbidelli} {et~al.}(2005){Morbidelli}, {Levison}, {Tsiganis}, \&
  {Gomes}}]{Morbidelli:2005}
{Morbidelli}, A., {Levison}, H.~F., {Tsiganis}, K., \& {Gomes}, R. 2005, \nat,
  435, 462

\bibitem[{{Mugnier} {et~al.}(2004){Mugnier}, {Fusco}, \&
  {Conan}}]{Mugnier:2004}
{Mugnier}, L.~M., {Fusco}, T., \& {Conan}, J.-M. 2004, Journal of the Optical
  Society of America A, 21, 1841

\bibitem[{{M{\"u}ller} \& {Blommaert}(2004)}]{Muller:2004}
{M{\"u}ller}, T.~G. \& {Blommaert}, J.~A.~D.~L. 2004, \aap, 418, 347

\bibitem[{{Nesvorn{\'y}} {et~al.}(2013){Nesvorn{\'y}}, {Vokrouhlick{\'y}}, \&
  {Morbidelli}}]{Nesvorny:2013}
{Nesvorn{\'y}}, D., {Vokrouhlick{\'y}}, D., \& {Morbidelli}, A. 2013, \apj,
  768, 45

\bibitem[{{Neveu} \& {Vernazza}(2019)}]{Neveu:2019}
{Neveu}, M. \& {Vernazza}, P. 2019, The Astrophysical Journal, 875, 30

\bibitem[{{Novakovic} \& {Radovic}(2019)}]{Novakovic_2019EPSC...13.1671N}
{Novakovic}, B. \& {Radovic}, V. 2019, in EPSC-DPS Joint Meeting 2019, Vol.
  2019, EPSC--DPS2019--1671

\bibitem[{{Nugent} {et~al.}(2016){Nugent}, {Mainzer}, {Bauer}, {Cutri},
  {Kramer}, {Grav}, {Masiero}, {Sonnett}, \& {Wright}}]{Nugent:2016}
{Nugent}, C.~R., {Mainzer}, A., {Bauer}, J., {et~al.} 2016, \aj, 152, 63

\bibitem[{{Nugent} {et~al.}(2015){Nugent}, {Mainzer}, {Masiero}, {Bauer},
  {Cutri}, {Grav}, {Kramer}, {Sonnett}, {Stevenson}, \&
  {Wright}}]{Nugent_2015ApJ...814..117N}
{Nugent}, C.~R., {Mainzer}, A., {Masiero}, J., {et~al.} 2015, \apj, 814, 117

\bibitem[{{O'Rourke} {et~al.}(2020){O'Rourke}, {M{\"u}ller}, {Biver},
  {Bockel{\'e}e-Morvan}, {Hasegawa}, {Valtchanov}, {K{\"u}ppers}, {Fornasier},
  {Campins}, {Fujiwara}, {Teyssier}, \& {Lim}}]{Orourke:2020}
{O'Rourke}, L., {M{\"u}ller}, T.~G., {Biver}, N., {et~al.} 2020, \apjl, 898,
  L45

\bibitem[{{Pajuelo} {et~al.}(2018){Pajuelo}, {Carry}, {Vachier}, {Marsset},
  {Berthier}, {Descamps}, {Merline}, {Tamblyn}, {Grice}, {Conrad}, {Storrs},
  {Timerson}, {Dunham}, {Preston}, {Vigan}, {Yang}, {Vernazza}, {Fauvaud},
  {Bernasconi}, {Romeuf}, {Behrend}, {Dumas}, {Drummond}, {Margot}, {Kervella},
  {Marchis}, \& {Girard}}]{Pajuelo:2018}
{Pajuelo}, M., {Carry}, B., {Vachier}, F., {et~al.} 2018, \icarus, 309, 134

\bibitem[{{Pilcher}(2010)}]{Pilcher2010e}
{Pilcher}, F. 2010, Minor Planet Bulletin, 37, 8

\bibitem[{{Pilcher}(2011)}]{Pilcher2011f}
{Pilcher}, F. 2011, Minor Planet Bulletin, 38, 50

\bibitem[{{Pilcher}(2012)}]{Pilcher2012c}
{Pilcher}, F. 2012, Minor Planet Bulletin, 39, 57

\bibitem[{{Pitjeva}(2013)}]{Pitjeva:2013}
{Pitjeva}, E.~V. 2013, Solar System Research, 47, 386

\bibitem[{{Quinn} {et~al.}(1991){Quinn}, {Tremaine}, \&
  {Duncan}}]{Quinn_1991AJ....101.2287Q}
{Quinn}, T.~R., {Tremaine}, S., \& {Duncan}, M. 1991, \aj, 101, 2287

\bibitem[{{Rambaux} {et~al.}(2017){Rambaux}, {Baguet}, {Chambat}, \&
  {Castillo-Rogez}}]{Rambaux:2017}
{Rambaux}, N., {Baguet}, D., {Chambat}, F., \& {Castillo-Rogez}, J.~C. 2017,
  \apjl, 850, L9

\bibitem[{{Rambaux} \& {Castillo-Rogez}(2020)}]{Rambaux:2020}
{Rambaux}, N. \& {Castillo-Rogez}, J.~C. 2020, \aap, 643, L10

\bibitem[{{Rambaux} {et~al.}(2015){Rambaux}, {Chambat}, \&
  {Castillo-Rogez}}]{Rambaux:2015}
{Rambaux}, N., {Chambat}, F., \& {Castillo-Rogez}, J.~C. 2015, \aap, 584, A127

\bibitem[{{Rivkin} \& {Emery}(2010)}]{Rivkin:2010}
{Rivkin}, A.~S. \& {Emery}, J.~P. 2010, \nat, 464, 1322

\bibitem[{{Santana-Ros} {et~al.}(2016){Santana-Ros}, {Marciniak}, \&
  {Bartczak}}]{Santana:2016}
{Santana-Ros}, T., {Marciniak}, A., \& {Bartczak}, P. 2016, Minor Planet
  Bulletin, 43, 205

\bibitem[{{Schober} {et~al.}(1980){Schober}, {Scaltriti}, {Zappala}, \&
  {Harris}}]{Schober1980}
{Schober}, H.~J., {Scaltriti}, F., {Zappala}, V., \& {Harris}, A.~W. 1980,
  \aap, 91, 1

\bibitem[{{Shevchenko} {et~al.}(1996){Shevchenko}, {Chiorny}, {Kalashnikov},
  {Krugly}, {Mohamed}, \& {Velichko}}]{Shevchenko1996}
{Shevchenko}, V.~G., {Chiorny}, V.~G., {Kalashnikov}, A.~V., {et~al.} 1996,
  \aaps, 115, 475

\bibitem[{{\v{S}idlichovsk{\'y}} \&
  {Nesvorn{\'y}}(1996)}]{Sidlichovsky_1996CeMDA..65..137S}
{\v{S}idlichovsk{\'y}}, M. \& {Nesvorn{\'y}}, D. 1996, Celestial Mechanics and
  Dynamical Astronomy, 65, 137

\bibitem[{{Siltala} \& {Granvik}(2020)}]{Siltala:2020}
{Siltala}, L. \& {Granvik}, M. 2020, \aap, 633, A46

\bibitem[{{Tanga} {et~al.}(2009){Tanga}, {Hestroffer}, {Delb{\`o}}, \&
  {Richardson}}]{Tanga:2009}
{Tanga}, P., {Hestroffer}, D., {Delb{\`o}}, M., \& {Richardson}, D.~C. 2009,
  \planss, 57, 193

\bibitem[{{Thalmann} {et~al.}(2008){Thalmann}, {Schmid}, {Boccaletti},
  {Mouillet}, {Dohlen}, {Roelfsema}, {Carbillet}, {Gisler}, {Beuzit}, {Feldt},
  {Gratton}, {Joos}, {Keller}, {Kragt}, {Pragt}, {Puget}, {Rigal}, {Snik},
  {Waters}, \& {Wildi}}]{Thalmann:2008}
{Thalmann}, C., {Schmid}, H.~M., {Boccaletti}, A., {et~al.} 2008, in Society of
  Photo-Optical Instrumentation Engineers (SPIE) Conference Series, Vol. 7014,
  Ground-based and Airborne Instrumentation for Astronomy II, ed. I.~S.
  {McLean} \& M.~M. {Casali}, 70143F

\bibitem[{{Usui} {et~al.}(2011){Usui}, {Kuroda}, {M{\"u}ller}, {Hasegawa},
  {Ishiguro}, {Ootsubo}, {Ishihara}, {Kataza}, {Takita}, {Oyabu}, {Ueno},
  {Matsuhara}, \& {Onaka}}]{Usui_2011PASJ...63.1117U}
{Usui}, F., {Kuroda}, D., {M{\"u}ller}, T.~G., {et~al.} 2011, \pasj, 63, 1117

\bibitem[{{Vasilyev} \& {Yagudina}(1999)}]{Vasilyev:1999}
{Vasilyev}, M. \& {Yagudina}, E. 1999, Transactions of the Institute of Applied
  Astronomy Russian Academy of Sciences, 4, 98

\bibitem[{{Vernazza} {et~al.}(2018){Vernazza}, {Bro{\v z}}, {Drouard}, {Hanu{\v
  s}}, {Viikinkoski}, {Marsset}, {Jorda}, {Fetick}, {Carry}, {Marchis},
  {Birlan}, {Fusco}, {Santana-Ros}, {Podlewska-Gaca}, {Jehin}, {Ferrais},
  {Bartczak}, {Dudzi{\'n}ski}, {Berthier}, {Castillo-Rogez}, {Cipriani},
  {Colas}, {Dumas}, {{\v D}urech}, {Kaasalainen}, {Kryszczynska}, {Lamy}, {Le
  Coroller}, {Marciniak}, {Michalowski}, {Michel}, {Pajuelo}, {Tanga},
  {Vachier}, {Vigan}, {Warner}, {Witasse}, {Yang}, {Asphaug}, {Richardson},
  {{\v S}eve{\v c}ek}, {Gillon}, \& {Benkhaldoun}}]{Vernazza:2018}
{Vernazza}, P., {Bro{\v z}}, M., {Drouard}, A., {et~al.} 2018, Astron.
  Astrophys., 618, A154

\bibitem[{{Vernazza} {et~al.}(2021){Vernazza}, {Ferrais}, {Jorda},
  {Hanu{\v{s}}}, {Carry}, {Marsset}, {Bro{\v{z}}}, {Fetick}, {Viikinkoski},
  {Marchis}, {Vachier}, {Drouard}, {Fusco}, {Birlan}, {Podlewska-Gaca},
  {Rambaux}, {Neveu}, {Bartczak}, {Dudzi{\'n}ski}, {Jehin}, {Beck}, {Berthier},
  {Castillo-Rogez}, {Cipriani}, {Colas}, {Dumas}, {{\v{D}}urech}, {Grice},
  {Kaasalainen}, {Kryszczynska}, {Lamy}, {Le Coroller}, {Marciniak},
  {Michalowski}, {Michel}, {Santana-Ros}, {Tanga}, {Vigan}, {Witasse}, {Yang},
  {Antonini}, {Audejean}, {Aurard}, {Behrend}, {Benkhaldoun}, {Bosch},
  {Chapman}, {Dalmon}, {Fauvaud}, {Hamanowa}, {Hamanowa}, {His}, {Jones},
  {Kim}, {Kim}, {Krajewski}, {Labrevoir}, {Leroy}, {Livet}, {Molina},
  {Montaigut}, {Oey}, {Payre}, {Reddy}, {Sabin}, {Sanchez}, \&
  {Socha}}]{Vernazza:2021}
{Vernazza}, P., {Ferrais}, M., {Jorda}, L., {et~al.} 2021, \aap, 654, A56

\bibitem[{{Vernazza} {et~al.}(2020){Vernazza}, {Jorda}, {{\v{S}}eve{\v{c}}ek},
  {Bro{\v{z}}}, {Viikinkoski}, {Hanu{\v{s}}}, {Carry}, {Drouard}, {Ferrais},
  {Marsset}, {Marchis}, {Birlan}, {Podlewska-Gaca}, {Jehin}, {Bartczak},
  {Dudzinski}, {Berthier}, {Castillo-Rogez}, {Cipriani}, {Colas}, {DeMeo},
  {Dumas}, {Durech}, {Fetick}, {Fusco}, {Grice}, {Kaasalainen}, {Kryszczynska},
  {Lamy}, {Le Coroller}, {Marciniak}, {Michalowski}, {Michel}, {Rambaux},
  {Santana-Ros}, {Tanga}, {Vachier}, {Vigan}, {Witasse}, {Yang}, {Gillon},
  {Benkhaldoun}, {Szakats}, {Hirsch}, {Duffard}, {Chapman}, \&
  {Maestre}}]{Vernazza:2020}
{Vernazza}, P., {Jorda}, L., {{\v{S}}eve{\v{c}}ek}, P., {et~al.} 2020, Nature
  Astronomy, 4, 136

\bibitem[{{Vernazza} {et~al.}(2015){Vernazza}, {Marsset}, {Beck}, {Binzel},
  {Birlan}, {Brunetto}, {Demeo}, {Djouadi}, {Dumas}, {Merouane}, {Mousis}, \&
  {Zanda}}]{Vernazza:2015}
{Vernazza}, P., {Marsset}, M., {Beck}, P., {et~al.} 2015, \apj, 806, 204

\bibitem[{{Viateau}(2000)}]{Viateau:2000}
{Viateau}, B. 2000, \aap, 354, 725

\bibitem[{{Viikinkoski}(2016)}]{Viikinkoski:2016}
{Viikinkoski}, M. 2016, PhD thesis, Tampere University of Technology

\bibitem[{{Viikinkoski} {et~al.}(2017){Viikinkoski}, {Hanu{\v{s}}},
  {Kaasalainen}, {Marchis}, \& {{\v{D}}urech}}]{Viikinkoski:2017}
{Viikinkoski}, M., {Hanu{\v{s}}}, J., {Kaasalainen}, M., {Marchis}, F., \&
  {{\v{D}}urech}, J. 2017, \aap, 607, A117

\bibitem[{{Viikinkoski} {et~al.}(2015{\natexlab{a}}){Viikinkoski},
  {Kaasalainen}, \& {Durech}}]{Viikinkoski:2015}
{Viikinkoski}, M., {Kaasalainen}, M., \& {Durech}, J. 2015{\natexlab{a}}, \aap,
  576, A8

\bibitem[{{Viikinkoski} {et~al.}(2015{\natexlab{b}}){Viikinkoski},
  {Kaasalainen}, {{\v{D}}urech}, {Carry}, {Marsset}, {Fusco}, {Dumas},
  {Merline}, {Yang}, {Berthier}, {Kervella}, \& {Vernazza}}]{Viikinkoski:2015b}
{Viikinkoski}, M., {Kaasalainen}, M., {{\v{D}}urech}, J., {et~al.}
  2015{\natexlab{b}}, \aap, 581, L3

\bibitem[{{Viikinkoski} {et~al.}(2018){Viikinkoski}, {Vernazza}, {Hanu{\v{s}}},
  {Le Coroller}, {Tazhenova}, {Carry}, {Marsset}, {Drouard}, {Marchis},
  {Fetick}, {Fusco}, {{\v{D}}urech}, {Birlan}, {Berthier}, {Bartczak}, {Dumas},
  {Castillo-Rogez}, {Cipriani}, {Colas}, {Ferrais}, {Grice}, {Jehin}, {Jorda},
  {Kaasalainen}, {Kryszczynska}, {Lamy}, {Marciniak}, {Michalowski}, {Michel},
  {Pajuelo}, {Podlewska-Gaca}, {Santana-Ros}, {Tanga}, {Vachier}, {Vigan},
  {Warner}, {Witasse}, \& {Yang}}]{Viikinkoski:2018}
{Viikinkoski}, M., {Vernazza}, P., {Hanu{\v{s}}}, J., {et~al.} 2018, \aap, 619,
  L3

\bibitem[{{Viswanathan} {et~al.}(2017){Viswanathan}, {Fienga}, {Gastineau}, \&
  {Laskar}}]{Viswanathan:2017}
{Viswanathan}, V., {Fienga}, A., {Gastineau}, M., \& {Laskar}, J. 2017, Notes
  Scientifiques et Techniques de l'Institut de Mecanique Celeste, 108

\bibitem[{{Vokrouhlick{\'y}} {et~al.}(2016){Vokrouhlick{\'y}}, {Bottke}, \&
  {Nesvorn{\'y}}}]{Vokrouhlicky:2016}
{Vokrouhlick{\'y}}, D., {Bottke}, W.~F., \& {Nesvorn{\'y}}, D. 2016, \aj, 152,
  39

\bibitem[{{Vokrouhlick{\'y}} {et~al.}(2010{\natexlab{a}}){Vokrouhlick{\'y}},
  {Nesvorn{\'y}}, {Bottke}, \& {Morbidelli}}]{Vokrouhlicky:2010}
{Vokrouhlick{\'y}}, D., {Nesvorn{\'y}}, D., {Bottke}, W.~F., \& {Morbidelli},
  A. 2010{\natexlab{a}}, \aj, 139, 2148

\bibitem[{{Vokrouhlick{\'y}} {et~al.}(2010{\natexlab{b}}){Vokrouhlick{\'y}},
  {Nesvorn{\'y}}, {Bottke}, \& {Morbidelli}}]{Vokrouhlicky_2010AJ....139.2148V}
{Vokrouhlick{\'y}}, D., {Nesvorn{\'y}}, D., {Bottke}, W.~F., \& {Morbidelli},
  A. 2010{\natexlab{b}}, \aj, 139, 2148

\bibitem[{{Weidenschilling} {et~al.}(1990){Weidenschilling}, {Chapman},
  {Davis}, {Greenberg}, \& {Levy}}]{Weidenschilling1990}
{Weidenschilling}, S.~J., {Chapman}, C.~R., {Davis}, D.~R., {Greenberg}, R., \&
  {Levy}, D.~H. 1990, Icarus, 86, 402

\bibitem[{{Weidenschilling} {et~al.}(1987){Weidenschilling}, {Chapman},
  {Davis}, {Greenberg}, {Levy}, \& {Vail}}]{Weidenschilling1987}
{Weidenschilling}, S.~J., {Chapman}, C.~R., {Davis}, D.~R., {et~al.} 1987,
  \icarus, 70, 191

\bibitem[{{Yang} {et~al.}(2020){Yang}, {Hanu{\v{s}}}, {Carry}, {Vernazza},
  {Bro{\v{z}}}, {Vachier}, {Rambaux}, {Marsset}, {Chrenko},
  {{\v{S}}eve{\v{c}}ek}, {Viikinkoski}, {Jehin}, {Ferrais}, {Podlewska-Gaca},
  {Drouard}, {Marchis}, {Birlan}, {Benkhaldoun}, {Berthier}, {Bartczak},
  {Dumas}, {Dudzi{\'n}ski}, {{\v{D}}urech}, {Castillo-Rogez}, {Cipriani},
  {Colas}, {Fetick}, {Fusco}, {Grice}, {Jorda}, {Kaasalainen}, {Kryszczynska},
  {Lamy}, {Marciniak}, {Michalowski}, {Michel}, {Pajuelo}, {Santana-Ros},
  {Tanga}, {Vigan}, \& {Witasse}}]{Yang:2020}
{Yang}, B., {Hanu{\v{s}}}, J., {Carry}, B., {et~al.} 2020, \aap, 641, A80

\bibitem[{{Yang} {et~al.}(2016){Yang}, {Wahhaj}, {Beauvalet}, {Marchis},
  {Dumas}, {Marsset}, {Nielsen}, \& {Vachier}}]{Yang:2016}
{Yang}, B., {Wahhaj}, Z., {Beauvalet}, L., {et~al.} 2016, \apjl, 820, L35

\bibitem[{{Zielenbach}(2011)}]{Zielenbach:2011}
{Zielenbach}, W. 2011, \aj, 142, 120

\end{thebibliography}

\begin{appendix}

\section{Observational circumstances}
\label{sec:appA}

Here, we provide the observational circumstances for the disk-resolved (AO images) and disk-integrated (optical light curves) data used to reconstruct the 3D shape of (65)~Cybele. Table~\ref{tab:ao} provides the circumstances for each VLT/SPHERE disk-resolved image obtained in this work. Table~\ref{tab:lcs} provides the circumstances for the new optical light curves acquired in this work, as well as previous light curves retrieved from the literature.

%Added by TeX Support
\onecolumn
%\scriptsize{
\begin{table*}
\caption{\label{tab:ao} VLT/SPHERE disk-resolved images obtained in the N$\_$R filter with the ZIMPOL camera.}
\centering
\begin{tabular}{rr rr rrr rrr}
\hline 
\multicolumn{1}{c} {Date} & \multicolumn{1}{c} {UT} & \multicolumn{1}{c} {Exp} & \multicolumn{1}{c} {Airmass} & \multicolumn{1}{c} {$\Delta$} & \multicolumn{1}{c} {$r$} & \multicolumn{1}{c} {$\alpha$} & \multicolumn{1}{c} {$D_\mathrm{a}$}  & \multicolumn{1}{c} {$\lambda_\mathrm{subE}$} & \multicolumn{1}{c} {$\beta_\mathrm{subE}$}\\
\multicolumn{1}{c} {} & \multicolumn{1}{c} {} & \multicolumn{1}{c} {(s)} & \multicolumn{1}{c} {} & \multicolumn{1}{c} {(AU)} & \multicolumn{1}{c} {(AU)} & \multicolumn{1}{c} {(\degr)} & \multicolumn{1}{c} {(\arcsec)}  & \multicolumn{1}{c} {(\degr)} & \multicolumn{1}{c} {(\degr)}\\\\
% &  &  &  &  &  &  &  & &  \\
\hline\hline
  2021$-$07$-$03 &     06:49:07 & 240 & 1.06 & 2.09 & 3.09 & 4.2 & 0.174 & 122.8 & $-$0.3\\
  2021$-$07$-$03 &     06:53:18 & 240 & 1.06 & 2.09 & 3.09 & 4.2 & 0.174 & 118.7 & $-$0.3\\
  2021$-$07$-$03 &     06:57:28 & 240 & 1.07 & 2.09 & 3.09 & 4.2 & 0.174 & 114.6 & $-$0.3\\
  2021$-$07$-$03 &     07:01:36 & 240 & 1.07 & 2.09 & 3.09 & 4.2 & 0.174 & 110.4 & $-$0.3\\
  2021$-$07$-$03 &     07:05:45 & 240 & 1.08 & 2.09 & 3.09 & 4.2 & 0.174 & 106.3 & $-$0.3\\
  2021$-$07$-$21 &     02:25:18 & 240 & 1.11 & 2.09 & 3.10 & 2.8 & 0.174 & 11.0 & $-$3.4\\
  2021$-$07$-$21 &     02:29:29 & 240 & 1.10 & 2.09 & 3.10 & 2.8 & 0.174 & 6.9 & $-$3.4\\
  2021$-$07$-$21 &     02:33:39 & 240 & 1.09 & 2.09 & 3.10 & 2.8 & 0.174 & 2.8 & $-$3.4\\
  2021$-$07$-$21 &     02:37:48 & 240 & 1.08 & 2.09 & 3.10 & 2.8 & 0.174 & 358.7 & $-$3.4\\
  2021$-$07$-$21 &     02:41:56 & 240 & 1.08 & 2.09 & 3.10 & 2.8 & 0.174 & 354.7 & $-$3.4\\
  2021$-$07$-$21 &     02:52:32 & 240 & 1.06 & 2.09 & 3.10 & 2.8 & 0.174 & 344.2 & $-$3.4\\
  2021$-$07$-$21 &     02:56:41 & 240 & 1.05 & 2.09 & 3.10 & 2.8 & 0.174 & 340.0 & $-$3.4\\
  2021$-$07$-$21 &     03:00:53 & 240 & 1.05 & 2.09 & 3.10 & 2.8 & 0.174 & 335.9 & $-$3.4\\
  2021$-$07$-$21 &     03:05:02 & 240 & 1.04 & 2.09 & 3.10 & 2.8 & 0.174 & 331.8 & $-$3.4\\
  2021$-$07$-$21 &     03:09:11 & 240 & 1.04 & 2.09 & 3.10 & 2.8 & 0.174 & 327.7 & $-$3.4\\
  2021$-$07$-$21 &     03:19:58 & 240 & 1.03 & 2.09 & 3.10 & 2.8 & 0.174 & 317.0 & $-$3.4\\
  2021$-$07$-$21 &     03:24:08 & 240 & 1.02 & 2.09 & 3.10 & 2.8 & 0.174 & 313.1 & $-$3.4\\
  2021$-$07$-$21 &     03:28:18 & 240 & 1.02 & 2.09 & 3.10 & 2.8 & 0.174 & 308.9 & $-$3.4\\
  2021$-$07$-$21 &     03:32:26 & 240 & 1.02 & 2.09 & 3.10 & 2.8 & 0.174 & 304.8 & $-$3.4\\
  2021$-$07$-$21 &     03:36:35 & 240 & 1.02 & 2.09 & 3.10 & 2.8 & 0.174 & 300.7 & $-$3.4\\
  2021$-$07$-$21 &     04:47:09 & 240 & 1.02 & 2.09 & 3.10 & 2.8 & 0.174 & 231.1 & $-$3.4\\
  2021$-$07$-$21 &     04:51:19 & 240 & 1.02 & 2.09 & 3.10 & 2.8 & 0.174 & 227.0 & $-$3.4\\
  2021$-$07$-$21 &     04:55:30 & 240 & 1.02 & 2.09 & 3.10 & 2.8 & 0.174 & 222.8 & $-$3.4\\
  2021$-$07$-$21 &     04:59:39 & 240 & 1.03 & 2.09 & 3.10 & 2.8 & 0.174 & 218.7 & $-$3.4\\
  2021$-$07$-$21 &     05:03:48 & 240 & 1.03 & 2.09 & 3.10 & 2.8 & 0.174 & 214.6 & $-$3.4\\
  2021$-$07$-$21 &     07:12:37 & 240 & 1.36 & 2.09 & 3.10 & 2.9 & 0.174 & 87.6 & $-$3.4\\
  2021$-$07$-$21 &     07:16:46 & 240 & 1.38 & 2.09 & 3.10 & 2.9 & 0.174 & 83.5 & $-$3.4\\
  2021$-$07$-$21 &     07:20:56 & 240 & 1.41 & 2.09 & 3.10 & 2.9 & 0.174 & 79.4 & $-$3.4\\
  2021$-$07$-$21 &     07:25:05 & 240 & 1.43 & 2.09 & 3.10 & 2.9 & 0.174 & 75.2 & $-$3.4\\
  2021$-$07$-$21 &     07:29:13 & 240 & 1.45 & 2.09 & 3.10 & 2.9 & 0.174 & 71.1 & $-$3.4\\
  2021$-$08$-$22 &     03:36:14 & 240 & 1.11 & 2.30 & 3.12 & 12.7 & 0.158 & 200.4 & $-$7.5\\
  2021$-$08$-$22 &     03:40:24 & 240 & 1.12 & 2.30 & 3.12 & 12.7 & 0.158 & 196.5 & $-$7.5\\
  2021$-$08$-$22 &     03:44:35 & 240 & 1.13 & 2.30 & 3.12 & 12.7 & 0.158 & 192.2 & $-$7.5\\
  2021$-$08$-$22 &     03:48:42 & 240 & 1.14 & 2.30 & 3.12 & 12.7 & 0.158 & 188.2 & $-$7.5\\
  2021$-$08$-$22 &     03:52:51 & 240 & 1.15 & 2.30 & 3.12 & 12.7 & 0.158 & 184.1 & $-$7.5\\

\hline
\end{tabular}
\tablefoot{
For each observation, we provide the epoch, the exposure time, the airmass, the distance to the Earth $\Delta$ and to the Sun $r$, the phase angle $\alpha$, the angular diameter $D_\mathrm{a}$ of Cybele, and the sub$-$Earth point longitude $\lambda_\mathrm{subE}$ and latitude $\beta_\mathrm{subE}$. The total exposure time of 240 $s$ corresponds to 1.1\% of the rotation period or 4\degr\, in the rotation phase.
    }
\end{table*}
\twocolumn
\onecolumn
%\scriptsize{
\begin{longtable}{rlr rrr l l}
\caption{\label{tab:lcs} Optical disk-integrated lightcurves of (65)~Cybele used for the \adam{} and \sage{} shape modelling.}\\
\hline 
\multicolumn{1}{c} {N} & \multicolumn{1}{c} {Epoch} & \multicolumn{1}{c} {$N_p$} & \multicolumn{1}{c} {$\Delta$} & \multicolumn{1}{c} {$r$} & \multicolumn{1}{c} {$\varphi$} & \multicolumn{1}{c} {Filter} & Reference \\
 &  &  & (AU) & (AU) & (\degr) &  &  \\
\hline\hline
\endfirsthead
\caption{continued.}\\
\hline
\multicolumn{1}{c} {N} & \multicolumn{1}{c} {Epoch} & \multicolumn{1}{c} {$N_p$} & \multicolumn{1}{c} {$\Delta$} & \multicolumn{1}{c} {$r$} & \multicolumn{1}{c} {$\varphi$} & \multicolumn{1}{c} {Filter} & Reference \\
 &  &  & (AU) & (AU) & (\degr) &  &  \\
\hline\hline
\endhead
\hline
\endfoot
\hline
     1  &  1977-08-31.2  &  56   &  2.23  &  3.24  &  2.2   &  V &  \citet{Schober1980}  \\
     2  &  1977-09-01.1  &  29   &  2.23  &  3.24  &  2.5   &  V &  \citet{Schober1980}  \\
     3  &  1977-09-02.2  &  65   &  2.24  &  3.24  &  2.9   &  V &  \citet{Schober1980}  \\
     4  &  1978-11-08.3  &  9    &  2.67  &  3.65  &  2.6   &  V &  \citet{Schober1980}  \\
     5  &  1982-02-17.2  &  10   &  2.81  &  3.23  &  17.0  &  V &  \citet{Weidenschilling1987} \\
     6  &  1982-02-18.3  &  12   &  2.79  &  3.23  &  16.9  &  V &  \citet{Weidenschilling1987} \\
     7  &  1982-02-19.3  &  15   &  2.77  &  3.23  &  16.8  &  V &  \citet{Weidenschilling1987} \\
     8  &  1982-02-20.3  &  6    &  2.76  &  3.23  &  16.8  &  V &  \citet{Weidenschilling1987} \\
     9  &  1982-05-21.3  &  16   &  2.23  &  3.16  &  8.6   &  V &  \citet{Weidenschilling1987} \\
    10  &  1982-05-22.3  &  5    &  2.23  &  3.16  &  8.9   &  V &  \citet{Weidenschilling1987} \\
    11  &  1982-05-23.2  &  9    &  2.24  &  3.16  &  9.2   &  V &  \citet{Weidenschilling1987} \\
    12  &  1982-06-22.3  &  6    &  2.50  &  3.14  &  16.3  &  V &  \citet{Weidenschilling1987} \\
    13  &  1982-06-23.2  &  11   &  2.51  &  3.13  &  16.5  &  V &  \citet{Weidenschilling1987} \\
    14  &  1982-06-24.4  &  8    &  2.53  &  3.13  &  16.7  &  V &  \citet{Weidenschilling1987} \\
    15  &  1983-05-21.2  &  6    &  2.58  &  3.09  &  17.7  &  V &  \citet{Weidenschilling1987} \\
    16  &  1983-06-30.4  &  23   &  2.19  &  3.11  &  9.2   &  V &  \citet{Weidenschilling1987} \\
    17  &  1983-09-14.2  &  18   &  2.46  &  3.16  &  14.9  &  V &  \citet{Weidenschilling1987} \\
    18  &  1983-09-19.4  &  11   &  2.52  &  3.16  &  15.7  &  V &  \citet{Weidenschilling1987} \\
    19  &  1983-10-11.2  &  9    &  2.82  &  3.18  &  17.9  &  V &  \citet{Weidenschilling1987} \\
    20  &  1983-10-12.4  &  11   &  2.84  &  3.18  &  18.0  &  V &  \citet{Weidenschilling1987} \\
    21  &  1983-10-13.3  &  25   &  2.85  &  3.18  &  18.0  &  V &  \citet{Weidenschilling1987} \\
    22  &  1983-10-15.4  &  5    &  2.89  &  3.18  &  18.1  &  V &  \citet{Weidenschilling1987} \\
    23  &  1983-11-11.4  &  10   &  3.29  &  3.20  &  17.5  &  V &  \citet{Weidenschilling1987} \\
    24  &  1983-11-13.2  &  12   &  3.31  &  3.21  &  17.4  &  V &  \citet{Weidenschilling1987} \\
    25  &  1983-11-15.3  &  9    &  3.34  &  3.21  &  17.2  &  V &  \citet{Weidenschilling1987} \\
    26  &  1984-07-05.2  &  7    &  3.37  &  3.43  &  17.2  &  V &  \citet{Weidenschilling1987} \\
    27  &  1984-11-21.2  &  22   &  2.83  &  3.56  &  12.1  &  V &  \citet{Weidenschilling1987} \\
    28  &  1985-10-25.2  &  26   &  3.03  &  3.77  &  11.2  &  V &  \citet{Weidenschilling1987} \\
    29  &  1986-01-19.2  &  38   &  3.11  &  3.79  &  11.8  &  V &  \citet{Weidenschilling1987} \\
    30  &  1986-01-20.2  &  27   &  3.12  &  3.79  &  12.0  &  V &  \citet{Weidenschilling1987} \\
    31  &  1987-02-07.2  &  44   &  2.72  &  3.69  &  3.7   &  V &  \citet{Weidenschilling1990} \\
    32  &  1988-04-26.2  &  63   &  2.39  &  3.29  &  9.2   &  V &  \citet{Weidenschilling1990} \\
    33  &  1989-07-02.2  &  28   &  2.08  &  3.08  &  3.7   &  V &  \citet{Hutton1990}    \\
    34  &  1994-02-12.0  &  16   &  2.56  &  3.48  &  6.9   &  V &  \citet{Shevchenko1996}\\
    35  &  1994-02-17.9  &  28   &  2.52  &  3.48  &  5.0   &  V &  \citet{Shevchenko1996}\\
    36  &  1994-04-11.0  &  61   &  2.62  &  3.43  &  11.3  &  C &  \citet{Lagerkvist1995} \\
    37  &  2007-04-06.0  &  18   &  2.34  &  3.32  &  4.7   &  C &  \citet{Franco2015} \\
    38  &  2007-04-07.0  &  25   &  2.35  &  3.32  &  5.1   &  C &  \citet{Franco2015} \\
    39  &  2009-07-31.4  &  166  &  2.51  &  3.33  &  11.9  &  R &  \citet{Pilcher2010e}\\
    40  &  2009-08-03.4  &  76   &  2.49  &  3.33  &  11.2  &  R &  \citet{Pilcher2010e}\\
    41  &  2009-08-16.3  &  258  &  2.40  &  3.34  &  7.5   &  R &  \citet{Pilcher2010e}\\
    42  &  2009-08-20.4  &  186  &  2.38  &  3.35  &  6.3   &  R &  \citet{Pilcher2010e}\\
    43  &  2009-08-23.4  &  467  &  2.37  &  3.35  &  5.3   &  R &  \citet{Pilcher2010e}\\
    44  &  2009-08-29.3  &  251  &  2.36  &  3.36  &  3.3   &  R &  \citet{Pilcher2010e}\\
    45  &  2009-09-13.2  &  186  &  2.37  &  3.37  &  1.9   &  R &  \citet{Pilcher2010e}\\
    46  &  2009-09-26.9  &  28   &  2.44  &  3.39  &  6.4   &  C &  \citet{Viikinkoski:2017} \\
    47  &  2009-09-27.0  &  36   &  2.44  &  3.39  &  6.4   &  C &  \citet{Viikinkoski:2017} \\
    48  &  2010-09-19.1  &  68   &  3.06  &  3.70  &  13.1  &  C &  \citet{Viikinkoski:2017} \\
    49  &  2010-09-24.4  &  261  &  3.00  &  3.70  &  12.3  &  R &  \citet{Pilcher2011f}\\
    50  &  2011-01-23.9  &  238  &  3.50  &  3.77  &  15.0  &  C &  \citet{Viikinkoski:2017} \\
    51  &  2011-01-25.9  &  246  &  3.54  &  3.77  &  15.1  &  C &  \citet{Viikinkoski:2017} \\
    52  &  2011-11-19.5  &  86   &  3.06  &  3.78  &  11.3  &  C &  \citet{Pilcher2012c}\\
    53  &  2011-11-22.4  &  324  &  3.03  &  3.78  &  10.8  &  C &  \citet{Pilcher2012c}\\
    54  &  2011-11-29.4  &  257  &  2.96  &  3.78  &  9.3   &  C &  \citet{Pilcher2012c}\\
    55  &  2011-12-05.4  &  394  &  2.90  &  3.78  &  7.8   &  C &  \citet{Pilcher2012c}\\
    56  &  2011-12-27.3  &  425  &  2.79  &  3.77  &  1.7   &  C &  \citet{Pilcher2012c}\\
    57  &  2012-01-03.3  &  354  &  2.79  &  3.77  &  1.3   &  C &  \citet{Pilcher2012c}\\
    58  &  2016-10-02.0  &  40   &  2.68  &  3.63  &  5.8   &  r &  This work$^1$     \\
    59  &  2016-10-31.1  &  216  &  2.68  &  3.65  &  3.6   &  r &  This work$^1$     \\
    60  &  2016-10-04.2  &  40   &  2.67  &  3.63  &  5.2   &  R &  This work$^1$   \\
    61  &  2016-10-07.2  &  216  &  2.66  &  3.63  &  4.3   &  r &  This work$^1$     \\
    62  &  2016-11-01.1  &  636  &  2.68  &  3.65  &  3.8   &  R &  This work$^1$   \\
    63  &  2016-11-20.1  &  407  &  2.82  &  3.67  &  9.1   &  r &  This work$^1$     \\
    64  &  2016-09-14.2  &  26   &  2.79  &  3.61  &  10.6  &  R &  This work$^1$   \\
    65  &  2016-09-15.1  &  17   &  2.79  &  3.61  &  10.3  &  R &  This work$^1$     \\
    66  &  2016-09-15.2  &  23   &  2.79  &  3.61  &  10.4  &  R &  This work$^1$   \\
    67  &  2016-09-22.1  &  79   &  2.73  &  3.62  &  8.6   &  R &  This work$^1$     \\
    68  &  2016-09-24.2  &  90   &  2.72  &  3.62  &  8.1   &  R &  This work$^1$  \\
    69  &  2016-09-28.2  &  117  &  2.70  &  3.62  &  7.0   &  V &  This work$^1$  \\
    70  &  2016-09-29.2  &  101  &  2.69  &  3.62  &  6.7   &  V &  This work$^1$  \\
    71  &  2016-09-05.2  &  141  &  2.88  &  3.60  &  12.5  &  r &  This work$^1$     \\
    72  &  2016-09-09.2  &  141  &  2.84  &  3.61  &  11.7  &  r &  This work$^1$     \\
    73  &  2017-10-25.2  &  299  &  3.16  &  3.81  &  12.5  &  r &  This work$^1$     \\
    74  &  2017-12-18.0  &  473  &  2.83  &  3.81  &  1.9   &  r &  This work$^1$     \\
    75  &  2021-04-18.4  &  279  &  2.81  &  3.05  &  19.1  &  Rc&  This work$^2$ \\
    76  &  2021-04-19.4  &  261  &  2.80  &  3.05  &  19.1  &  Rc&  This work$^2$ \\
    77  &  2021-05-11.4  &  167  &  2.52  &  3.06  &  17.6  &  Rc&  This work$^2$ \\
    78  &  2021-05-12.4  &  249  &  2.51  &  3.06  &  17.5  &  Rc&  This work$^2$ \\
    79  &  2021-05-16.4  &  183  &  2.46  &  3.06  &  17.0  &  Rc&  This work$^2$ \\
    80  &  2021-05-18.4  &  143  &  2.43  &  3.06  &  16.7  &  Rc&  This work$^2$ \\
    81  &  2021-05-26.4  &  588  &  2.35  &  3.07  &  15.3  &  Rc&  This work$^2$ \\
    82  &  2021-05-28.2  &  193  &  2.33  &  3.07  &  14.9  &  Rc&  This work$^2$ \\
    83  &  2021-06-25.4  &  430  &  2.11  &  3.08  &  7.1   &  Rc&  This work$^2$ \\
    84  &  2021-06-26.2  &  148  &  2.11  &  3.08  &  6.7   &  Rc&  This work$^2$ \\
    85  &  2021-06-28.2  &  425  &  2.10  &  3.08  &  6.1   &  Rc&  This work$^2$ \\
    86  &  2021-06-28.3  &  269  &  2.10  &  3.08  &  6.0   &  Rc&  This work$^2$ \\
    87  &  2021-07-14.1  &  240  &  2.08  &  3.09  &  1.2   &  Rc&  This work$^2$ \\
    88  &  2021-07-16.1  &  173  &  2.08  &  3.09  &  1.4   &  Rc&  This work$^2$ \\
    89  &  2021-07-20.2  &  360  &  2.08  &  3.10  &  2.4   &  Rc&  This work$^2$ \\
    90  &  2021-07-04.1  &  304  &  2.08  &  3.09  &  4.0   &  Rc&  This work$^2$ \\
    91  &  2021-07-05.2  &  241  &  2.08  &  3.09  &  3.6   &  Rc&  This work$^2$ \\
    92  &  2021-08-10.3  &  547  &  2.19  &  3.11  &  9.4   &  Rc&  This work$^2$ \\
   \hline
\end{longtable}
\tablefoot{
The table contains the epoch, the number of individual measurements $N_p$, the asteroid's distances to the Earth $\Delta$ and to the Sun $r$, the phase angle $\varphi$, the photometric filter and the reference.
     $^1$Gaia-GOSA (Gaia-Ground-based Observational Service for Asteroids, \url{www.gaiagosa.eu}),
     $^2$TRAPPIST.
    }

\twocolumn

\section{Mass estimates of binary asteroids}
\label{sec:appBC}

The same method of mass determination 
used for (65)~Cybele was applied to a number of binary and triple asteroids in order to estimate the accuracy of the method by comparing our derived values to the mass obtained from the orbital study of the companions. 
Figure\,\ref{fig:mass_binaries} presents the result of the analysis for eight different binaries and triple systems. 
The average discrepancy between the methods is found to be $\sim$12\% of the mass value, which we adopted as our uncertainty on the mass of (65)~Cybele (Section\,\ref{sec:density}). We note that in the case of (121)~Hermione and (130)~Elektra, a better fit is obtained by applying a 2-$\sigma$ clipping instead of 3-$\sigma$ clipping of the measurements. 
Mass estimates derived from studies of planetary ephemeris and orbital deflections during close encounters are from: 

\begin{itemize}
    \item (22)~Kalliope: \citet{Kochetova:2004, Folkner:2009, Zielenbach:2011, Fienga:2013, Fienga:2019, Fienga:2020, Goffin:2014}.
    \item (31)~Euphrosyne: \citet{Kochetova:2004, Baer:2008, Fienga:2009, Fienga:2013, Fienga:2019, Folkner:2009, Baer:2011, Zielenbach:2011, Kuchynka:2013, Goffin:2014, Kochetova:2014, Viswanathan:2017}.
    \item (41)~Daphne: \citet{Fienga:2009, Fienga:2011, Fienga:2014, Fienga:2019, Folkner:2009, Konopliv:2011, Zielenbach:2011, Kuchynka:2013, Pitjeva:2013, Goffin:2014, Kochetova:2014, Siltala:2020}.
    \item (45)~Eugenia: \citet{Vasilyev:1999, Krasinsky:2001, Aslan:2007, Ivantsov:2007, Ivantsov:2008, Folkner:2009, Goffin:2014, Zielenbach:2011, Fienga:2014, Fienga:2019, Fienga:2020}.
    \item (87)~Sylvia: \citet{Vasilyev:1999, Krasinsky:2001, Aslan:2007, Ivantsov:2007, Ivantsov:2008,  Zielenbach:2011, Folkner:2014, Goffin:2014, Fienga:2019}
    \item (107)~Camilla: \citet{Ivantsov:2007, Fienga:2011, Fienga:2013, Fienga:2019, Fienga:2020, Zielenbach:2011, Folkner:2014, Goffin:2014, Viswanathan:2017}
    \item (121)~Hermione: \citet{Viateau:2000, Zielenbach:2011, Folkner:2014, Goffin:2014,    Kretlow:2014, Baer:2017, Viswanathan:2017,     Fienga:2019}.
    \item (130)~Elektra:
    \citet{Zielenbach:2011, Fienga:2011, Fienga:2019, Fienga:2020, Goffin:2014, Viswanathan:2017}.
\end{itemize}

We note that we did not use mass estimates from \citet{Folkner:2014} as no uncertainties on the measurements were reported by the authors. Mass estimates of (22)~Kalliope, (45)~Eugenia, and (87)~Sylvia from \citet{Viswanathan:2017} were discarded as being largely inconsistent with other estimates for the same objects. 
Mass estimates derived from orbital fitting of the satellite(s) are from: 

\begin{itemize}
    \item (22)~Kalliope: \citet{Ferrais:2022}
    \item (31)~Euphrosyne: \citet{Yang:2020}
    \item (41)~Daphne: \citet{Carry:2019}
    \item (45)~Eugenia: Bro\v z et al. (private comm.)
    \item (87)~Sylvia: \citet{Carry:2021}
    \item (107)~Camilla: \citet{Pajuelo:2018}
    \item (121)~Hermione: Ferrais et al. (in prep.)
    \item (130)~Elektra: \citet{Yang:2016}.
\end{itemize}

%%%%%%%%%% Start Figure %%%%%%%%%%%%%%%%%%%%%%%%
   \begin{figure*}
   \centering
   \includegraphics[width=0.495\hsize, trim={0 0 0 0}, clip]{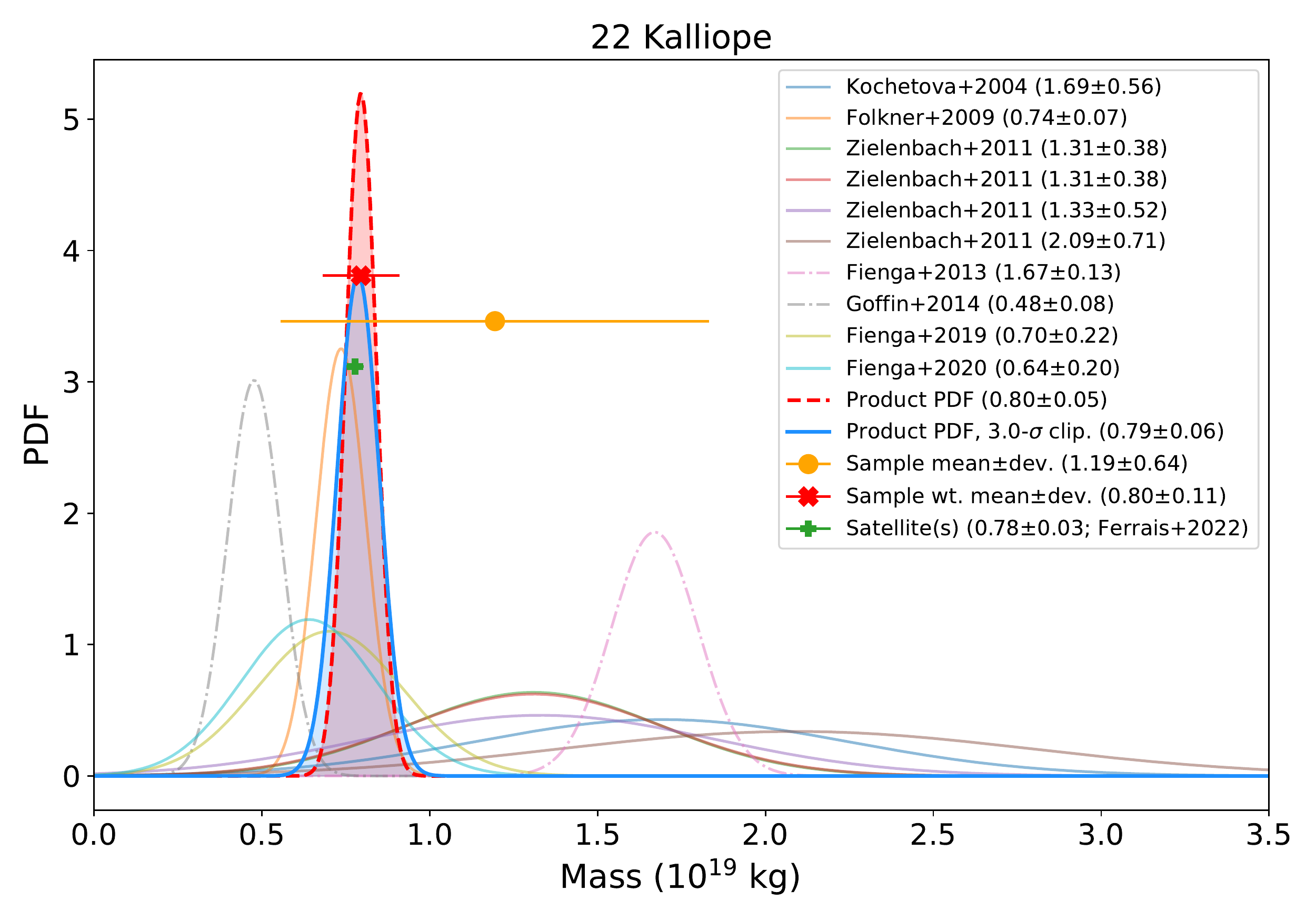}
   \includegraphics[width=0.495\hsize, trim={0 0 0 0}, clip]{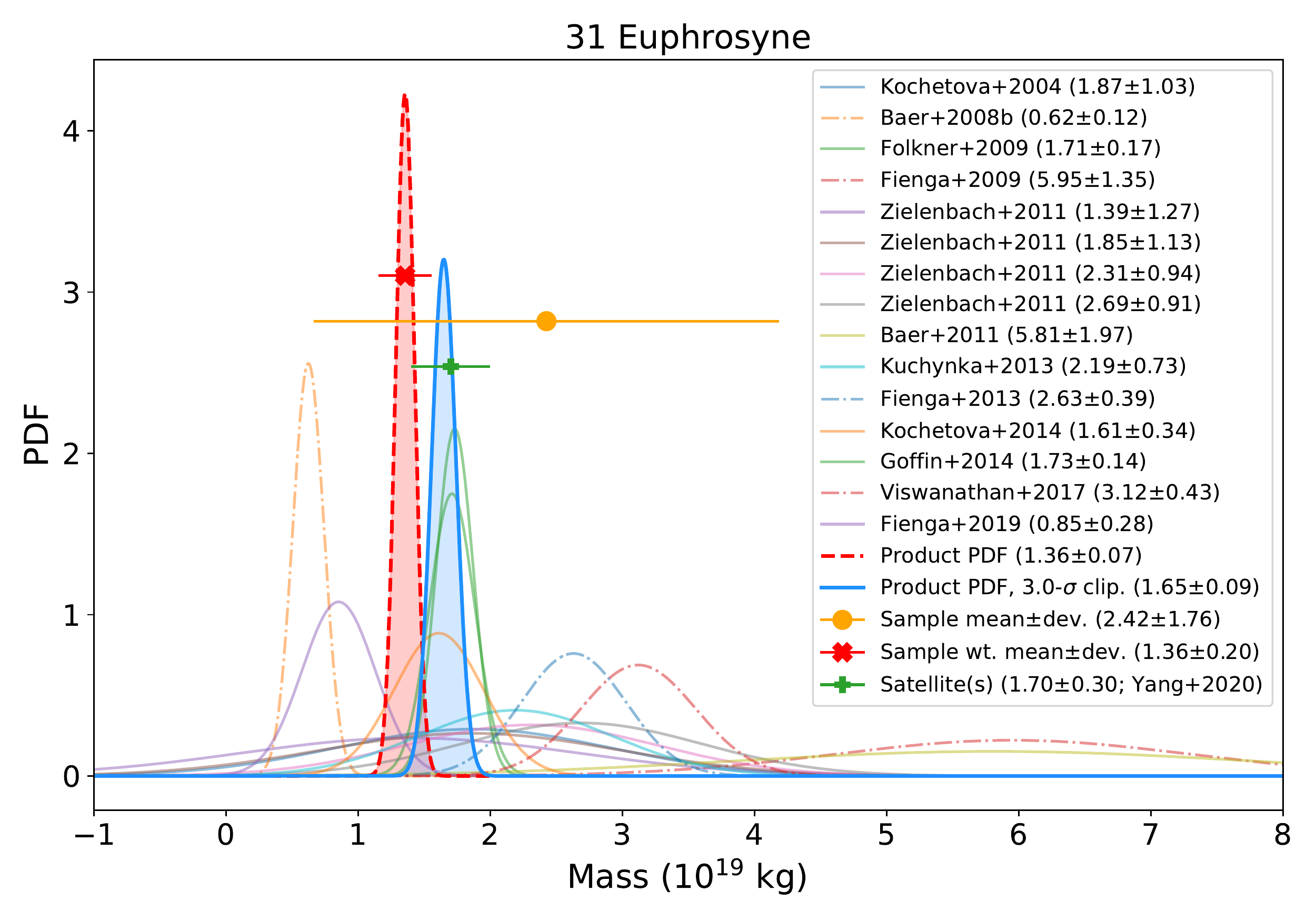}
   \includegraphics[width=0.495\hsize, trim={0 0 0 0}, clip]{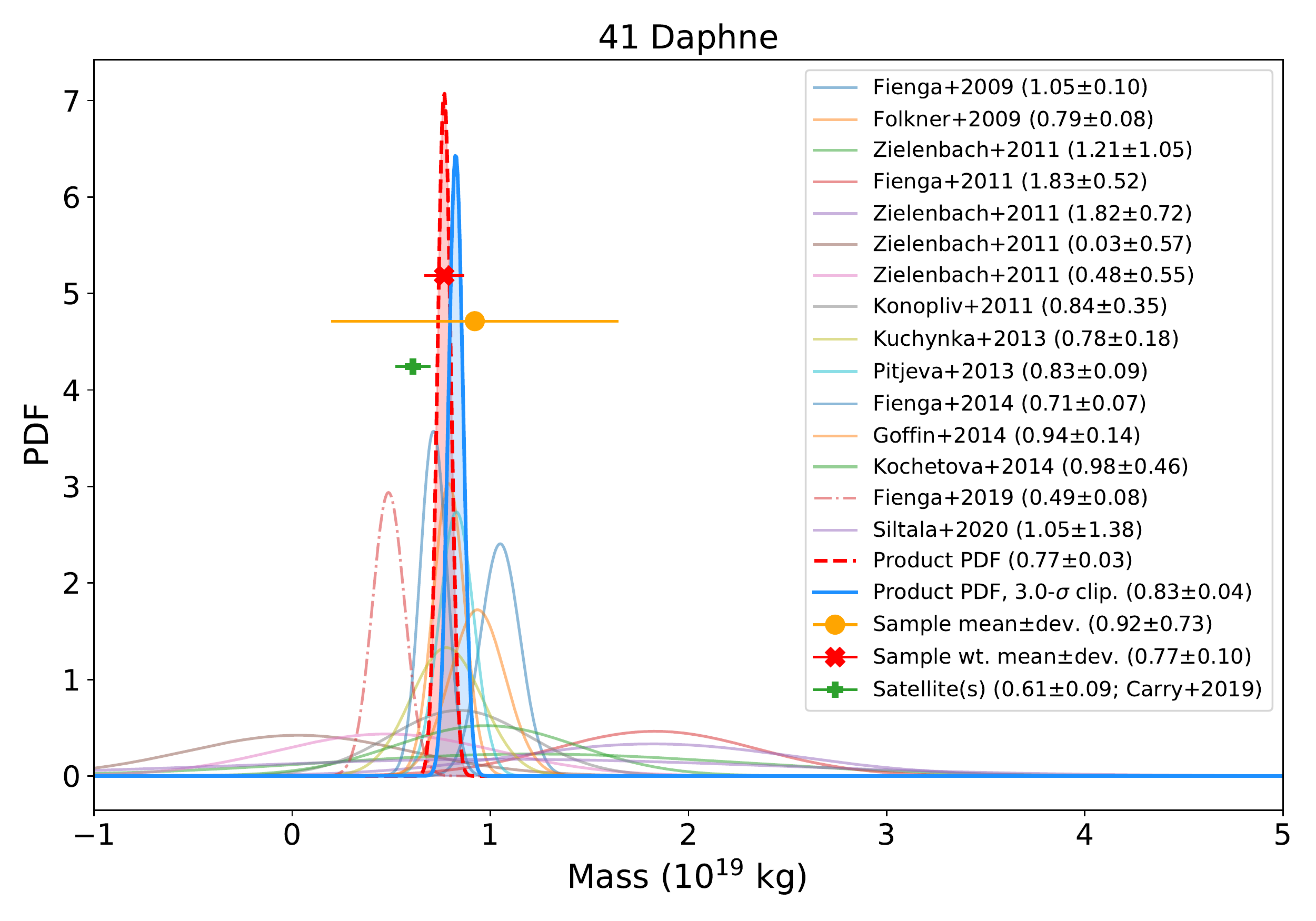}
   \includegraphics[width=0.495\hsize, trim={0 0 0 0}, clip]{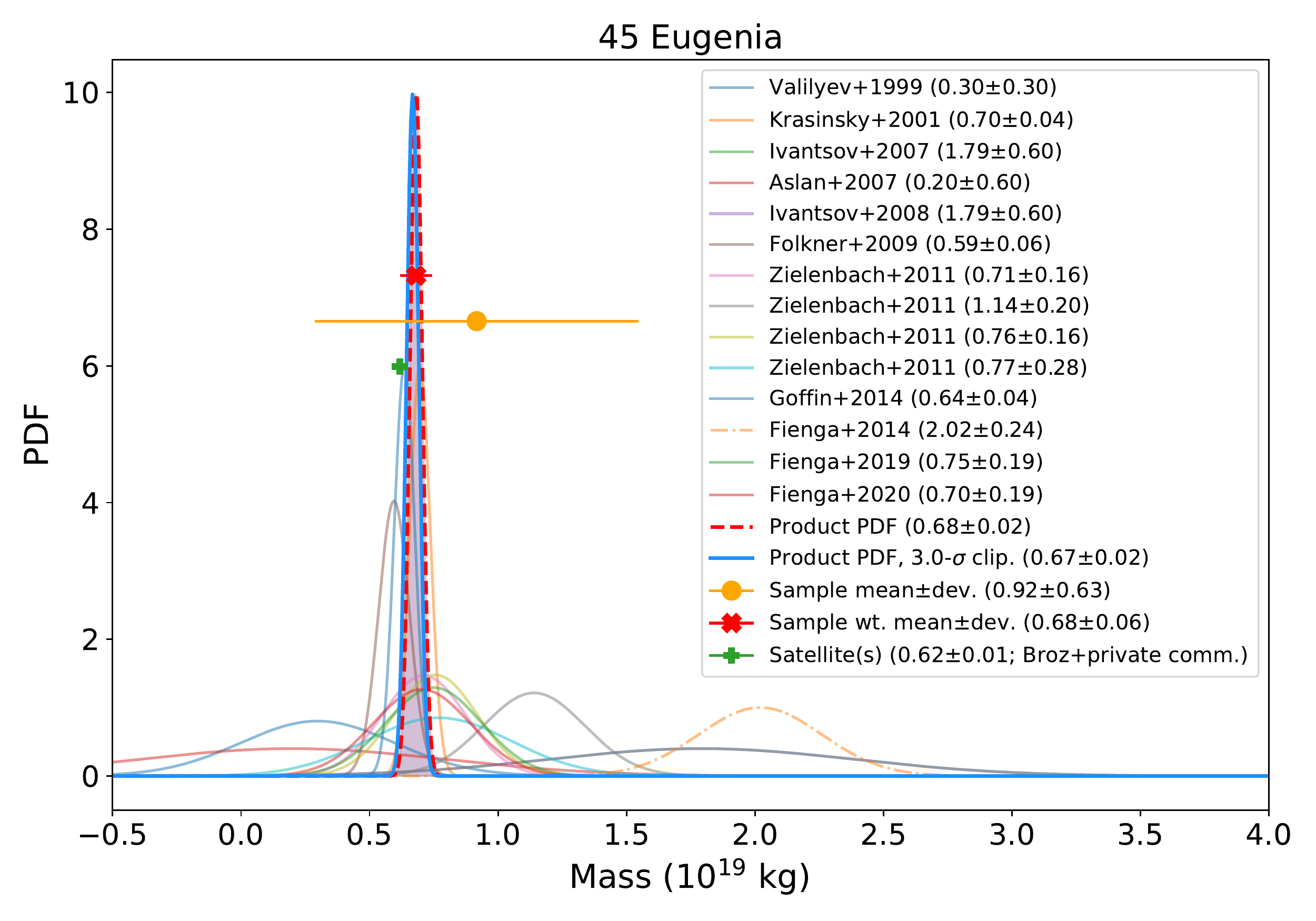}
   \includegraphics[width=0.495\hsize, trim={0 0 0 0}, clip]{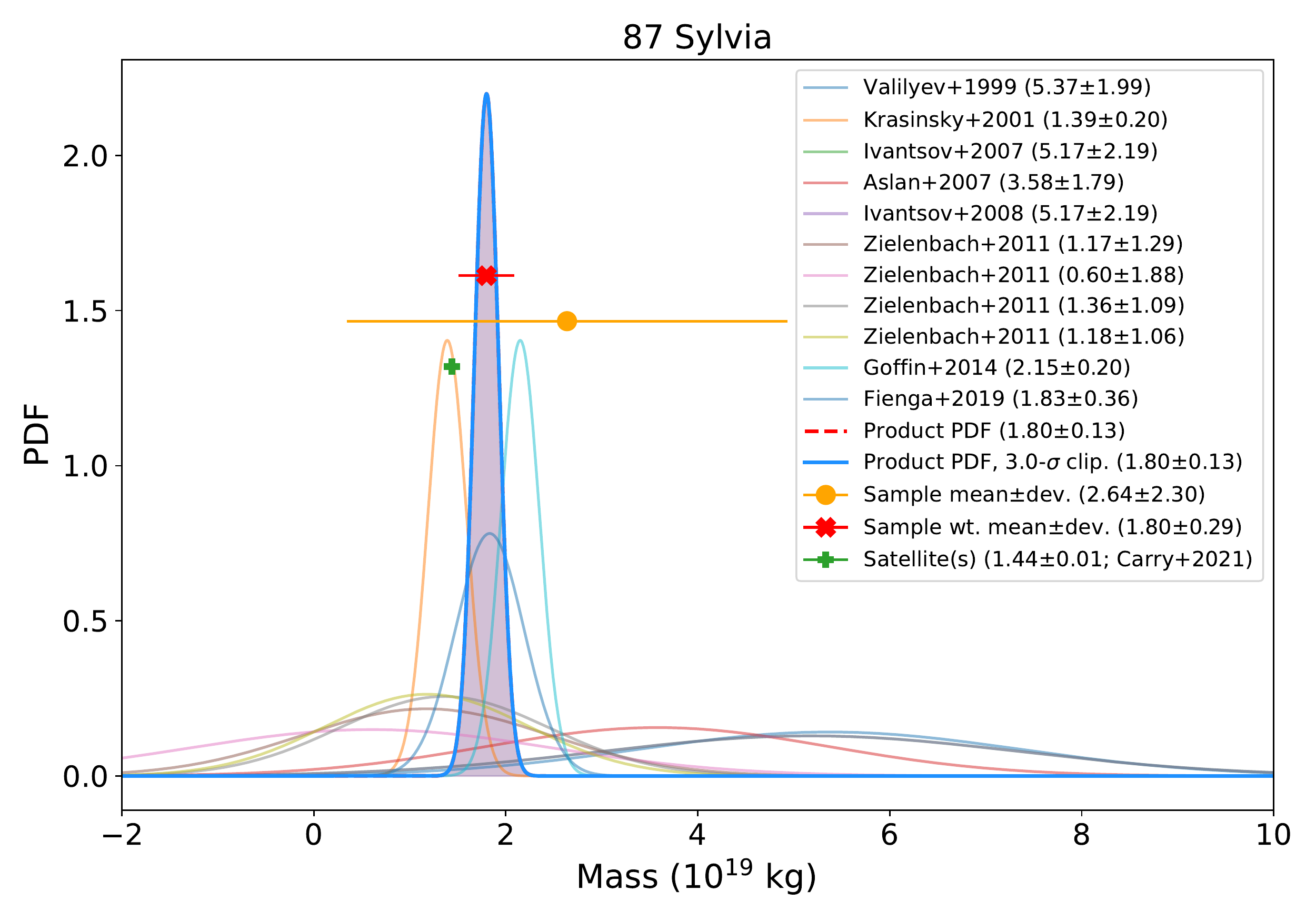}
   \includegraphics[width=0.495\hsize, trim={0 0 0 0}, clip]{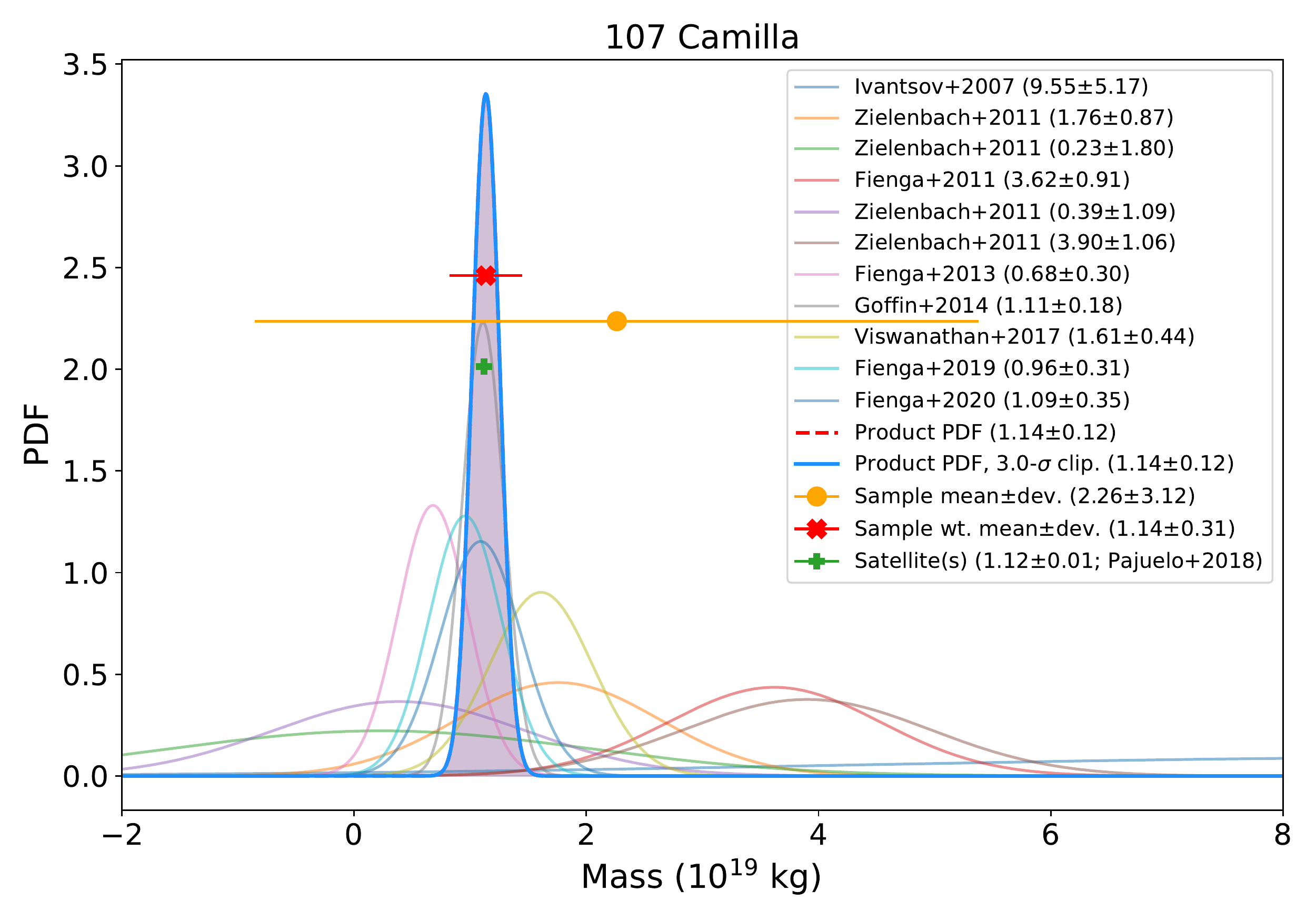}
      \caption{Same as Fig.\,\ref{fig:masses} but for large asteroids harbouring one or two satellites. The green circle corresponds to the asteroid mass derived from the orbital study of the companion(s). The average agreement between this method and the {sigma-clipped product PDF} 
      is 12\% of the asteroid mass.}
         \label{fig:mass_binaries}
   \end{figure*}
%%%%%%%%%% End Figure %%%%%%%%%%%%%%%%%%%%%%%%

%%%%%%%%%% Start Figure %%%%%%%%%%%%%%%%%%%%%%%%
   \begin{figure*}
   \ContinuedFloat
   \centering
   \includegraphics[width=0.495\hsize, trim={0 0 0 0}, clip]{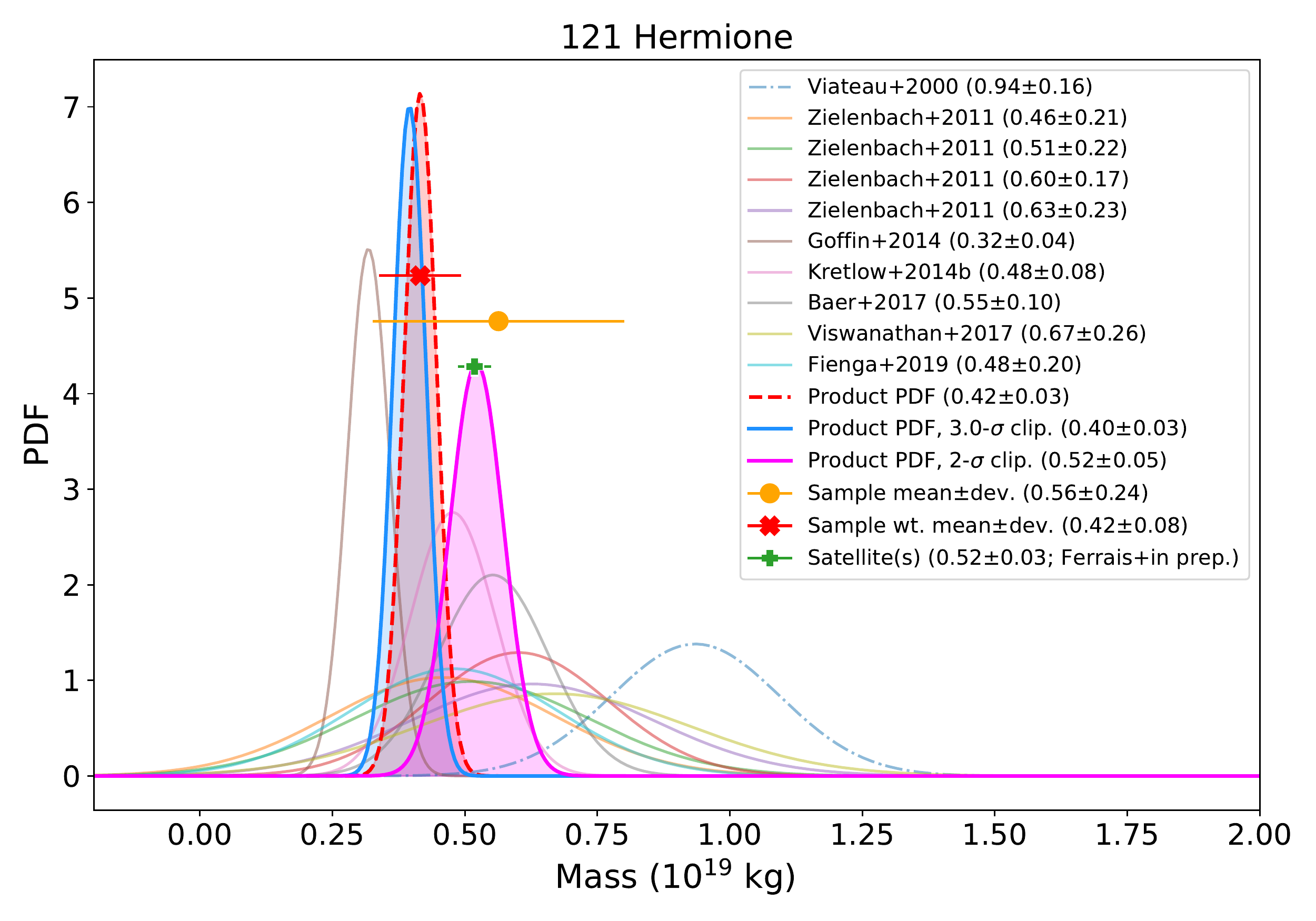}
   \includegraphics[width=0.495\hsize, trim={0 0 0 0}, clip]{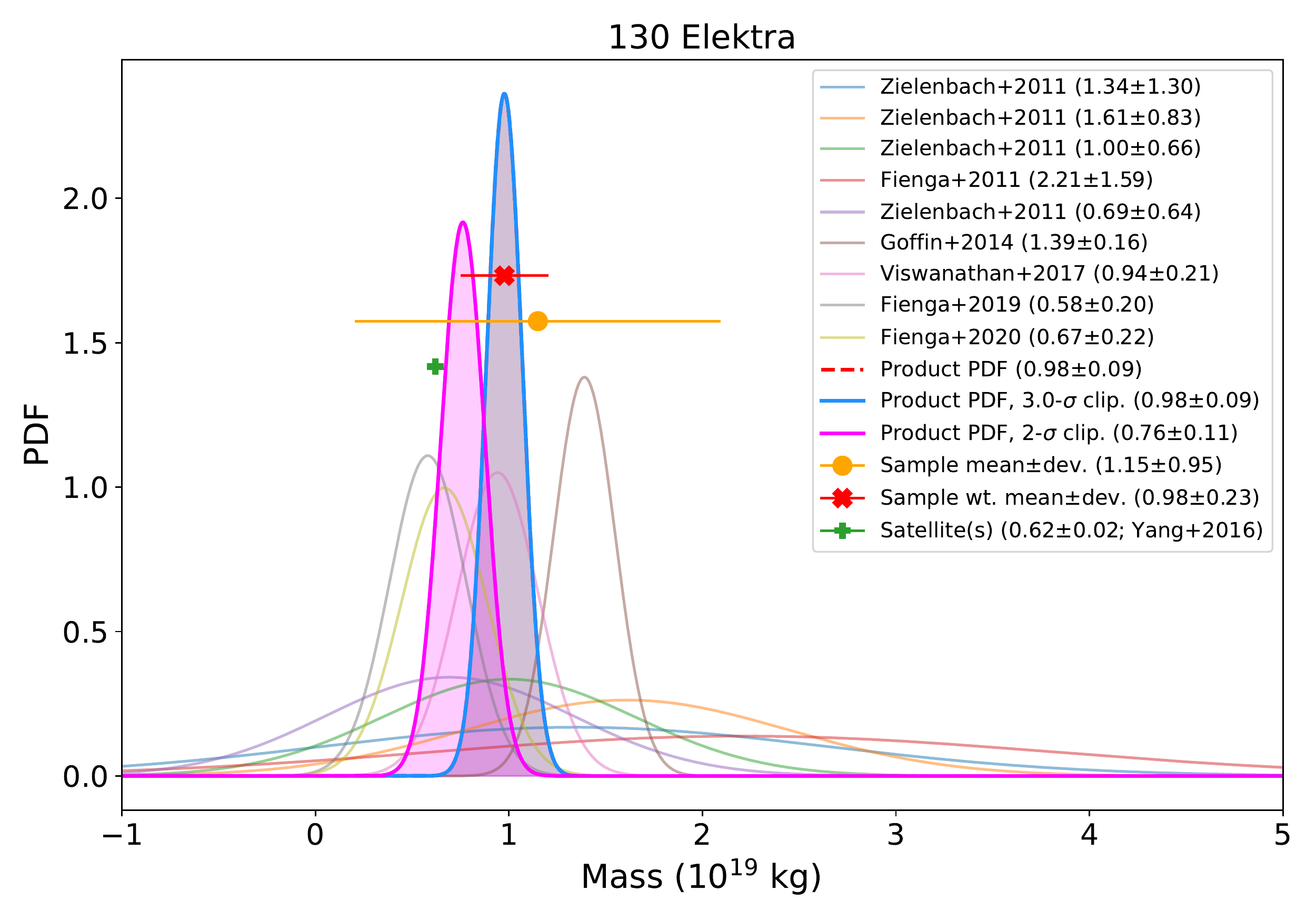}
      \caption{\textit{(continued)}}
         \label{fig:mass_binaries}
   \end{figure*}
%%%%%%%%%% End Figure %%%%%%%%%%%%%%%%%%%%%%%%

\section{Complementary information about the \sage~model}
\label{sec:appC}

\begin{figure*}
\centering
\includegraphics[width=9cm]{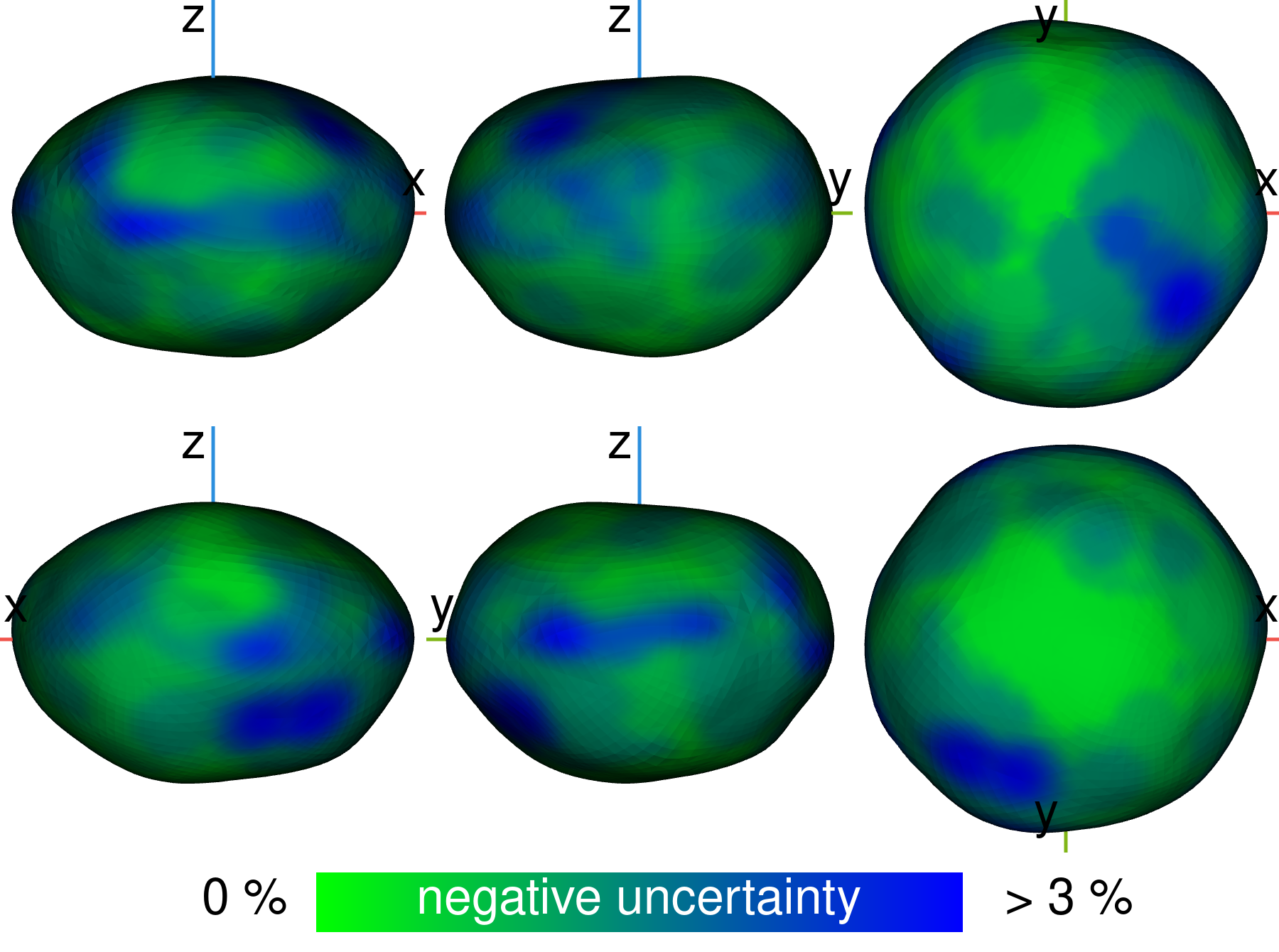}
\includegraphics[width=9cm]{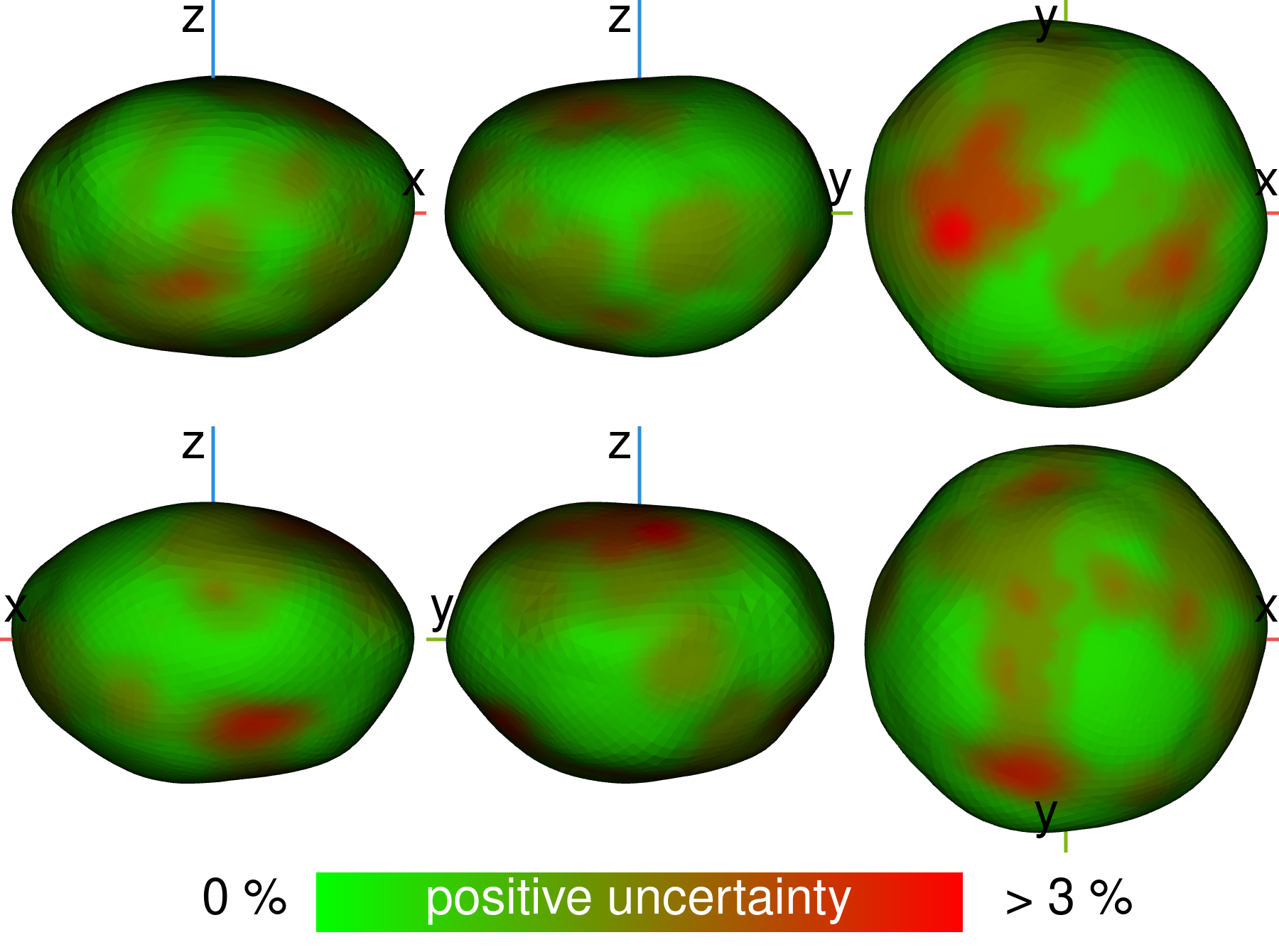}
\caption{Projections of (65) Cybele SAGE model with colours representing
upper (red) and lower (blue) local surface uncertainties. The colours correspond to the level of deviation of a given vertex from the nominal position in the clone population (Section\,\ref{sec:shape}).}
\label{sage_proj}
\end{figure*}

In Fig.\,\ref{sage_proj}, we show the equatorial and polar projections of the \sage{} model of (65)~Cybele with colours representing local surface uncertainties. 
The overall size uncertainty was obtained by fitting projections of the population of accepted clones (Section\,\ref{sec:shape}) to individual AO images. 
Specifically, the equivalent sphere diameter $D_{i,c}$ of each clone was obtained by comparing each $i$-th image from the complete set $I$, taken under an aspect angle $\xi_i$ with an angular resolution $\delta_i$, to the corresponding projection of the clone. Once all clones were compared to the $i$-th image, a range of diameters between $D_{i}^{min}$ and $D_{i}^{max}$ was derived.

\begin{figure*}
\centering
\includegraphics[width=9cm]{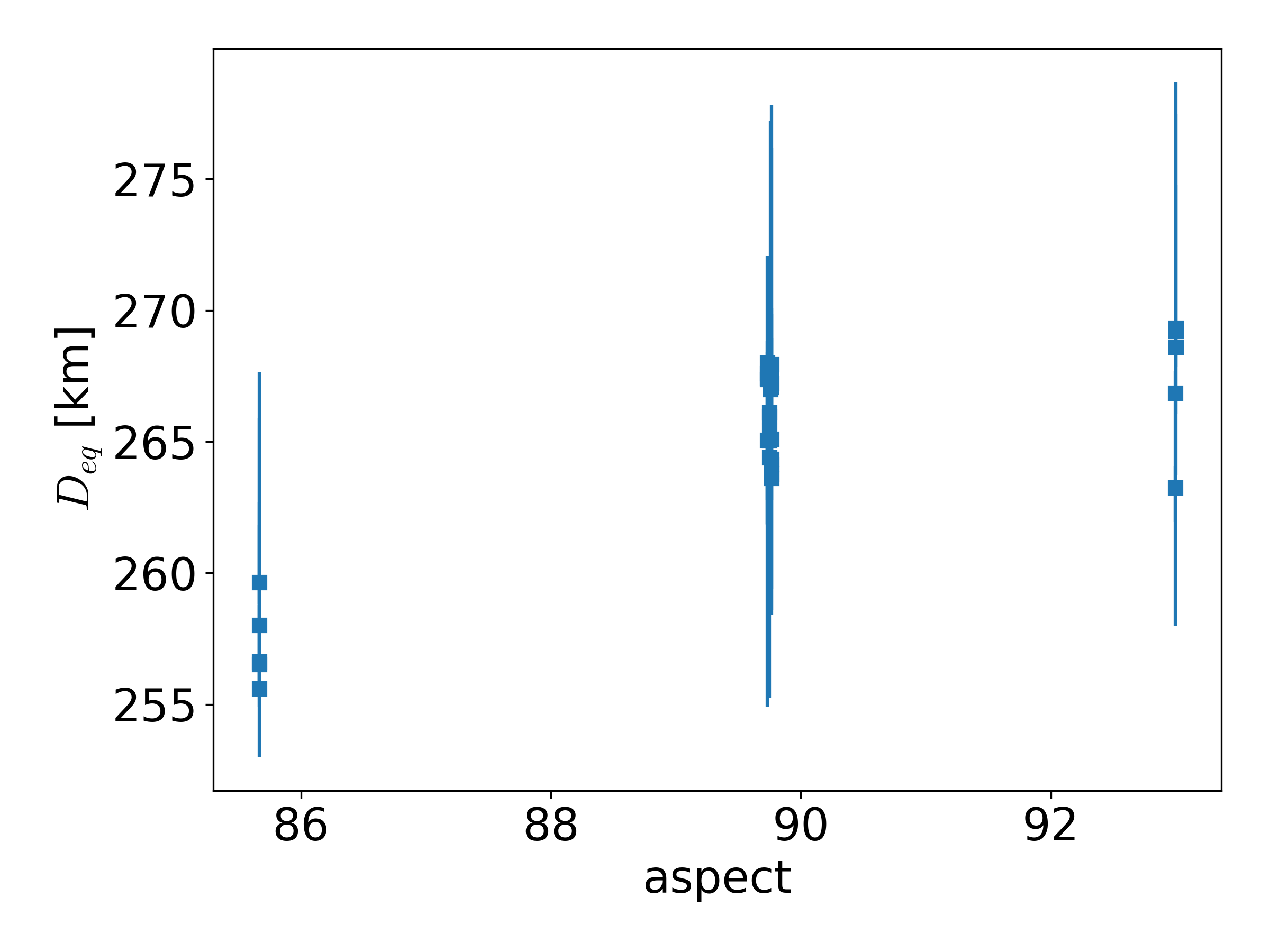}
\caption{Measured diameters with uncertainties obtained for the \sage{} model of (65)~Cybele by contour fitting the AO images, as a function of aspect angle.}
\label{sage_diam_aspect}
\end{figure*}

The final diameter $D$ was calculated by grouping images into subsets with similar values of aspect angles $\Xi_j$. In the case of Cybele, we established three subsets: $\Xi_1 = [80^{\circ},87^{\circ}]$, $\Xi_2 = [89^{\circ}, 91^{\circ}]$, $\Xi_3 = [92^{\circ}, 94^{\circ}]$ (see Fig.\,\ref{sage_diam_aspect}). For each subset of images $I_j$  
(where the index $j$ indicates that the images have aspect angles from the set $\Xi_j$), the weighted average $D_j$ was computed such that:
\begin{equation}
    D_j=\frac{\sum_{i}1/\delta_{i} D_{i}}{\sum_{i}1/\delta_{i}}, \text{\,\hspace{1cm}where\,\hspace{1cm}}  \xi_{i}\in\Xi_{j}
\label{eq1}
.\end{equation}
Finally, to get diameter $D$, another average was computed such that:
\begin{equation}
    D=\frac{\sum_{j}1/\overline{\delta_{j}} D_{j}}{\sum_{j}1/\overline{\delta_{j}}},
\end{equation}
where $\overline{\delta_j}$ is the average image resolution in the subset $I_j$ . By computing $D_i$ = $D_{i,nom}$ in Eq.~\ref{eq1}, we obtain the nominal diameter value. When performing calculations for all of the clones ($D_i = D_{i,c}$), we obtain a set of diameters from which error bars can be extracted, i.e. the minimum $D^{min}$ and maximum $D^{max}$ values in the set. 

\section{Image residuals}
\label{sec:appB}

\begin{figure*}%[!t]
\begin{center}
\resizebox{.98\hsize}{!}{\includegraphics{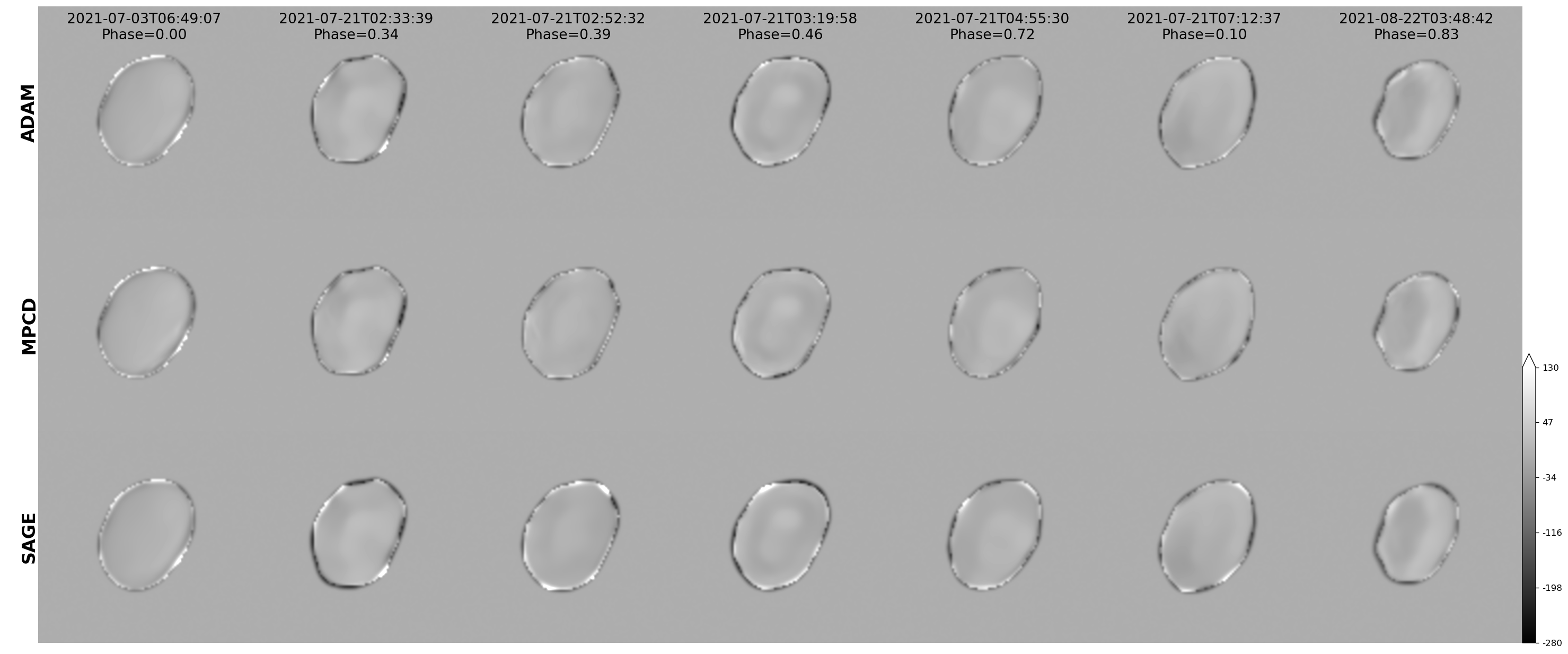}}
\end{center}
\caption{\label{fig:mosaic_res} Residuals in units of instrumental noise between the observed SPHERE images and the corresponding synthetic images from the three shape models.}
\end{figure*}

Residual images between the observed and synthetic images of (65)~Cybele are presented in Fig.~\ref{fig:mosaic_res} for the three shape models of Cybele presented in this work. 

\section{Elevation map}
\label{sec:elev}

\begin{figure*}%[!t]
\begin{center}
\resizebox{.9\hsize}{!}{\includegraphics{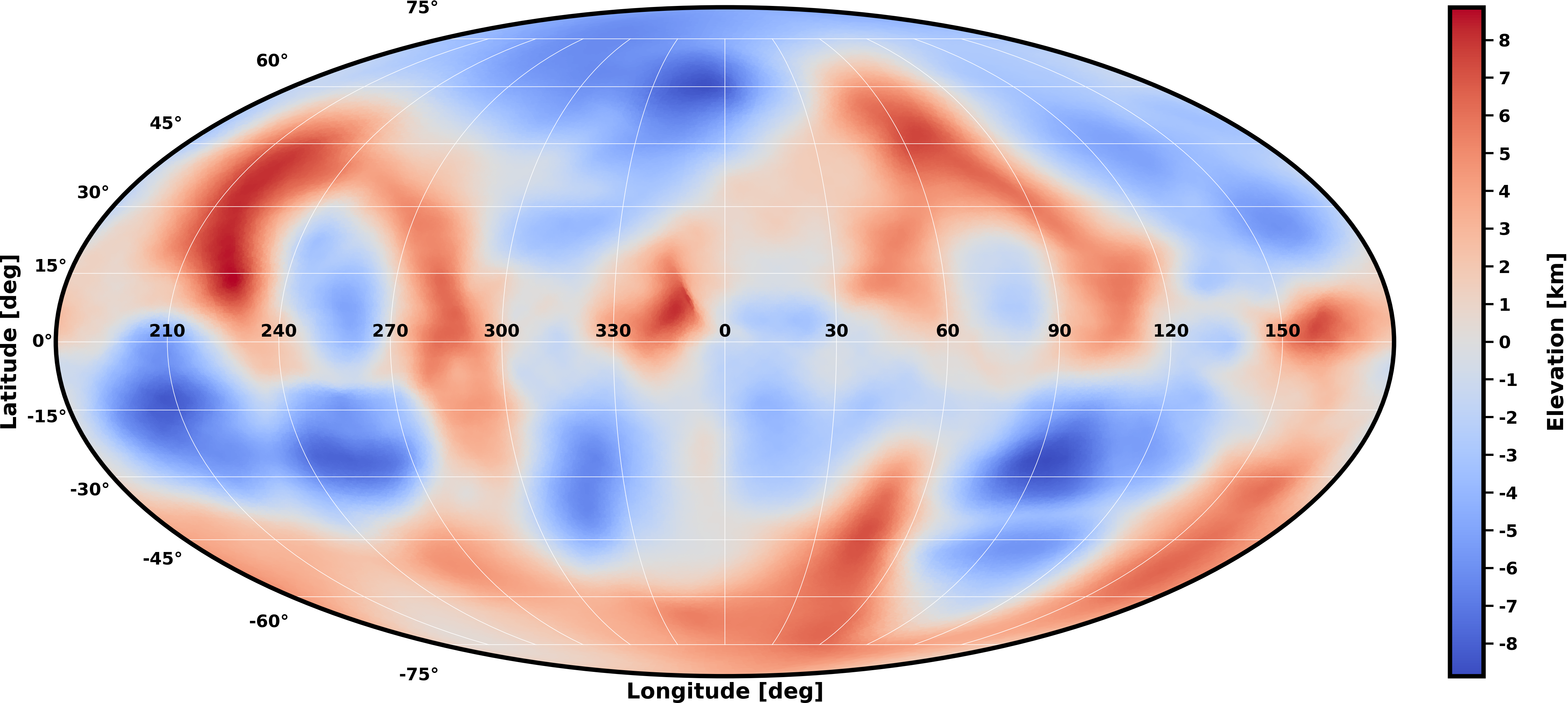}}
\end{center}
\caption{\label{fig:elev} Elevation map of (65)~Cybele, calculated as the local radius of the \adam{} shape model, minus the radius of a best-fit ellipsoid.}
\end{figure*}

In Fig.~\ref{fig:elev}, we present the elevation map of (65)~Cybele obtained by subtracting a best-fit ellipsoid to the \adam{} shape model. The map highlights the hemispherical asymmetry of the asteroid, with the north pole being flatter than the south pole. Additional variations in elevation may be due to impact features.

\section{Rotation state}
\label{sec:w_L}

\begin{figure*}
\centering
\includegraphics[width=0.5\hsize, trim={3.8cm 1.8cm 3.5cm 1.4cm}, clip]{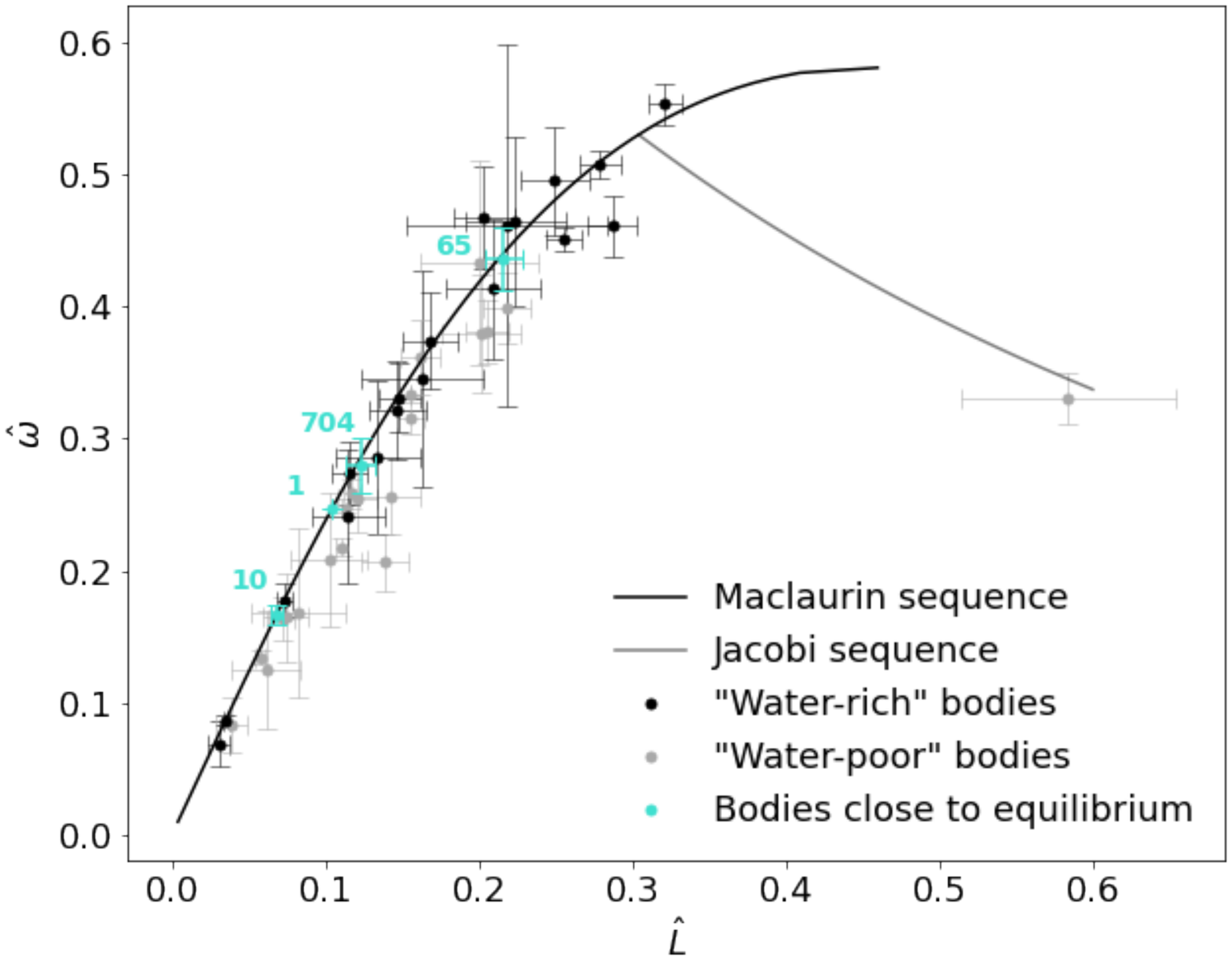}
\caption{Normalised angular velocity ($\hat{\omega}$) versus specific angular momentum ({\it \^L}) of asteroids observed with VLT/SPHERE \citep{Vernazza:2021}, including (65)~Cybele (this work). Expected sequences for Maclaurin (black line) and Jacobi (grey line) ellipsoids are shown for comparison. 
Objects close to an equilibrium shape: (1)~Ceres, (10)~Hygiea \citep{Vernazza:2020}, (65)~Cybele (this work) and (704)~Interamnia \citep{Hanus:2020}, are highlighted in light blue. With the exception of (216)~Kleopatra (the only data point located at the right-hand side of the plot; \citealt{Marchis:2021}), all of the objects are close to the Maclaurin sequence in this parameter space.}
\label{fig:w_L}
\end{figure*}

In Fig.\,\ref{fig:w_L}, we compare the specific angular momentum and the normalised angular velocity of Cybele and additional asteroids observed with SPHERE with the expected values for Maclaurin and Jacobi ellipsoids following \citet{Vernazza:2021}.

%%%%%%%%%% Start Figure %%%%%%%%%%%%%%%%%%%%%%%%
%%%%%%%%%% End Figure %%%%%%%%%%%%%%%%%%%%%%%%

\end{appendix}

\end{document}